\documentclass[final,authoryear,3p,times]{elsarticle}
\usepackage{graphicx,color,colortbl}
\usepackage{enumerate}
\usepackage[colorlinks]{hyperref}
\usepackage{subfigure}
\usepackage{booktabs}
\usepackage{amsmath,amssymb,amsthm,bm}
\usepackage{mathtools}
\usepackage[dvipsnames]{xcolor}
\usepackage{ulem}
\usepackage{tikz}
\usetikzlibrary{automata,positioning}
\usepackage{afterpage}
\usepackage{calc}% http://ctan.org/pkg/calc
\usepackage{ifthen}
\usepackage[ruled, vlined, linesnumbered]{algorithm2e}
\usepackage{algpseudocode}
\usepackage{caption}
\usepackage{siunitx}
\usepackage{booktabs}
\usepackage{placeins}
\usepackage{float}
\usepackage{verbatim}
\usepackage{cancel}

\usepackage{bigfoot}
\DeclareNewFootnote[para]{default}

\usepackage{multirow}

\edef\svtheparindent{\the\parindent}
\usepackage{parskip}
\parindent=\svtheparindent\relax

\newcommand*\argumentbullet{\bullet}

\newlength{\boxwidth}
\def\btheorem{\begin{theorem}}
\def\etheorem{\end{theorem}}
\def\blemma{\begin{lemma}}
\def\elemma{\end{lemma}}
\def\bproposition{\begin{proposition}}
\def\eproposition{\end{proposition}}
\def\bcorollary{\begin{corollary}}
\def\ecorollary{\end{corollary}}
\def\bdefinition{\begin{definition}}
\def\edefinition{\end{definition}}
\def\bexample{\begin{example}}
\def\eexample{\end{example}}
\def\bremark{\begin{remark}}
\def\eremark{\end{remark}}

\newcommand{\be}{\begin{equation}}
\newcommand{\ee}{\end{equation}}
\newcommand{\beq}{\begin{eqnarray}}
\newcommand{\eeq}{\end{eqnarray}}
\newcommand{\bem}{\begin{multline}}
\newcommand{\eem}{\end{multline}}
\newcommand{\ba}{\begin{align}}
\newcommand{\ea}{\end{align}}

\journal{CMAME}

\begin{document}

\begin{frontmatter}

\title{Enhanced Floating Isogeometric Analysis\\}

\author[eth]{Helge C. Hille}
\author[tud]{Siddhant Kumar}
\author[eth]{Laura De Lorenzis\corref{cor1}}

\cortext[cor1]{Correspondence: ldelorenzis@ethz.ch}

\address[eth]{Department of Mechanical and Process Engineering, ETH Z\"{u}rich, 8092 Z\"{u}rich, Switzerland}
\address[tud]{Department of Materials Science and Engineering, Delft University of Technology, 2628 CD Delft, The Netherlands}

\begin{abstract}
The numerical simulation of additive manufacturing techniques promises the acceleration of costly experimental procedures to identify suitable process parameters. We recently proposed Floating Isogeometric Analysis (FLIGA), a new computational solid mechanics approach, which is mesh distortion-free in one characteristic spatial direction. FLIGA emanates from Isogeometric Analysis and its key novel aspect is the concept of deformation-dependent ``floating'' of individual B-spline basis functions along one parametric axis of the mesh. Our previous work showed that FLIGA not only overcomes the problem of mesh distortion associated to this direction, but is also ideally compatible with material point integration and enjoys a stability similar to that of conventional Lagrangian mesh-based methods. These features make the method applicable to the simulation of large deformation problems with history-dependent constitutive behavior, such as additive manufacturing based on polymer extrusion. In this work, we enhance the first version of FLIGA by \textit{(i)} a novel quadrature scheme which further improves the robustness against mesh distortion, \textit{(ii)} a procedure to automatically regulate floating of the basis functions (as opposed to the manual procedure of the first version), and \textit{(iii)} an adaptive refinement strategy. We demonstrate the performance of enhanced FLIGA on relevant numerical examples including a selection of viscoelastic extrusion problems.
\end{abstract}


\end{frontmatter}

%%%%%%%%%%%%%%%%%%%%%%%%%%%%%%%%%%%%%%%%%%%%%%%%%%%%%%%%%%%%%%%%%%%%%%%%%%%%%%%%%%%%%%%%%%%%%%%%%%%%%%%%%%%%%%%%%%%%%%%

\section{Introduction}\label{sec:Introduction}
Additive manufacturing (AM) is a wide-ranging class of modern fabrication technologies in which material is successively added to build the final part, see \cite{GIBSON2020} for an introduction to the field. Among others, a prominent concept for the controlled addition of material is polymer extrusion. Polymer extrusion has long been a state-of-the-art technique also for the manufacturing of plastic profiles, where molten polymer is forced through a die to create a continuous strand. In its application to AM, on a smaller scale, geometrically simple strands are deposited under movement of a heated robotic nozzle, thereby building products by adding strand to strand, and then layer to layer. The technique is especially attractive as it facilitates the customized fabrication of parts with complex topology and/or multi-material arrangement. However, the quality of the solidified end product is heavily dependent on the suitable choice of the process parameters, which is often specific to the individual printing task. This motivates the need for process simulations, to avoid costly and time-consuming trial-and-error experimental procedures.

The Lagrangian and Eulerian viewpoints of motion in continuum mechanics are usually preferred to describe solid and fluid behavior, respectively. When it comes to the numerical approximation, the finite element method (FEM) and Isogeometric Analysis (IGA) \citep{HUGHES20054135} are often combined with the Lagrangian viewpoint for the discretization of solid mechanics problems. Lagrangian FEM is highly sensitive to mesh distortion for very large deformations and therefore inapplicable to the simulation of fluid behavior. While showing a higher degree of robustness \citep{LIPTON2010357}, standard (Lagrangian) IGA also suffers from the same issue. Eulerian formulations of FEM or IGA, or beyond, are inherently free from mesh distortion, but typically come along with additional efforts since a stabilization of the advective terms as well as a special treatment of moving boundaries are required, see e.g. \cite{AKKERMAN20114137}. A challenge arises for problems involving both very large (fluid-like) deformations \textit{and} solid behavior. Polymer extrusion is one of such cases, as the viscoelastic deformation response of most polymers lies somewhere in between fluid and solid behavior. In our previous paper \citep{HKDL} we discussed the available computational techniques to handle such cases; as follows, we only mention them very briefly along with a few areas requiring further improvements. The re-generation of undistorted meshes, also known as remeshing, is computationally expensive and its success is highly dependent on suitable techniques to control the errors of history data projection \citep{WIECKOWSKI20044417}. Adoption of remeshing in IGA is complex and a subject of ongoing research \citep{Fueder2015IsogeometricFE, Shamanskiy2020, HENNIG2020}. Though avoiding costly mesh generations, most meshless methods \citep{BELYTSCHKO19963,Chen2017,PASETTO2021} are still considered less efficient than their mesh-based competitors \citep{XIAO2020,Kumar2020nme,MATSUDA2022} and stabilization plays an important role in case of extreme deformations \citep{BELYTSCHKO2000,KUMAR2019858}. In many cases, accurate quadrature of meshless basis functions is challenging; moreover, for incompressible material behavior which is relevant for polymers, the construction of mixed discretizations to avoid locking is not trivial \citep{GOH2018575}. The well-founded theory of mesh-based analysis and the flexibility of meshless methods have motivated some hybrid approaches, aiming to benefit from the complementing advantages of both settings, see e.g. \cite{ArroyoOrtiz2006,ROSOLEN201395, CARDOSO2014, MILLAN2015712, FATHI2021}. Finally, mesh-based approaches combining Lagrangian and Eulerian concepts have also been devised \citep{HIRT1974, HU1996}, but with increased methodical complexity, which remains again a limitation to their further establishment \citep{AUBRAM2015}.

Beyond the original goal to circumvent the expensive mesh generation in FEM, the adoption of basis functions stemming from Computer Aided Design (CAD) in IGA \citep{HUGHES20054135} proved beneficial for the analysis itself in many applications. Mostly due to the higher and tunable continuity of isogeometric basis functions at element boundaries, IGA was shown to lead to advantages in e.g., structural dynamics \citep{COTTRELL20065257}, the solution of higher-order partial differential equations \citep{GOMEZ20084333}, and contact mechanics \citep{DeLorenzis2014}, among many others. As mentioned, like FEM, Lagrangian IGA also suffers from mesh distortion; the basis functions are constructed first on a reference domain and a distorted mapping to the physical space in case of large deformations negatively affects the numerical solution \citep{LIPTON2010357}. Yet, there is one significant difference between FEM and IGA. While the reference (parent) domain in FEM collects a few basis functions which are mapped many times to different \textit{elements}, the reference (parametric) domain in IGA collects many basis functions and is mapped only a few times to different \textit{patches}. In IGA, a patch is again subdivided into (so-called Bézier) elements, so that the construction of the basis functions can be alternatively viewed as \textit{local} or \textit{elementwise} (like in FEM) or \textit{global} or \textit{patchwise} (an alternative typical of IGA). While the first perspective was adopted in the isogeometric counterpart of FEM remeshing techniques \citep{Fueder2015IsogeometricFE, Shamanskiy2020}, the alternative one can be considered the underlying viewpoint of the new strategy to overcome mesh distortion that we recently proposed and denoted as Floating IGA (FLIGA) \citep{HKDL}.

\begin{figure}[t!]
\centering
\includegraphics[trim={6.25cm 4.15cm 4.75cm 2.75cm},clip,width=0.97\textwidth]{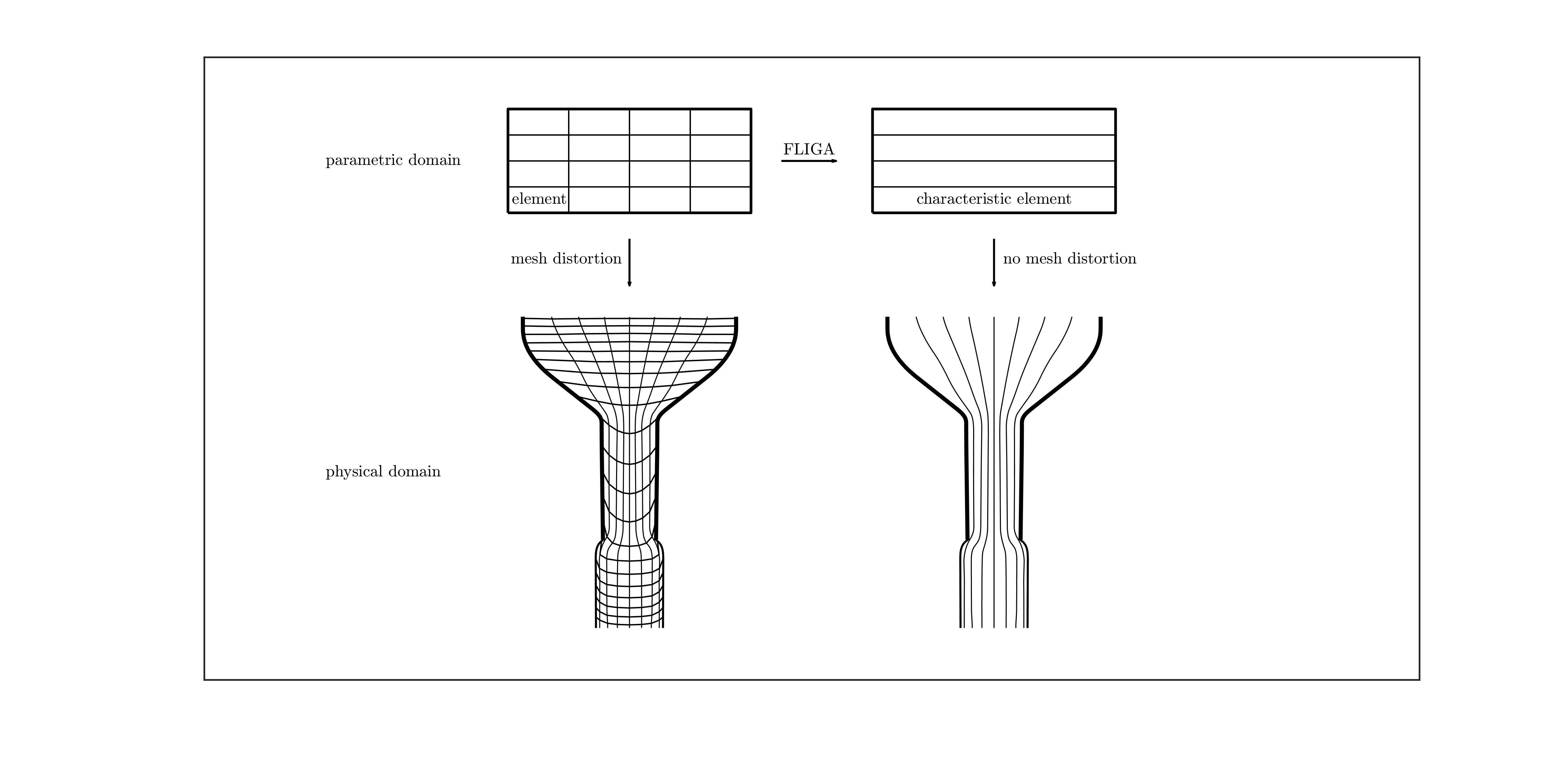}
\caption{Element concepts in IGA and FLIGA.}
\label{fig:elementviewpoint}
\end{figure}

\begin{figure}[t!]
\centering
\includegraphics[trim={0cm 0cm 0cm 0.5cm},clip,width=10cm]{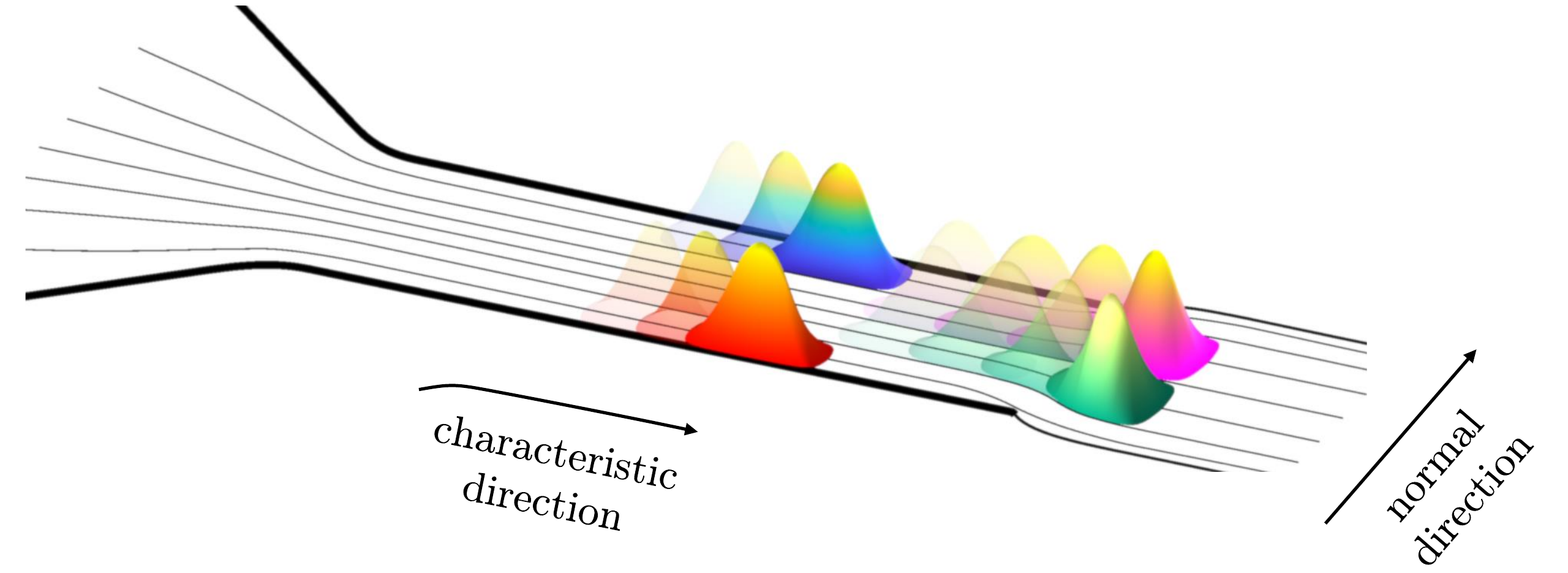}
\caption{Exemplary ``floating'' movement of four basis functions along the characteristic elements in the physical space.}
\label{fig:floatingillustration}
\end{figure}

FLIGA approaches the numerical solution of a Lagrangian model from a global (patchwise) perspective. An important starting assumption is that, in the problem at hand, extreme deformations occur along one (possibly curved) physical axis, which can be mapped to one corresponding parametric direction, denoted as the \textit{characteristic direction}. The key idea is to comprise all Bézier elements along the characteristic direction in a so-called \textit{characteristic element}, as illustrated in \autoref{fig:elementviewpoint} on the exemplary geometry of a material extrusion problem, while preserving local support of its basis functions. The abolition of Bézier element boundaries in such direction confers a meshless character to the basis functions limited to this direction, while the mesh is preserved in the other parametric direction (also denoted as \textit{normal direction}). The problem of mesh distortion along the characteristic elements is then solved by the mere update of the corresponding univariate B-Spline basis functions. The movement of basis functions that we denote as \textit{floating} is schematically illustrated in \autoref{fig:floatingillustration} for the extrusion example and provides the basis for a new class of boundary-fitted spline approximation spaces, which are distortion-free along one parametric direction.

In our recent paper \citep{HKDL}, we developed FLIGA following the isoparametric concept and demonstrated its performance on the Taylor-Couette flow problem and on a polymer extrusion problem, both featuring viscoelastic material behavior and extreme deformations. In this paper, we aim at improving upon three aspects of the initial formulation. Firstly, in \cite{HKDL} we could solve the Taylor-Couette flow problem up to over five full turns of the cylinder, after which we observed a rapid increase of the numerical error. We suspected the reason to be related to the quadrature scheme, i.e. to the accumulation of quadrature error. Secondly, floating of the basis functions was regulated by the movement of the so-called floating regulation points, which was connected to the control point displacements through a manually design level-set function. This manual design was based on a qualitatively reasonable \textit{a priori} guess about the displacements, which was not too difficult to obtain for the considered problems but may easily become impractical or even impossible in more general cases. Finally, while FLIGA solved the issue of shear-driven distortion, the floating of one basis function was still associated to its elongational deformation. To prevent issues stemming from an excessive dilation for severe floating, we used very fine meshes, which led to a high computational cost.

In this paper, we address these three points and propose three corresponding enhancements. For the first problem, a new numerical quadrature concept is introduced to improve the numerical stability. The second issue is solved via an automated procedure for floating regulation based on the \textit{actual} (and not on an \textit{a priori} guessed) deformation. To address the third issue, we devise a simple adaptive refinement strategy for FLIGA, which reduces the computational cost.
% In fact, FLIGA endowed with such strategy may be considered as an alternative technique to achieve local refinement in IGA even for problems where extreme deformations do not necessarily appear.
A number of numerical examples demonstrate the effect of these enhancements.

The remainder of this paper is organized as follows: In \autoref{sec:bsplines}, after reviewing the concept of floating basis functions, we illustrate the three proposed enhancements. In \autoref{sec:analysis}, we focus on the Lagrangian modeling of viscoelasticity and its standard IGA discretization, and eventually propose the enhanced version of FLIGA incorporating the new approaches at the previous section. Finally, \autoref{sec:results} discusses the numerical examples and conclusions are drawn in \autoref{sec:conclusions}.

\section{Floating B-Splines}\label{sec:bsplines}

At the heart of FLIGA are \textit{floating} B-Splines, which we proposed in \cite{HKDL} as a generalization of conventional B-Splines \citep{Schoenberg1946, PiegTill96}. %The standard B-Spline theory that has evolved since (see \cite{PiegTill96} for a summary) is fundamental also to the new basis functions.%\footnote{In our current estimation, all the reported concepts should be extendable to NURBS-based designs too (though not yet tested).} 
In this section, we illustrate the design of floating B-Splines in a further generalized fashion compared to \cite{HKDL} and we introduce some new operations on them. These concepts are later applied to analysis in \autoref{ssec:FLIGA}, which outlines the enhanced version of FLIGA.

\subsection{Basis construction}
\label{ssec:construction}
A univariate standard B-Spline basis is constructed on a parametric domain with coordinate $\xi$, here $\hat{\Omega}_{\xi}=[0,1]$, by means of the Cox-de Boor recursion formula \citep{COX1972,DEBOOR1972}
\begin{align}
\begin{split}
p&=0:\quad \hat{N}_{i,0}\left(\xi\right)=
\begin{cases}
1, & \textrm{for}\ \xi_{i} \leq \xi < \xi_{i+1},\\
0, & \textrm{otherwise},\\
\end{cases}\\
p&\geq 1:\quad \hat{N}_{i,p}\left(\xi\right)=\frac{\xi-\xi_i}{\xi_{i+p}-\xi_i}\hat{N}_{i,p-1}\left(\xi\right)+\frac{\xi_{i+p+1}-\xi}{\xi_{i+p+1}-\xi_{i+1}}\hat{N}_{i+1,p-1}\left(\xi\right),
\end{split}
\label{eq:coxdeboor}
\end{align}
where we define $\frac{0}{0}=0$. The non-decreasing sequence of real numbers denoted as  knots, $\xi_i$, constitutes the knot vector $\Xi$ and if the first (last) $p+1$ knots are all $0$ ($1$) we call the knot vector ``open''. Here $p$ is the polynomial order of the B-Spline basis, which tunes its continuity at the knots. At a unique (not repeated) knot, continuity is $\mathcal{C}^{p-1}$; each repetition of a knot decreases the order of continuity at the knot by one; between the knots, continuity is $\mathcal{C}^{\infty}$. In this paper, we always use open knot vectors and unique inner knots. Each knot span is also denoted as an \textit{element}.

\begin{figure}[t!]
\centering
\includegraphics[trim={4cm 2.5cm 4cm 2.5cm},clip,width=0.7\textwidth]{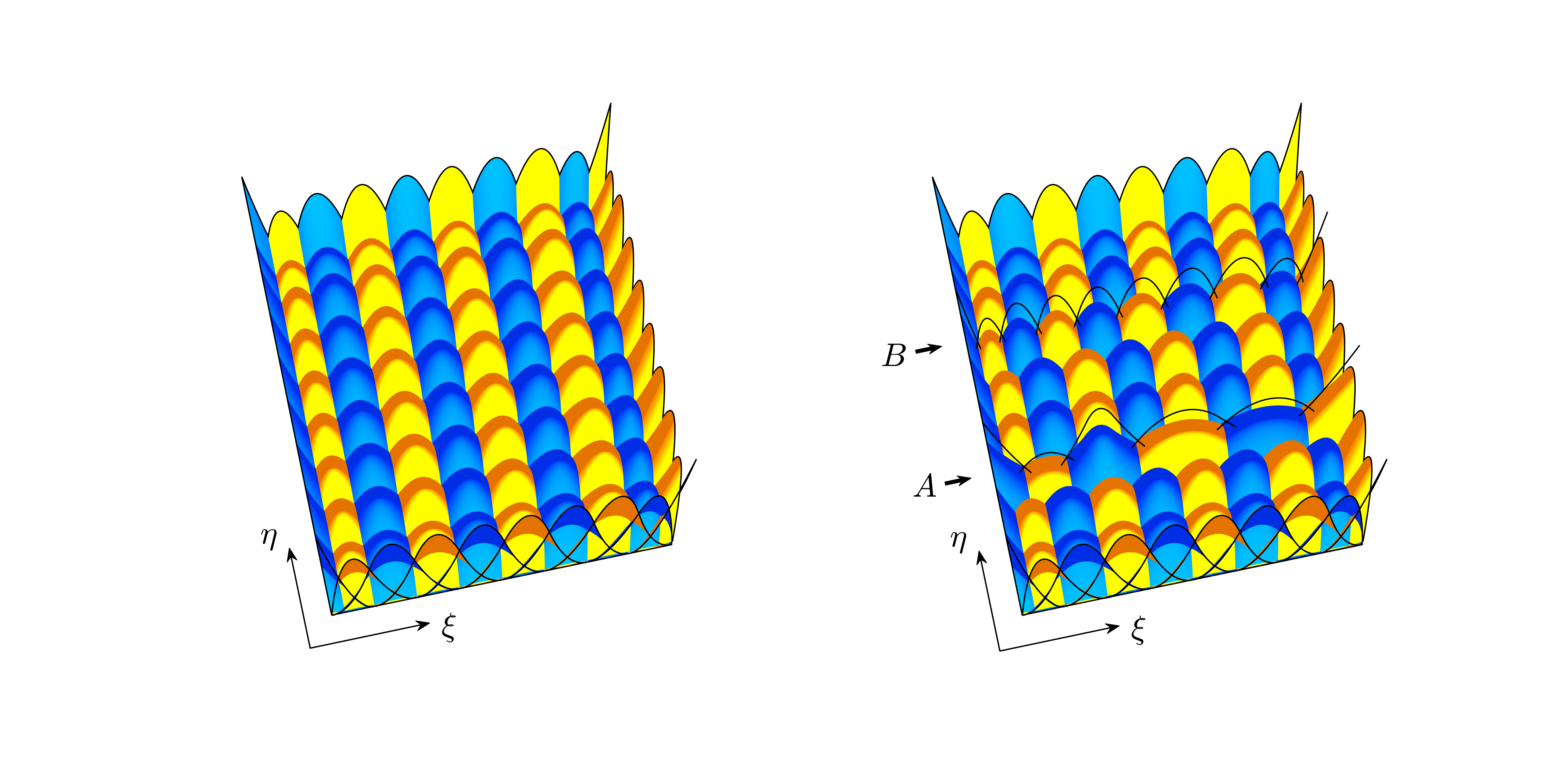}
\caption{Examples of classical (left) and floating (right) tensor product structure.}
\label{fig:tensorproducts}
\end{figure}

Let us now define two different types of bivariate B-spline bases (trivariate bases are not treated in this paper). To this end, a patch is spanned on a square parametric domain with coordinates $\xi$ and $\eta$, here $\hat{\Omega}=\hat{\Omega}_{\xi}\times\hat{\Omega}_{\eta}=[0, 1]\times[0, 1]$. The first type of bivariate B-Spline basis has the classical tensor product structure (the one used in IGA); this is obtained by combining one univariate basis in $\eta$ with only \textit{one} univariate basis in $\xi$, i.e.
\begin{equation}
\hat{B}^{TP}_{ij}\left(\xi,\eta\right)=\hat{N}_{i}\left(\xi\right)\hat{M}_{j}\left(\eta\right)\qquad i=1,...,I; j=1,...,J,
\label{eq:classicaltensorproduct}
\end{equation}
with $I$ and $J$ as the total number of basis functions in the two parametric directions. The basis $\left\lbrace\hat{M}_{j}\right\rbrace_{j=1,...,J}$ is constructed similarly to basis $\left\lbrace\hat{N}_{i}\right\rbrace_{i=1,...,I}$ knowing the order $q$ and the knot vector $H$ along coordinate $\eta$. The second type of bivariate B-Spline basis has the \textit{floating} tensor product structure (the one used in FLIGA); here we associate a different univariate basis in $\xi$ to each $j$
\begin{equation}
\hat{B}_{ij}\left(\xi,\eta\right)=\hat{N}_{ij}\left(\xi\right)\hat{M}_{j}\left(\eta\right)\qquad i=1,...,I_j; j=1,...,J.\\
\label{eq:floatingtensorproduct2}
\end{equation}
Note that no summation over repeated indices is implied in (\autoref{eq:floatingtensorproduct2}) (throughout the paper, to avoid any confusion we will write all summation symbols explicitly). We denote $\left\lbrace\hat{N}_{ij}\right\rbrace_{i=1,...,I_j}$ for each fixed $j=1,...,J$ as \textit{characteristic basis} and $\left\lbrace\hat{M_j}\right\rbrace_{j=1,...,J}$ as \textit{normal basis}. This floating tensor product structure allows us to construct bases as depicted in \autoref{fig:tensorproducts}. In the figure, starting from a classical bivariate tensor product basis we only modify the characteristic bases for $j=A,B$, i.e. $\left\lbrace\hat{N}_{iA}\right\rbrace_{i=1,...,I_A}$ and $\left\lbrace\hat{N}_{iB}\right\rbrace_{i=1,...,I_B}$ for better clarity (in practice, we will typically use different characteristic bases for all $j$). Note that, unlike in our first FLIGA paper \citep{HKDL}, here the number of basis functions along the characteristic direction, $I_j$, can vary for different $j$, a property crucial for local refinement (see \autoref{ssec:refinement}). E.g. in \autoref{fig:tensorproducts} it is $I_A=6$ and $I_B=10$. \autoref{fig:tensorproducts} also shows that the element boundaries of the characteristic bases are not required to stay aligned for the floating tensor product structure. In fact, the element concept has been relaxed and element division is only kept in the $\eta$ direction, associated to the normal basis. This corresponds to the concept of characteristic elements schematized in \autoref{fig:elementviewpoint}.

The next question which arises is how to construct the different characteristic bases in the floating tensor product structure. To this end, for each $j=1,...,J$ we construct a so-called \textit{parent} basis $\left\lbrace\tilde{N}_{ij}\left(\tilde{\xi}\right)\right\rbrace_{i=1,...,I_j}$ with order $p$ from parent knot vector $\tilde{\Xi}_j$ on the parent domain $\tilde{\Omega}=\left[0, 1\right]$ along coordinate $\tilde{\xi}$. Then, for each $j=1,...,J$ we define a scalar mapping $\mathcal{G}_j:\tilde{\Omega}\rightarrow\hat{\Omega}_{\xi}$ which transforms a given $\tilde{\xi}$ to
\begin{equation}
\xi_j\left(\argumentbullet;\mathcal{H}\right)=\mathcal{G}_j\left(\tilde{\xi};\mathcal{H}\right)=\sum^{I_j}_{i=1} h_{ij}\tilde{N}_{ij}\left(\tilde{\xi}\right).
\label{eq:floatingmaps}
\end{equation}
The inverse mapping $\mathcal{G}^{-1}_j: 
\hat{\Omega}_{\xi}\rightarrow \tilde{\Omega}$ transforms back a given $\xi$ to
\begin{equation}
\tilde{\xi}_j\left(\argumentbullet;\mathcal{H}\right)=\mathcal{G}^{-1}_j\left(\xi;\mathcal{H}\right).
\label{eq:inversefloatingmaps}
\end{equation}
We denote $\mathcal{G}_j$ as \textit{floating maps}. The scalar linear coefficients $h_{ij}$ are termed \textit{floating regulation points} and collected in the set
\begin{equation}
\mathcal{H}=\left\lbrace h_{ij}\right\rbrace_{i=1,...,I_j;j=1,...,J}.
\end{equation}
The characteristic basis functions are derived as the push-forwards of the respective parent basis functions
\begin{equation}
\hat{N}_{ij}\left(\xi;\mathcal{H}\right)=\tilde{N}_{ij}\left(\mathcal{G}^{-1}_{j}\left(\xi;\mathcal{H}\right)\right)=\tilde{N}_{ij}\left(\tilde{\xi}_j\left(\argumentbullet;\mathcal{H}\right)\right).
\label{eq:CharacteristicBases}
\end{equation}
Let us highlight that, by construction, a parent basis function $\tilde{N}_{ij}\left(\tilde{\xi}\right)$ has no dependency on $\mathcal{H}$, but a characteristic basis function $\hat{N}_{ij}\left(\xi;\mathcal{H}\right)$ does. In this sense, the modification of the floating regulation points allows to control the decoupled floating of the characteristic bases. With the above parent concept, (\autoref{eq:floatingtensorproduct2}) now reads
\begin{equation}
\hat{B}_{ij}\left(\bm{\xi};\mathcal{H}\right)=\hat{N}_{ij}\left(\xi;\mathcal{H}\right)\hat{M}_{j}\left(\eta\right)\qquad i=1,...,I_j; j=1,...,J,\\
\label{eq:floatingtensorproduct3}
\end{equation}
where $\bm{\xi}=\left(\xi,\eta\right)^T$. Unless stated otherwise, in the following equations we continue to indicate if a quantity or function has a parameter-dependency on the floating regulation points. 

\subsection{Physical map}

\begin{figure}[t!]
\centering
\includegraphics[trim={5.5cm 7.25cm 4.6cm 7cm},clip,width=1.0\textwidth]{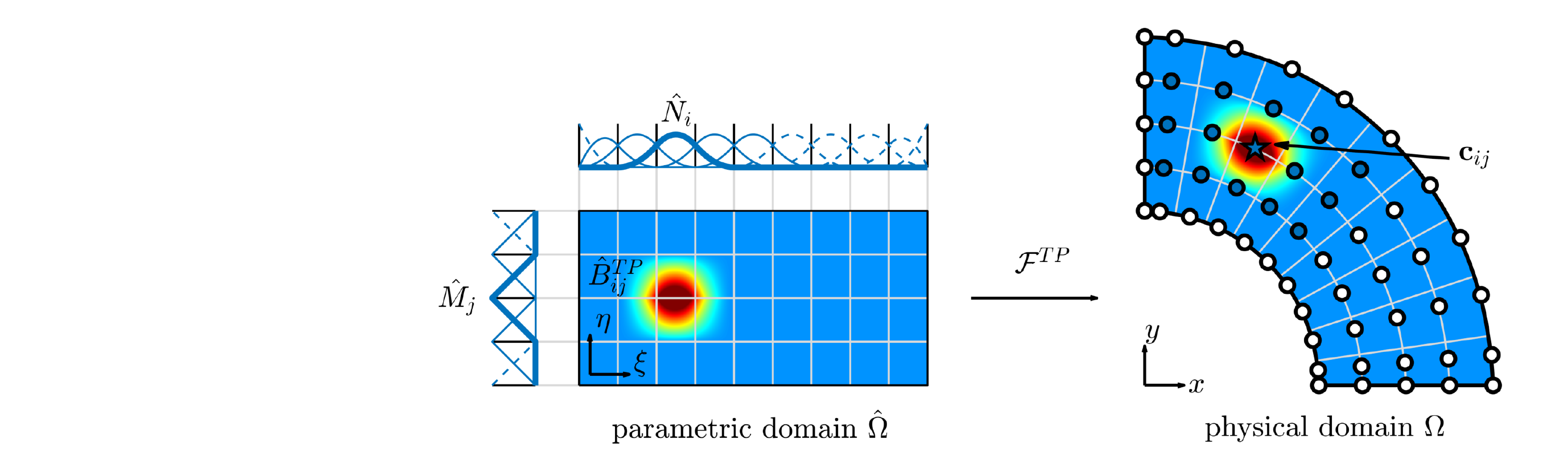}
\caption{Summary of the construction procedure for classical B-Spline surfaces.}
\label{fig:constructionsummary1}
\centering
\includegraphics[trim={5.5cm 4.75cm 4.6cm 4cm},clip,width=1.0\textwidth]{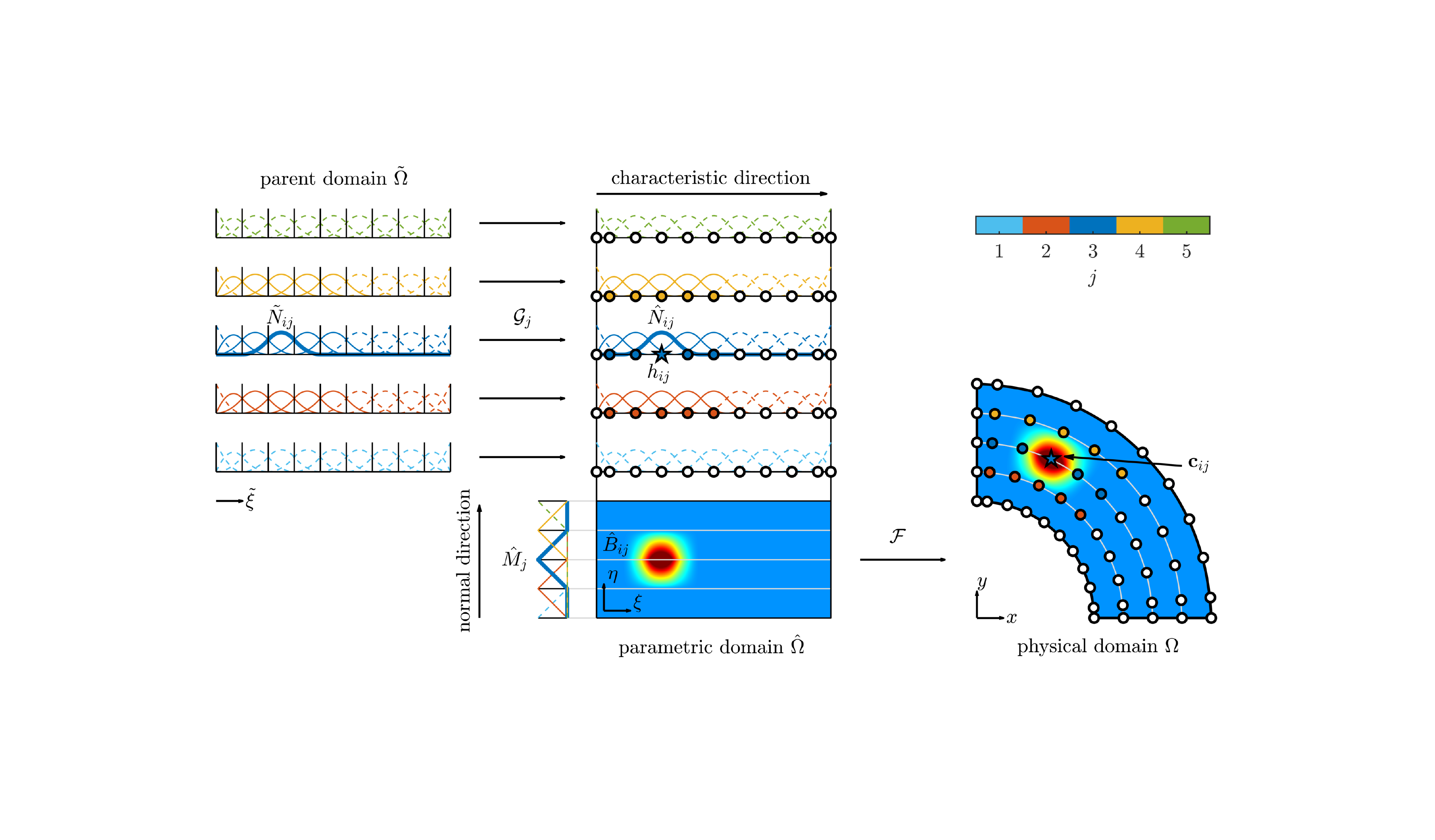}
\caption{Summary of the construction procedure for floating B-Spline surfaces.}
\label{fig:constructionsummary2}
\end{figure}

Standard B-Spline free-form surfaces are constructed from a linear combination of a function basis and so-called control points, mapping the parametric to the physical space, $\mathcal{F}^{TP}: \hat{\Omega}\rightarrow\Omega$
\begin{equation}
\bm{x}=\mathcal{F}^{TP}\left(\bm{\xi}\right)=\sum_{j=1}^J \sum_{i=1}^I \bm{c}_{ij}\hat{B}^{TP}_{ij}\left(\bm{\xi}\right),
\label{eq:physicalmapiga}
\end{equation}
where $\bm{c}_{ij}$ are the position vectors of the control points. The basis functions can be pushed-forward to a given $\bm{x}$ in physical space by
\begin{equation}
B^{TP}_{ij}\left(\bm{x}\right)=\hat{B}^{TP}_{ij}\left(\left(\mathcal{F}^{TP}\right)^{-1}\left(\bm{x}\right)\right)=\hat{B}^{TP}_{ij}\left(\bm{\xi}\right).
\label{eq:physicalpushforwardiga}
\end{equation}

Similarly, for a floating B-Spline basis we employ the following mapping, $\mathcal{F}: \hat{\Omega}\rightarrow\Omega$, from the parametric to the physical space
\begin{equation}
\bm{x}\left(\argumentbullet;\mathcal{H}\right)=\mathcal{F}\left(\bm{\xi};\mathcal{H}\right)=\sum_{j=1}^J \sum_{i=1}^{I_j} \bm{c}_{ij}\hat{B}_{ij}\left(\bm{\xi};\mathcal{H}\right).
\label{eq:physicalmapfliga}
\end{equation}
The floating basis functions are pushed-forward to a given $\bm{x}$ in physical space by
\begin{equation}
B_{ij}\left(\bm{x};\mathcal{H}\right)=\hat{B}_{ij}\left(\mathcal{F}^{-1}\left(\bm{x};\mathcal{H}\right);\mathcal{H}\right)=\hat{B}_{ij}\left(\bm{\xi}\left(\argumentbullet;\mathcal{H}\right);\mathcal{H}\right).
\end{equation}
Note that in general, continuity in the mapping is limited at $\bm{\xi}$ if just \textit{one} of all supported basis functions has limited continuity. However, continuity of the map is always at least $\mathcal{C}^{p-1}$ when traveling along lines of constant $\eta$, and $\mathcal{C}^{q-1}$ when traveling along lines of constant $\xi$ (recall that we assumed that inner knots are not repeated).

We summarize the entire construction concept of classical and floating tensor product B-Splines in \autoref{fig:constructionsummary1} and \autoref{fig:constructionsummary2}, respectively, where two exemplary bivariate bases with $p=2$, $q=1$ are mapped to a free-form geometry in physical space each. The colored circles in the physical domain indicate the control points (in \autoref{fig:constructionsummary2}, the colored circles in the parametric domain denote the floating regulation points) connected to the control point $\mathbf{c}_{ij}$ (floating regulation point $h_{ij}$), for $i=4$ and $j=3$, which is highlighted with a star symbol. For the floating tensor product, different $j$ indices are referred to by a different color and the color legend of \autoref{fig:constructionsummary2} equally applies to the following figures on this tensor product. Until now, the exemplary surfaces and their parametrizations are the same for both concepts, as all floating regulation points are aligned. However, the floating B-Spline representation allows for a richer modification as will be seen next.

\subsection{Effect of floating regulation}
\label{ssec:floatingeffect}

\begin{figure}[t!]
\centering
\includegraphics[trim={5.5cm 7.25cm 4.0cm 7cm},clip,width=1.0\textwidth]{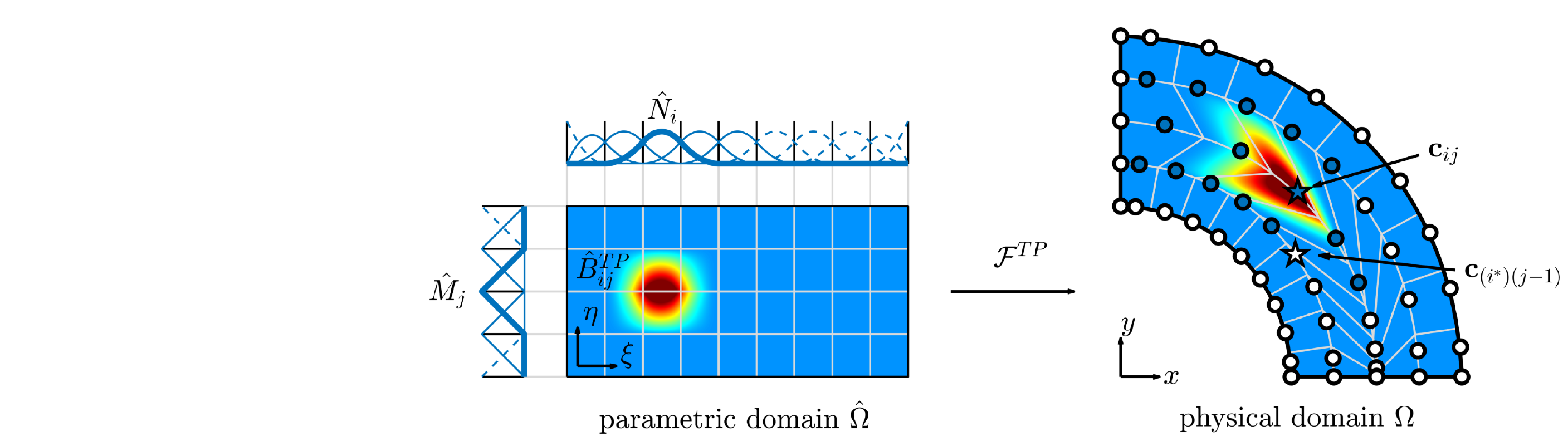}
\caption{Mesh distortion in IGA.}
\label{fig:meshdistortion1}
\centering
\includegraphics[trim={5.5cm 4.75cm 4.0cm 4cm},clip,width=1.0\textwidth]{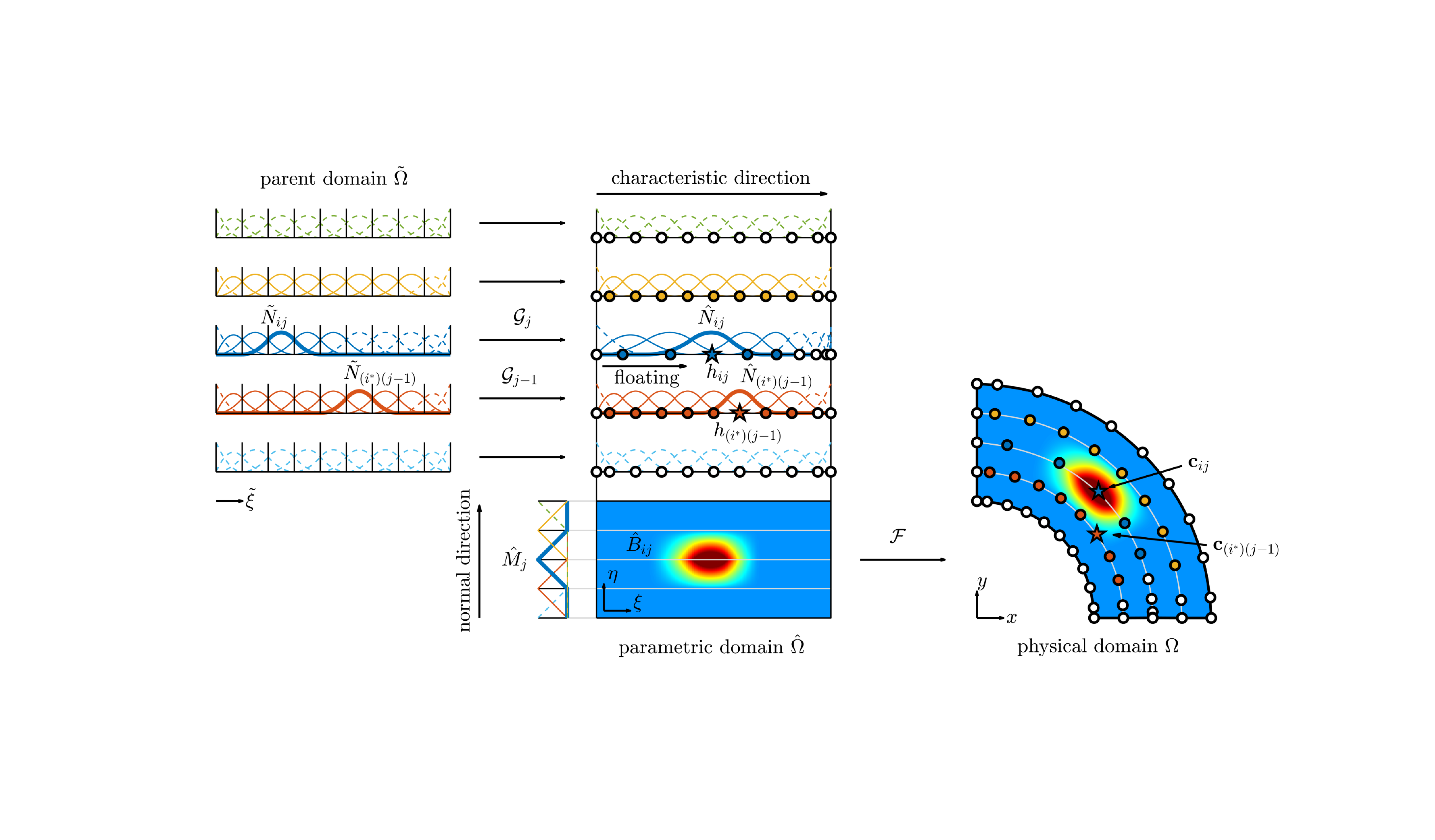}
\caption{Overcoming shear mesh distortion in FLIGA.}
\label{fig:meshdistortion2}
\end{figure}

When using classical B-Spline representations for Lagrangian large deformation analysis, the extreme displacement of control points can result in a distorted mesh. \autoref{fig:meshdistortion1} visualizes again the tensor product mesh from \autoref{fig:constructionsummary1}, but for a deformed control point set, resulting in significant mesh distortion. Note that here control point connectivities are rigid; e.g., the control point $\bm{c}_{ij}$ still connects to the same (now displaced) control points as in the previous configuration. Even though $\bm{c}_{ij}$ has moved close to $\bm{c}_{(i^*)(j-1)}$ ($i^*=7$, white star symbol), these two remain unconnected.

For floating B-Splines, two control points $\bm{c}_{ij}$ and $\bm{c}_{(i^*)(j-1)}$ with a given $i^* \in \left\lbrace 1,2,...,I_{j-1}\right\rbrace$ are connected if the associated $\hat{N}_{ij}$ and $\hat{N}_{(i^*)(j-1)}$ are both supported within a sub-range of $\hat{\Omega}_{\xi}$. Hence, changing the floating maps (which is possible by tuning the floating regulation points, recall (\autoref{eq:floatingmaps})) allows to unconnect control points that are far from each other in physical space and/or connect those which are close neighbors. Such adjustment reparametrizes the physical domain and reduces the shear distortion in the physical mapping $\mathcal{F}$. As a consequence we observe the ``floating'' movement of the basis functions as initially illustrated in \autoref{fig:floatingillustration}. In \autoref{fig:meshdistortion2} we demonstrate this effect on the same displaced control point set as in \autoref{fig:meshdistortion1}, but with a suitable deformation-dependent choice of the floating maps, resulting in an update of connectivity. E.g., the control points $\bm{c}_{ij}$ and $\bm{c}_{(i^*)(j-1)}$ ($i^*=7$, red star symbol) are now connected. The question of how to automatically obtain robust choices of floating regulation points is addressed in \autoref{ssec:updates}.

\subsection{Enhanced features}
\label{ssec:normalbasis}

The construction of the basis functions has been presented with a level of generality that now allows us to introduce three new features of floating B-Splines with respect to \cite{HKDL}:
\begin{itemize}
\item a modified quadrature (\autoref{ssec:quadrature});
\item the automated choice of floating regulation points (\autoref{ssec:updates});
\item local $h$-refinement (\autoref{ssec:refinement}).
\end{itemize}
At this point, we  also add one restriction, namely, the choice of the polynomial order of the \textit{normal} basis as $q=1$. This choice may be surprising, but will be shown in the next subsection to lead to important advantages for numerical quadrature in case of history-dependent material behavior. The B-Spline order $p$ of the \textit{characteristic} bases remains tunable and hence we preserve the smooth floating character along the characteristic direction, see \autoref{fig:meshdistortion2}.

\subsubsection{Numerical quadrature}
\label{ssec:quadrature}

Conventional quadrature in IGA makes use of the tensor product structure and combines quadrature coordinates of the univariate bases to obtain quadrature points in parametric space. In the Lagrangian setting, the quadrature points are mapped to consistent material points in undeformed and deformed physical configurations. The renowned Gauss-Legendre rule was the first quadrature coordinate stencil suggested in IGA \citep{HUGHES20054135} and, owing to its high robustness and accuracy for regular integrands, established itself as one of the standards in the field. Other rules have gained noteworthy attention as well, e.g. for cost critical applications and the quadrature of trimmed elements \citep{HUGHES2010,AURICCHIO2012,ADAM2015,FAHRENDORF2018,ZOU2022,MEMER2022}.

Conversely, the situation for floating tensor product B-Splines is more challenging, as the parametrization of a patch changes during large deformation analysis in order to remove mesh distortion. In \cite{HKDL}, we proposed to advect the Gauss-Legendre points of the initial mesh, i.e. to treat them as material points in physical space (a procedure borrowed from meshless analysis). This strategy enjoys the advantages of the fully Lagrangian viewpoint of motion, however it also has a few shortcomings, most notably:
\begin{itemize}
\item The physical material points must be mapped back to the parametric domain to compute basis function values.
\item The quadrature locations in parent/parametric space change over time following the deformation by an \textit{a priori} unpredictable relation and thereby lose Gauss point character.
\item The parent/parametric quadrature weights are also lost and the physical quadrature weights must be updated explicitly under the assumption of affine deformations.
\end{itemize}
Ultimately, this may result in accumulation of quadrature error and thus numerical instability in the long run. To address these shortcomings, after some preliminary considerations, we propose an improved Lagrangian quadrature scheme for floating B-Splines exploiting the underlying floating tensor product structure.

\paragraph{Preliminaries}
Let us assume that a  quadrature point stays fixed in the parametric space, $\bm{\xi}^*=\left(\xi^*,\eta^*\right)^T$, like in IGA. For a floating B-Spline basis, the physical coordinate of this point reads (\autoref{eq:physicalmapfliga})
\begin{equation}
\bm{x}^*\left(\argumentbullet;\mathcal{H}\right)=\sum_{j=1}^J \sum_{i=1}^{I_j} \bm{c}_{ij}\hat{B}_{ij}\left(\bm{\xi^*};\mathcal{H}\right)=\sum_{j=1}^J \sum_{i=1}^{I_j} \bm{c}_{ij}\hat{N}_{ij}\left(\xi^*;\mathcal{H}\right)\hat{M}_{j}\left(\eta^*\right),
\end{equation}
where we used (\autoref{eq:floatingtensorproduct3}). Pulling the characteristic bases back to the parent space with (\autoref{eq:CharacteristicBases}), we obtain
\begin{equation}
\bm{x}^*\left(\argumentbullet;\mathcal{H}\right)=\sum_{j=1}^J \sum_{i=1}^{I_j} \bm{c}_{ij}\tilde{N}_{ij}\left(\tilde{\xi}^*_j\left(\argumentbullet;\mathcal{H}\right)\right)\hat{M}_{j}\left(\eta^*\right).
\end{equation}
As the evaluation locations of the static parent bases depend on the floating regulation points $\mathcal{H}$, upon floating, we change the physical position corresponding to $\bm{\xi}^*$ and hence lose the material point character of the quadrature point.

Let us now assume that a quadrature point stays fixed in the \textit{parent} space, at coordinate $\left(\tilde{\xi}^*,\eta^*\right)^T$, and that it is mapped to the parametric space  
by a certain $\mathcal{G}_s$ with $s\in\left\lbrace1,2,...,J\right\rbrace$, such that with (\autoref{eq:floatingmaps})
\begin{equation}
\bm{\xi}^*_s\left(\argumentbullet;\mathcal{H}\right)=\left(\xi^*_s\left(\argumentbullet;\mathcal{H}\right),\eta^*\right)^T=\left(\mathcal{G}_s\left(\tilde{\xi}^*;\mathcal{H}\right),\eta^*\right)^T.
\label{eq:quadintro}
\end{equation}
Mapping to the physical space through (\autoref{eq:physicalmapfliga}) and using (\autoref{eq:floatingtensorproduct3}) gives
\begin{equation}
\bm{x}^*_s\left(\argumentbullet;\mathcal{H}\right)=\mathcal{F}\left(\bm{\xi}^*_s\left(\argumentbullet;\mathcal{H}\right);\mathcal{H}\right) = \sum_{j=1}^J \sum_{i=1}^{I_j} \bm{c}_{ij}\hat{B}_{ij}\left(\bm{\xi}^*_s\left(\argumentbullet;\mathcal{H}\right);\mathcal{H}\right)=\sum_{j=1}^J \sum_{i=1}^{I_j} \bm{c}_{ij}\hat{N}_{ij}\left(\xi^*_s\left(\argumentbullet;\mathcal{H}\right);\mathcal{H}\right)\hat{M}_{j}\left(\eta^*\right).
\end{equation}
After pulling the characteristic bases back to the parent space with (\autoref{eq:CharacteristicBases}), the physical coordinate is alternatively expressed as
\begin{equation}
\bm{x}^*_s\left(\argumentbullet;\mathcal{H}\right)=\sum_{j=1}^J \sum_{i=1}^{I_j} \bm{c}_{ij}\tilde{N}_{ij}\left(\tilde{\xi}^*_{sj}\left(\argumentbullet;\mathcal{H}\right)\right)\hat{M}_{j}\left(\eta^*\right)=\sum_{\substack{j=1 \\j \neq s}}^J \sum_{i=1}^{I_j} \bm{c}_{ij}\tilde{N}_{ij}\left(\tilde{\xi}^*_{sj}\left(\argumentbullet;\mathcal{H}\right)\right)\hat{M}_{j}\left(\eta^*\right)+\sum_{i=1}^{I_s} \bm{c}_{is}\tilde{N}_{is}\left(\tilde{\xi}^*\right)\hat{M}_{s}\left(\eta^*\right),
\label{eq:quadpullback}
\end{equation}
with $\tilde{\xi}^*_{sj}\left(\argumentbullet;\mathcal{H}\right) = \mathcal{G}^{-1}_j\left(\xi^*_s\left(\argumentbullet;\mathcal{H}\right);\mathcal{H}\right)$ and, in particular, $\tilde{\xi}^*_{ss} = \mathcal{G}^{-1}_s\left(\xi^*_s\left(\argumentbullet;\mathcal{H}\right);\mathcal{H}\right)=\tilde{\xi^*}$, see (\autoref{eq:inversefloatingmaps}) and (\autoref{eq:quadintro}).
From this equation we can see that the physical quadrature point coordinate becomes independent of the floating regulation points if a Kronecker-delta property w.r.t. $s$ holds for $\hat{M}_j$  at $\eta^*$, 
in which case
\begin{equation}
\bm{x}^*_s=\sum_{i=1}^{I_s} \bm{c}_{is}\tilde{N}_{is}\left(\tilde{\xi}^*\right).
\end{equation}
For B-Splines we can easily realize a Kronecker-delta property of $\hat{M}_j$ by setting $q=1$ and adopting a Gauss-Lobatto quadrature scheme in the normal direction.

\paragraph{A new quadrature scheme}
Let us now apply these concepts to construct a quadrature scheme in which the physical location of quadrature points is independent of the floating regulation, i.e. quadrature points have material point character. We set $q=1$ (which is always assumed in the following) and choose the Gauss-Lobatto quadrature scheme in $\eta$, as it places the quadrature points $\eta^l$, $l=1,...,2J-2$, at the Kronecker-delta positions of the normal basis. At inner knots in $H$, this 1D Gauss-Lobatto scheme gives \textit{two} quadrature points (one from each adjacent knot span) and the total number of Gauss-Lobatto points is therefore larger than the number of characteristic bases. Hence, we formulate an index map deriving the index $s$ of the supported characteristic basis at $\eta^l$ from the Gauss-Lobatto point index $l$
\begin{equation}
s = \mathcal{S}(l) = 1 + \dfrac{l-(l\%2)}{2},
\label{eq:lsmap}
\end{equation}
where $\%$ is the modulo operator.

\begin{figure}[t!]
\centering
\includegraphics[trim={5.75cm 5.1cm 4.25cm 4cm},clip,width=1.0\textwidth]{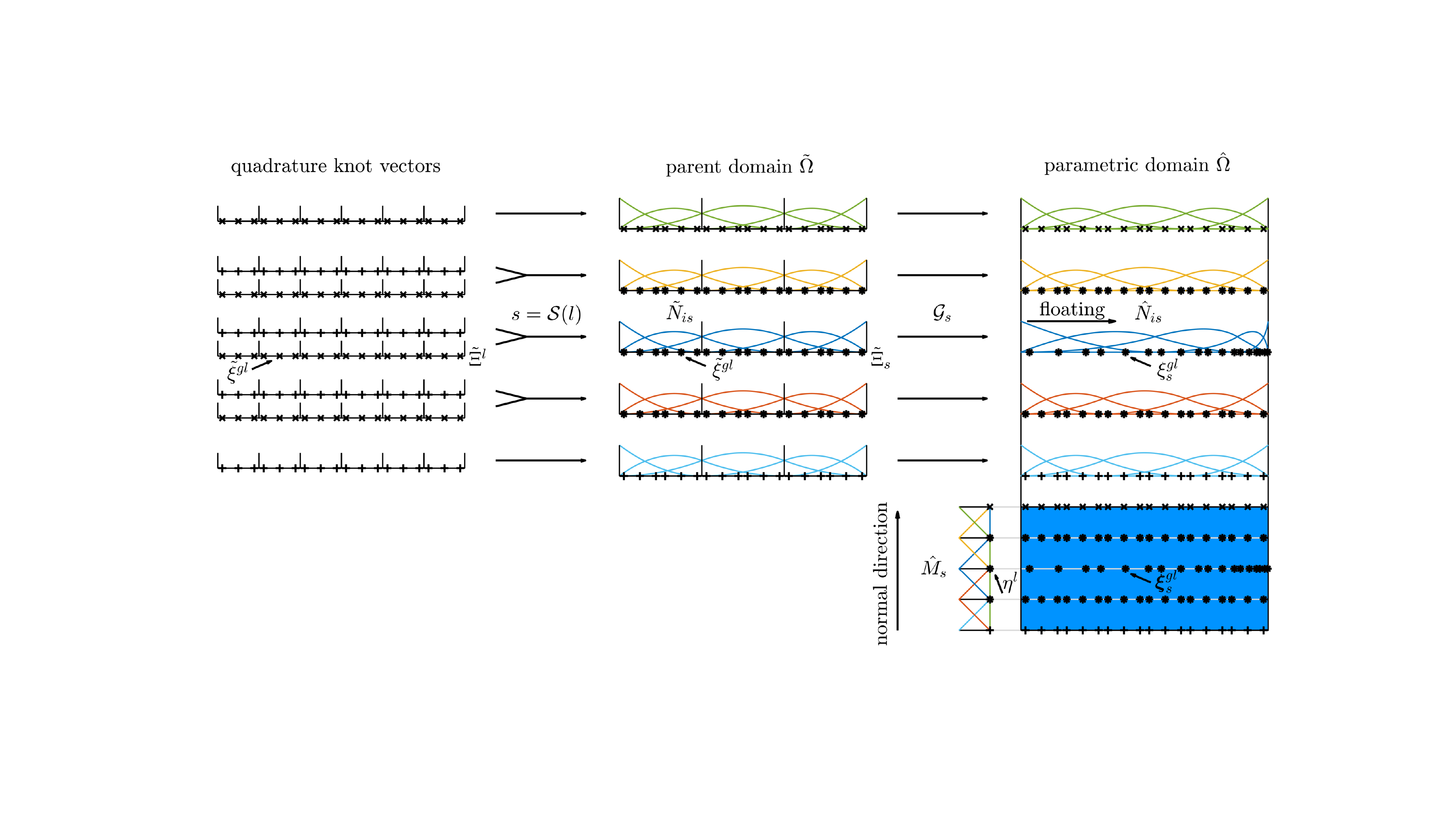}
\caption{A novel quadrature stencil for FLIGA.}
\label{fig:quadratureconcept1}
\end{figure}

As a result of the above choices, all quadrature points are positioned at the boundaries of the characteristic elements, along which we perform quadrature at Gauss-Legendre\footnote{In principle, other quadrature rules for isogeometric analysis might as well be used, see e.g. the introductory references of the section.} positions of the respectively supported parent basis functions $\tilde{N}_{is}$ (which are constant apart from refinement in \autoref{ssec:refinement}). However, instead of working directly with the Gauss-Legendre points of the parent knot spans, we evaluate the Gauss-Legendre points of potentially refined quadrature knot vectors $\tilde{\Xi}^l\supseteq \tilde{\Xi}_{\mathcal{S}(l)}$ as supersets of the parent knot vectors. This strategy grants us flexibility to adjust the point density of the quadrature set, which will be crucial in \autoref{ssec:refinement} within the context of adaptive refinement. In that context we will need the dependency of the quadrature knot vectors on $l$ (and not just $j$) in order to perform manipulations dependent on which of the two adjacent characteristic elements a quadrature point belongs to.

Thus, we obtain the quadrature locations starting from combining the $G$-point Gauss-Legendre schemes $\mathcal{Q}^{Leg,G,l}=\mathcal{Q}^{Leg,G}\left(\tilde{\Xi}^l\right)$ for the quadrature knot vectors $\tilde{\Xi}^l$ (leading to a number of quadrature points $n^{QP,l}$ for each $l$) with the $2$-point Gauss-Lobatto scheme $\mathcal{Q}^{Lob,2}=\mathcal{Q}^{Lob,2}\left(H\right)$ for the normal knot vector $H$, i.e.
\begin{equation}
\tilde{\bm{\xi}}^{gl}=\left\lbrace\tilde{\xi}^{gl},\eta^l\right\rbrace,
\label{eq:constrefquadpoint}
\end{equation}
with fixed $\tilde{\xi}^{gl}\in\mathcal{Q}^{Leg,G,l}$, $g=1,...,n^{QP,l}$,  and fixed $\eta^l \in \mathcal{Q}^{Lob,2}$, $l=1,...,2J-2$. Note that $\tilde{\xi}^{gl}$ corresponds to $\tilde{\xi}^*$ in the above preliminary considerations.

The map of $\tilde{\xi}^{gl}$ to the parametric space reads
\begin{equation}
\xi^{gl}_{s}\left(\argumentbullet;\mathcal{H}\right) =\mathcal{G}_{s}\left(\tilde{\xi}^{gl};\mathcal{H}\right),
\label{eq:quadratureGmap}
\end{equation}
with $s=\mathcal{S}(l)$ given by (\autoref{eq:lsmap}), and the quadrature point vector reads (\autoref{eq:quadintro})
\begin{equation}
\bm{\xi}^{gl}_{s}\left(\argumentbullet;\mathcal{H}\right)=\left( \xi^{gl}_{s}\left(\argumentbullet;\mathcal{H}\right), \eta^l\right)^T.
\end{equation}

\autoref{fig:quadratureconcept1} illustrates the novel concepts which we have introduced so far. We start the quadrature set construction on the left of the figure, where we highlight the quadrature point $\tilde{\xi}^{gl}$ ($l=4$) that is the $g=5$-th member of the Gauss-Legendre point set $\mathcal{Q}^{Leg,G}\left(\tilde{\Xi}^l\right)$ (using a $G=3$-point rule). Here, $\tilde{\Xi}^l$ is twice as dense as $\tilde{\Xi}_s$ (s=3); as we do not have adaptive refinement yet, all other quadrature knot vectors are identical. Note that the quadrature point is marked by a $\times$ symbol to emphasize its association to an \textit{even} $l$, whereas for quadrature points associated with an \textit{odd} $l$ we use a $+$ symbol; in this way we distinguish between points with different $l$ but the same $s$. We push forward the quadrature parent coordinate by $\mathcal{G}_s$, obtaining the characteristic parametric coordinate $\xi^{gl}_{s}$ which is floating-dependent. Lastly, the combination of $\xi^{gl}_{s}$ with $\eta^l$ follows the idea of a \textit{floating} tensor product (i.e., $\xi^{gl}_{s}$ depends on $l$, cf. (\autoref{eq:floatingtensorproduct2})) and gives the parametric quadrature point vector $\bm{\xi}^{gl}_{s}$.
 
Let us now discuss the pull back of $\bm{\xi}^{gl}_{s}$ to parent coordinates, cf. (\autoref{eq:quadpullback}). The pull back with the floating map for $j=s$ is trivial, as due to (\autoref{eq:quadratureGmap})
\begin{equation}
\tilde{\xi}^{gl}_{ss}=\mathcal{G}^{-1}_s\left(\xi^{gl}_{s}\left(\argumentbullet;\mathcal{H}\right);\mathcal{H}\right)=\tilde{\xi}^{gl},
\label{eq:inversefloatingmapfors}
\end{equation}
and in the following, we will always simply write $\tilde{\xi}^{gl}$. For the case $j\neq s$, let us first introduce $n$ as the index of the other normal basis function contained in the knot span of $\eta^l$ (than the one with index $s$)
\begin{equation}
n = \mathcal{N}(l) = \mathcal{S}(l)+\beta, \qquad \begin{cases}\beta = 1,& \mathrm{if}\ l\ \mathrm{is\ odd},\\ \beta = -1, &\mathrm{if}\ l\ \mathrm{is\ even,}\end{cases}
\label{eq:lnmap}
\end{equation}
i.e., the index of the \textit{neighbor} characteristic basis. Note that, due to compact support in normal direction ($\hat{M}_{j}\left(\eta^l\right)=\dfrac{\partial}{\partial \eta}\hat{M}_{j}\left(\eta^l\right)=0\ \mathrm{for}\ (j\neq s\wedge j\neq n)$), it suffices to compute the pull back using only the neighbor floating map $\mathcal{G}^{-1}_n$,
\begin{equation}
\tilde{\xi}^{gl}_{sn}\left(\argumentbullet;\mathcal{H}\right) = \mathcal{G}^{-1}_n\left(\xi^{gl}_{s}\left(\argumentbullet;\mathcal{H}\right);\mathcal{H}\right).
\label{eq:neighborpullback}
\end{equation}
In \cite{HKDL} we formulated a Newton method for the solution of scalar nonlinear equations of the type of (\autoref{eq:neighborpullback}).

Let us denote
\begin{equation}
\tilde{\bm{\xi}}^{gl}_{sn}\left(\argumentbullet;\mathcal{H}\right):=\left\lbrace\tilde{\xi}^{gl},\tilde{\xi}^{gl}_{sn}\left(\argumentbullet;\mathcal{H}\right),\eta^l\right\rbrace,
\label{eq:constrefquadpointwithneighbor}
\end{equation}
as the set of all quadrature point coordinates, and highlight that the neighbor parent coordinate $\tilde{\xi}^{gl}_{sn}$, hence the set, depends on the floating regulation points. Writing $\tilde{\bm{\xi}}^{gl}$ without subscript $sn$, we exclude $\tilde{\xi}^{gl}_{sn}$ (\autoref{eq:constrefquadpoint}).

\paragraph{Computing basis functions and their gradients}
Let us now compute the basis functions at quadrature point $\bm{\xi}^{gl}_{s}$ following \autoref{ssec:construction} as
\begin{align}
\begin{split}
\hat{B}_{ij}\left(\bm{\xi}^{gl}_{s}\left(\argumentbullet;\mathcal{H}\right);\mathcal{H}\right) = \hat{N}_{ij}\left(\xi^{gl}_{s}\left(\argumentbullet;\mathcal{H}\right);\mathcal{H}\right)\ \hat{M}_j\left(\eta^l\right) = \tilde{N}_{ij}\left(\tilde{\xi}^{gl}_{sj}\left(\argumentbullet;\mathcal{H}\right)\right)\ \hat{M}_j\left(\eta^l\right) &= \begin{cases}
\tilde{N}_{is}\left(\tilde{\xi}^{gl}\right)\hat{M}_s\left(\eta^l\right), &\mathrm{for}\ j=s,\\
\tilde{N}_{in}\left(\tilde{\xi}^{gl}_{sn}\left(\argumentbullet;\mathcal{H}\right)\right)\hat{M}_n\left(\eta^l\right), &\mathrm{for}\ j=n,\\
0, & \mathrm{otherwise},
\end{cases}\\
&=
\begin{cases}
\tilde{N}_{is}\left(\tilde{\xi}^{gl}\right), &\mathrm{for}\ j=s,\\
0, & \mathrm{otherwise},
\end{cases}\\
&=\hat{B}_{ij}\left(\bm{\tilde{\xi}}^{gl}\right),
\end{split}
\label{eq:quadbasisfunction}
\end{align}
considering the Kronecker-delta property of $\hat{M}_j$. Note that, in the final expression, all univariate basis function evaluations are of Gauss-Legendre character. Basis function gradients with respect to $\bm{\xi}$ at the quadrature point are
\begin{align}
\begin{split}
\dfrac{\partial}{\partial\bm{\xi}}\hat{B}_{ij}\left(\bm{\xi}^{gl}_{s}\left(\argumentbullet;\mathcal{H}\right);\mathcal{H}\right)
&=\begin{pmatrix}
\dfrac{\partial}{\partial \xi}\hat{N}_{ij}(\xi^{gl}_{s}\left(\argumentbullet;\mathcal{H}\right);\mathcal{H})\ \hat{M}_j(\eta^l)\\
\hat{N}_{ij}(\xi^{gl}_s\left(\argumentbullet;\mathcal{H}\right);\mathcal{H})\ \dfrac{\partial}{\partial \eta}\hat{M}_j(\eta^l)\\
\end{pmatrix}
=\begin{pmatrix}
J_j\left(\tilde{\xi}^{gl}_{sj}\left(\argumentbullet;\mathcal{H}\right);\mathcal{H}\right)^{-1}\ \dfrac{\partial}{\partial \tilde{\xi}} \tilde{N}_{ij}\left(\tilde{\xi}^{gl}_{sj}\left(\argumentbullet;\mathcal{H}\right)\right)\ \hat{M}_j(\eta^l)\\
\tilde{N}_{ij}\left(\tilde{\xi}^{gl}_{sj}\left(\argumentbullet;\mathcal{H}\right)\right)\ \dfrac{\partial}{\partial \eta} \hat{M}_{j}\left(\eta^l\right)\\
\end{pmatrix}.\\
&=
\begin{cases}
\begin{pmatrix}
J_s\left(\tilde{\xi}^{gl};\mathcal{H}\right)^{-1}\ \dfrac{\partial}{\partial\tilde{\xi}}\tilde{N}_{is}\left(\tilde{\xi}^{gl}\right)\\
\tilde{N}_{is}\left(\tilde{\xi}^{gl}\right)\ \dfrac{\partial}{\partial\eta}\hat{M}_s\left(\eta^l\right)
\end{pmatrix}, &\mathrm{for}\ j=s,\\
\begin{pmatrix}
0\\
\tilde{N}_{in}\left(\tilde{\xi}^{gl}_{sn}\left(\argumentbullet;\mathcal{H}\right)\right)\ \dfrac{\partial}{\partial\eta}\hat{M}_n\left(\eta^l\right)
\end{pmatrix}, &\mathrm{for}\ j=n,\\
\begin{pmatrix}
0\\
0
\end{pmatrix}, &\mathrm{otherwise},
\end{cases}\\
&=
\dfrac{\partial}{\partial\bm{\xi}}\hat{B}_{ij}\left(\tilde{\bm{\xi}}^{gl}_{sn}\left(\argumentbullet;\mathcal{H}\right);\mathcal{H}\right),
\end{split}
\label{eq:dBdxi}
\end{align}
with the scalar Jacobian $J_s$ of the supported floating map (\autoref{eq:floatingmaps}) derived from
\begin{equation}
J_j\left(\tilde{\xi};\mathcal{H}\right)=\dfrac{\partial}{\partial \tilde{\xi}}\mathcal{G}_j\left(\tilde{\xi};\mathcal{H}\right)=\sum^{I_j}_{i=1}h_{ij} \dfrac{\partial}{\partial \tilde{\xi}} \tilde{N}_{ij}\left(\tilde{\xi}\right).
\label{eq:Gjacobian}
\end{equation}
Let us clarify that, while the neighbor normal basis function is not supported in terms of function value, $\hat{M}_n\left(\eta^l\right)=0$, it is supported in terms of the gradient $\dfrac{\partial}{\partial\eta}\hat{M}_n\left(\eta^l\right)\neq 0$. In the final expression of (\autoref{eq:dBdxi}), including (\autoref{eq:Gjacobian}), all univariate basis function evaluations are of either Gauss-Legendre or Gauss-Lobatto character, except for the evaluation of $\tilde{N}_{in}\left(\tilde{\xi}^{gl}_{sn}\left(\argumentbullet;\mathcal{H}\right)\right)$ appearing for $j=n$.

Basis function gradients with respect to $\bm{x}$ are
\begin{align}
\dfrac{\partial}{\partial\bm{x}}\hat{B}_{ij}\left(\tilde{\bm{\xi}}^{gl}_{sn}\left(\argumentbullet;\mathcal{H}\right);\mathcal{H}\right)= \bm{J}\left(\tilde{\bm{\xi}}^{gl}_{sn}\left(\argumentbullet;\mathcal{H}\right);\mathcal{H}\right)^{-T}\ \dfrac{\partial}{\partial\bm{\xi}}\hat{B}_{ij}\left(\tilde{\bm{\xi}}^{gl}_{sn}\left(\argumentbullet;\mathcal{H}\right);\mathcal{H}\right),
\end{align}
with the Jacobian of the physical map (\autoref{eq:physicalmapfliga})
\begin{align}
\bm{J}\left(\tilde{\bm{\xi}}^{gl}_{sn}\left(\argumentbullet;\mathcal{H}\right);\mathcal{H}\right)= \sum^J_{j=1}\sum^{I_j}_{i=1}\bm{c}_{ij}\left(\left.\dfrac{\partial}{\partial\bm{\xi}}\hat{B}_{ij}\left(\tilde{\bm{\xi}}^{gl}_{sn}\left(\argumentbullet;\mathcal{H}\right);\mathcal{H}\right)\right.^T\right).
\label{eq:jacobimatrix}
\end{align}

With the basis functions at hand, the physical quadrature point position is derived ((\autoref{eq:physicalmapfliga}) and (\autoref{eq:quadbasisfunction})) as
\begin{equation}
\bm{x}^{gl}_{s}\left(\argumentbullet;\mathcal{H}\right)=\mathcal{F}\left(\bm{\xi}^{gl}_{s}\left(\argumentbullet;\mathcal{H}\right);\mathcal{H}\right) = \sum^J_{j=1}\sum^{I_j}_{i=1} \bm{c}_{ij}\hat{B}_{ij}\left(\bm{\xi}^{gl}_{s}\left(\argumentbullet;\mathcal{H}\right);\mathcal{H}\right) = \sum^J_{j=1}\sum^{I_j}_{i=1} \bm{c}_{ij}\hat{B}_{ij}\left(\tilde{\bm{\xi}}^{gl}\right) = \sum^{I_s}_{i=1} \bm{c}_{is}\tilde{N}_{is}\left(\tilde{\xi}^{gl}\right)=\bm{x}^{gl}_{s},
\label{eq:quadLag}
\end{equation}
which shows dependency only on the control points (not the floating regulation points) confirming the Lagrangian character.

\begin{figure}[t!]
    \centering{\includegraphics[trim={5.5cm 4.75cm 4.6cm 4cm},clip,width=1.0\textwidth]{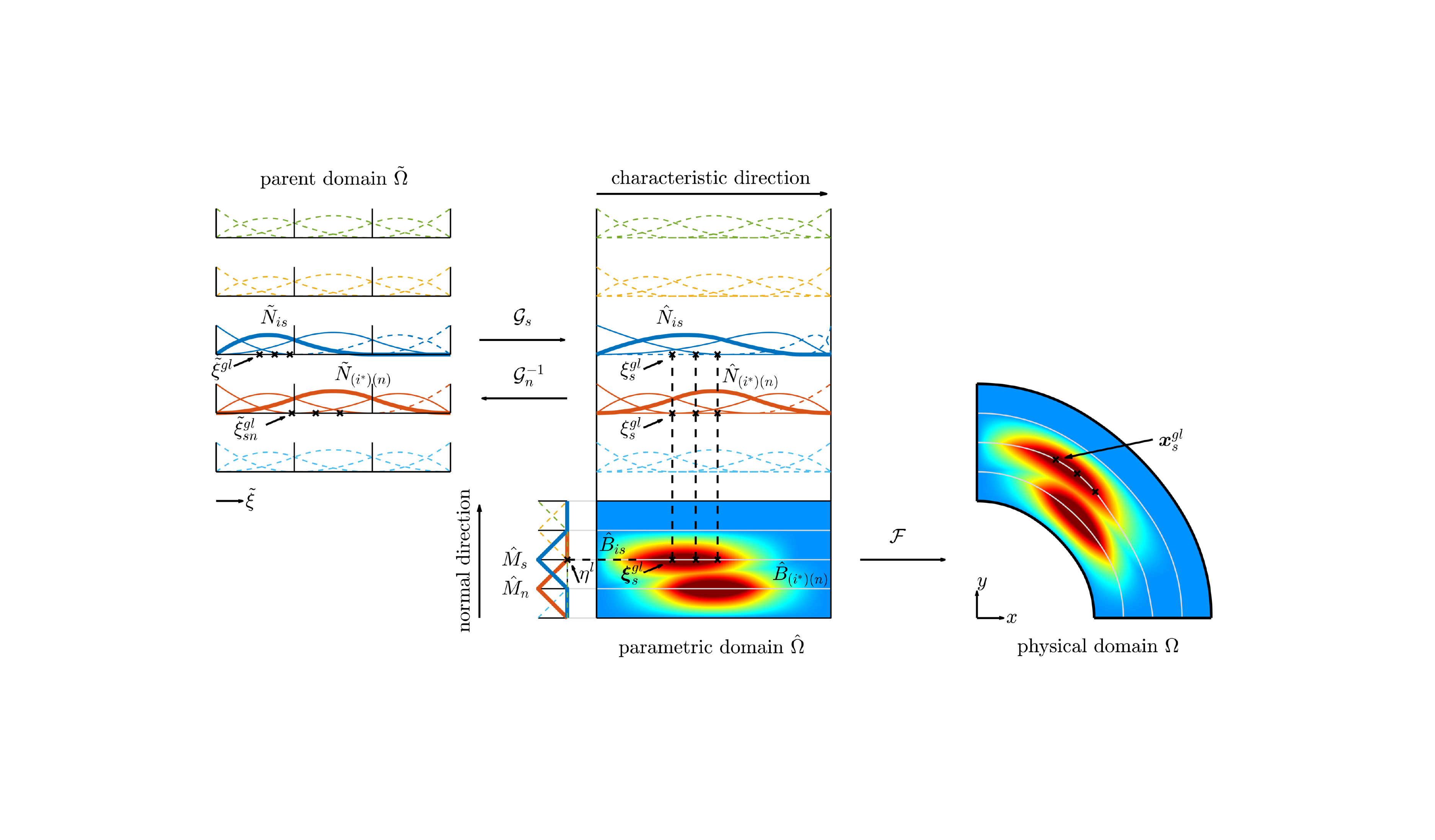} }
    \caption{Evaluation point characteristics for the novel quadrature scheme.}
    \label{fig:quadratureillustration}
\end{figure}

The evaluation of basis functions is demonstrated in \autoref{fig:quadratureillustration} for two exemplary functions with $i=2,s=3$ (blue) and $i^*=3,n=2$ (red), respectively, proceeding the example of \autoref{fig:quadratureconcept1}. Most notably the figure shows how the parent coordinates $\tilde{\xi}^{gl}$ and $\tilde{\xi}^{gl}_{sn}$ are different due to floating regulation. On a qualitative basis, we find that all function evaluations of $\tilde{N}_{is},\hat{M}_s,\hat{M}_n$ at $G=3$ exemplary points of a given quadrature knot span are indeed of Gauss-Legendre or Gauss-Lobatto character; however, not those of $\tilde{N}_{(i^*)(n)}$.

\paragraph{Connectivities} 
Due to the local support of the basis functions, the implementation needs the connectivities at a quadrature point. With $\tilde{\xi}^{gl} \in \left[\tilde{\xi}_k\right., \left.\tilde{\xi}_{k+1}\right]$ for $\tilde{\xi}_k,\tilde{\xi}_{k+1}\in\tilde{\Xi}_s$ and $\tilde{\xi}_k\neq\tilde{\xi}_{k+1}$, the connectivities of $\tilde{\xi}^{gl}$ (alternatively expressed as $\tilde{\xi}^{gl}_{ss}$, (\autoref{eq:inversefloatingmapfors})) to the supported parent basis $\left\lbrace\tilde{N}_{is}\right\rbrace_{i=1,...,I_s}$ are
\begin{equation}
\mathcal{N}^{gl}_{ss} = \left\lbrace k-p,\ k-p+1,\ ...,\ k \right\rbrace.
\end{equation}
Note that these are independent of the floating regulation points. Instead, with
\begin{equation}
\tilde{\xi}^{gl}_{sn}\left(\argumentbullet;\mathcal{H}\right) \in \left[\tilde{\xi}_k\right., \left.\tilde{\xi}_{k+1}\right],
\end{equation}
for $\tilde{\xi}_k,\tilde{\xi}_{k+1}\in\tilde{\Xi}_n$ and $\tilde{\xi}_k\neq\tilde{\xi}_{k+1}$, the connectivities of $\tilde{\xi}^{gl}_{sn}$ to the neighbor parent basis $\left\lbrace\tilde{N}_{in}\right\rbrace_{i=1,...,I_n}$ are
\begin{equation}
\mathcal{N}^{gl}_{sn}\left(\argumentbullet;\mathcal{H}\right) = \left\lbrace k\left(\argumentbullet;\mathcal{H}\right)-p,\ k\left(\argumentbullet;\mathcal{H}\right)-p+1,\ ...,\ k\left(\argumentbullet;\mathcal{H}\right) \right\rbrace,
\end{equation}
and update upon floating. Then, with (\autoref{eq:lsmap}) and (\autoref{eq:lnmap}) the constant connectivities of $\eta^l$ to the normal basis $\left\lbrace\hat{M}_j\right\rbrace_{j=1,...,J}$, sorted in ascending order, are
\begin{equation}
\mathcal{M}^l = \mathrm{sort}\left(\left\lbrace s,\ n \right\rbrace\right).
\label{eq:connectivitiesM}
\end{equation}
Hence, the updating connectivities of $\tilde{\bm{\xi}}^{gl}_{sn}\left(\argumentbullet;\mathcal{H}\right)$ to the bivariate basis $\left\lbrace\hat{B}_{ij}\right\rbrace_{i=1,...,I_j;j=1,...,J}$ are easily derived from the floating tensor product structure as the set of tuples
\begin{equation}
    \mathcal{B}^{gl}_{sn}\left(\argumentbullet;\mathcal{H}\right) =\left\lbrace(i,j)\ \left|\rule{0cm}{0.4cm}\right.\ i\in \mathcal{N}^{gl}_{sj}\left(\argumentbullet;\mathcal{H}\right), j\in \mathcal{M}^l\right\rbrace.
\end{equation}
As with the classical tensor product structure, the total number of supported basis functions at $\tilde{\bm{\xi}}^{gl}_{sn}$ is $(p+1)\cdot(q+1)$, i.e., $(p+1)\cdot 2$. For implementation purposes, the translation of a tuple $(i,j)$ into a single running index $m=1,...,M$ is most convenient by
\begin{equation}
m=\left(\sum^{j-1}_{\gamma=1} I_\gamma\right) + i,
\label{eq:runningindex}
\end{equation}
where $\sum^0_{\gamma=1}I_{\gamma}:=0$ and $I_{\gamma}$ is the number of basis functions in the characteristic basis with index $\gamma$. This expression allows to fill two arrays for associating $m\leftarrow(i,j)$ and $(i,j)\leftarrow m$.

\paragraph{Weights} 
The parametric weights $w^{gl}_{s}$ are obtained from the weights of the Gauss-Legendre schemes $\mathcal{W}^{Leg,G,l}=\mathcal{W}^{Leg,G}\left(\tilde{\Xi}^l\right)$ and of the Gauss-Lobatto scheme $\mathcal{W}^{Lob,2}=\mathcal{W}^{Lob,2}\left(H\right)$ with (\autoref{eq:Gjacobian}) by
\begin{equation}
w^{gl}_{s}\left(\argumentbullet;\mathcal{H}\right) = J_{s}\left(\tilde{\xi}^{gl};\mathcal{H}\right)\ \tilde{w}^{gl} \ w^l,
\end{equation}
where $\tilde{w}^{gl} = \tilde{w}_{g}\in\mathcal{W}^{Leg,G,l}$ and $w^l \in \mathcal{W}^{Lob,2}$. The physical weight is obtained as
\begin{align}
W^{gl}_{s}\left(\argumentbullet;\mathcal{H}\right) =\mathrm{det}\left(\bm{J}\left(\tilde{\bm{\xi}}^{gl}_{sn}\left(\argumentbullet;\mathcal{H}\right);\mathcal{H}\right)\right) \ w^{gl}_{s}\left(\argumentbullet;\mathcal{H}\right).
\end{align}
Finally, numerical approximation of a physical integral follows as
\begin{equation}
\int_{\Omega}\left[\bullet\right] \mathrm{d}\Omega \approx \sum^{2J-2}_{l=1}\sum^{n^{QP,l}}_{g=1} \left[\bullet\right]_{\bm{x}=\bm{x}^{gl}_{s}} W^{gl}_{s}\left(\argumentbullet;\mathcal{H}\right).
\end{equation}

\paragraph{Summary}
Summarizing the new strategy, we highlight that the quadrature point $\tilde{\bm{\xi}}^{gl}_{sn}$
\begin{itemize}
\item is Lagrangian, i.e. moves as material point;
\item has a constant parent coordinate $\tilde{\xi}^{gl}$ hence is part of a $G$-point Gauss-Legendre scheme for the associated $p$-th order supported characteristic basis $\left\lbrace\tilde{N}_{is}\right\rbrace_{i=1,...,I_s}$;
\item has a floating-dependent neighbor parent coordinate $\tilde{\xi}^{gl}_{sn}$ hence is \textit{not} part of a Gauss-Legendre scheme for the associated $p$-th order neighbor characteristic basis $\left\lbrace\tilde{N}_{in}\right\rbrace_{i=1,...,I_n}$;
\item has a constant normal coordinate $\eta^l$ hence is part of a $2$-point Gauss-Lobatto scheme for the $1$-st order normal basis $\left\lbrace\hat{M}_j\right\rbrace_{j=1,...,J}$;
\item has constant parent and normal weights $\tilde{w}^{gl}, w^l$.
\end{itemize}

\subsubsection{Automated floating regulation}
\label{ssec:updates}

\begin{figure}[bt]
\centering
\subfigure[\centering Manual design of static levels.]{\includegraphics[trim={0cm 2cm 0cm 1.5cm},clip,width=5cm]{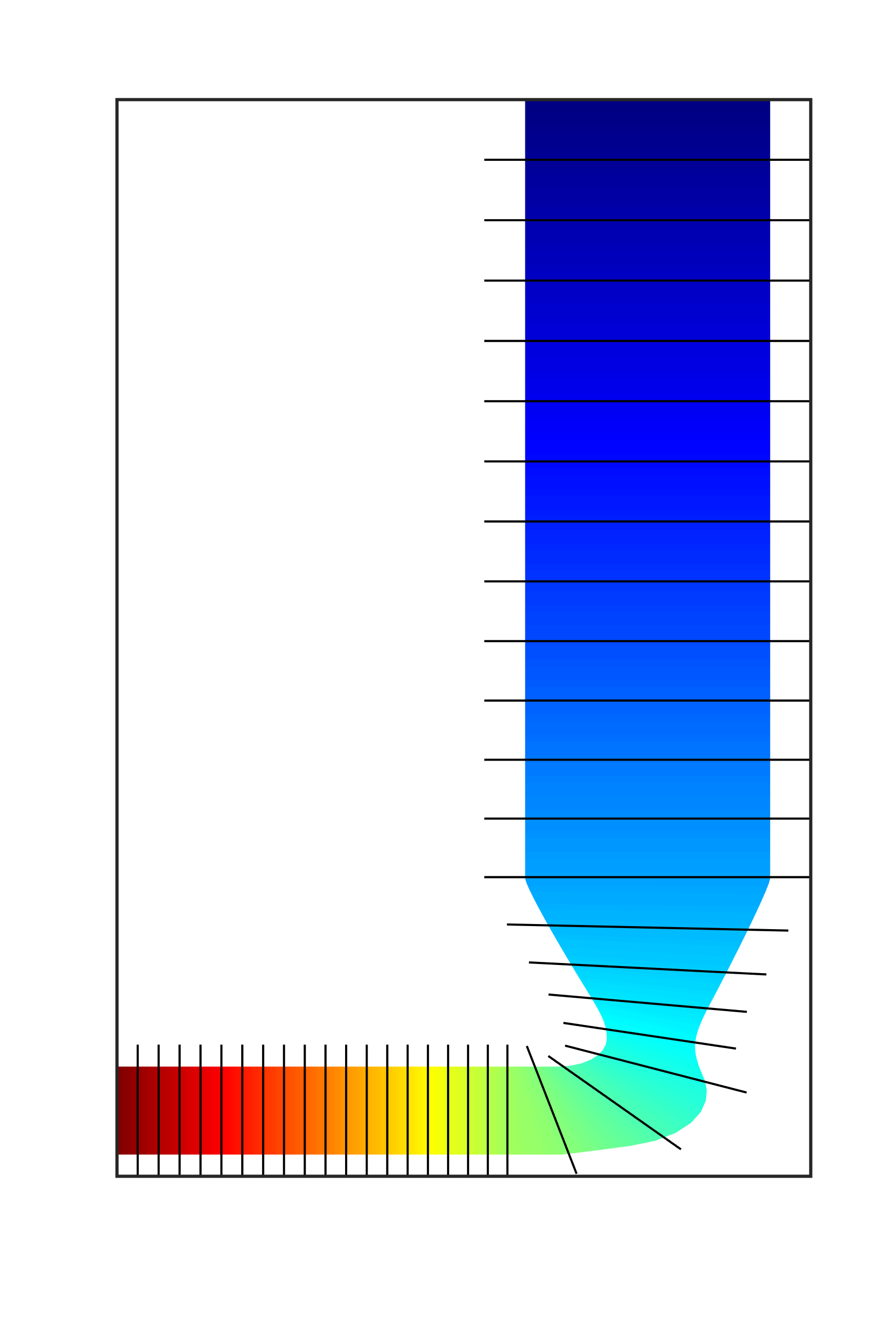} }
\qquad
\subfigure[\centering Automated PDE-based parametrization.]{\includegraphics[trim={0cm 2cm 0cm 1.5cm},clip,width=5cm]{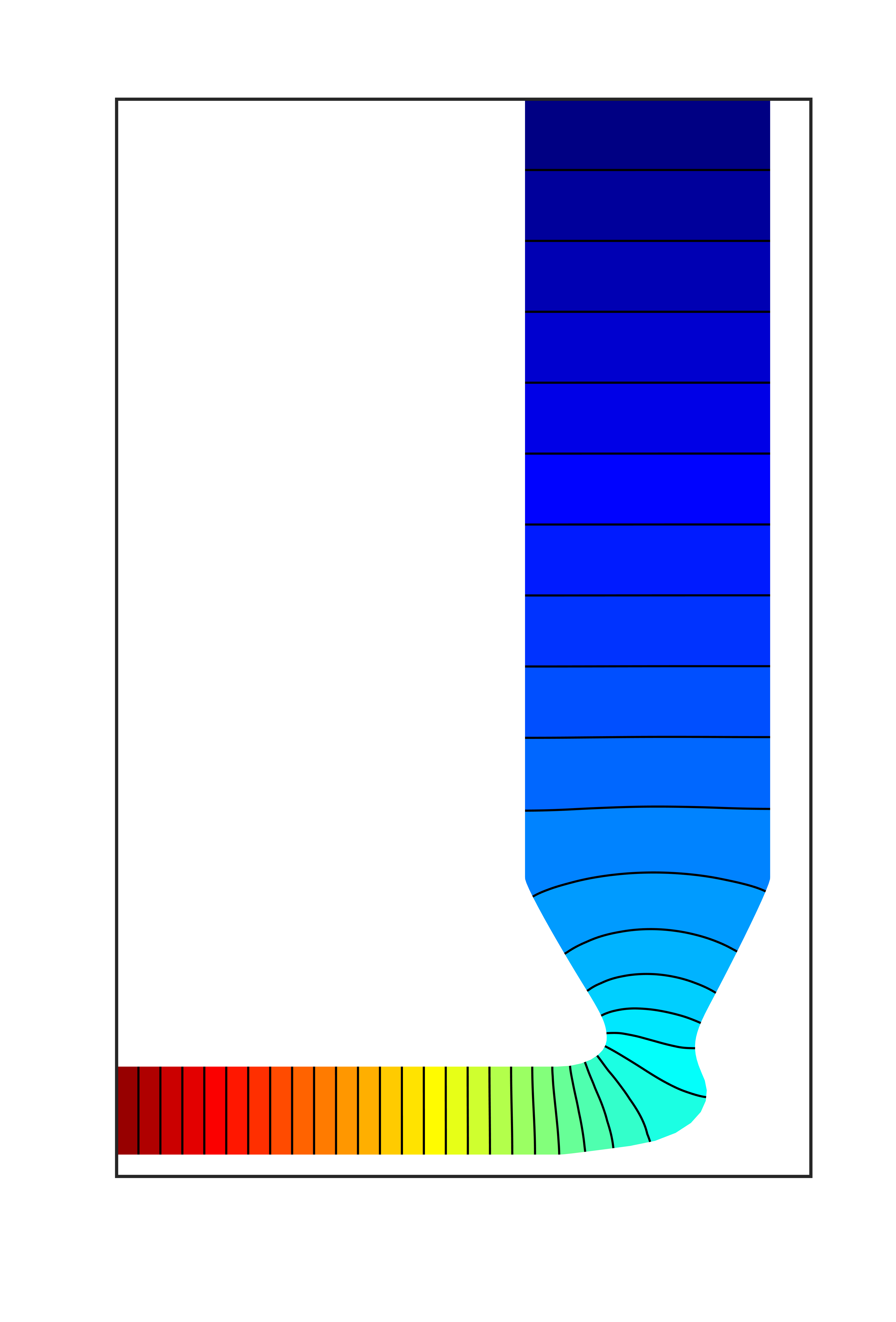} }
    \caption{Schematics of the old (a) and new (b) concept for the undistorted parametrization of physical space w.r.t. $\xi$. Colors indicate values of $\xi$.}
    \label{fig:levelfunctions}
\end{figure}

In \cite{HKDL} we proposed the concept of a problem-specific level function to determine suitable floating regulation point positions for the reduction of mesh distortion. This function was designed so as to express the desired association of characteristic and physical coordinates, $\xi$ and $\bm{x}$, respectively, and floating regulation points were then chosen so as to establish this relation. The level function construction was manual and based on an \textit{a priori} guess about the deformations which is generally difficult to obtain. The significant construction effort suggested the use of only one level function to be applied to all load steps (and a local blending to force the levels in the rectangular shape of the parametric domain). We graphically summarize the old procedure in \autoref{fig:levelfunctions} (a) on a simple example of extrusion, where the qualitative path of material flow is easy to predict. 

In this paper, we replace this manual method with an automated approach, where the characteristic coordinate field is adjusted by recurrently solving a partial differential equation (PDE) on different deformed configurations; an exemplary time step result is illustrated in \autoref{fig:levelfunctions} (b). In fact, the concept of PDE-based parametrization has been investigated before to determine analysis-suitable \textit{control} point positions in IGA, see \cite{HINZ2018} and references therein. In our case however, an important requirement on the reparametrization procedure is to preserve the Lagrangian character of the quadrature points introduced in \autoref{ssec:quadrature} which are dependent on the control points (\autoref{eq:quadLag}). For this reason, the control points are kept constant throughout the following procedure.

We start with introducing a floating B-Spline ansatz (indicated by superscript $h$) which expresses the parametric coordinate $\xi$ at a physical point $\bm{x}$ in terms of the control and floating regulation points,
\begin{equation}
\xi^h\left(\bm{x};\mathcal{H}\right) = \sum^M_{m=1} h_m B_m\left(\bm{x};\mathcal{H}\right)=\xi\left(\argumentbullet;\mathcal{H}\right),
\label{eq:xiansatz}
\end{equation}
see \ref{xiansatz} for a proof. Here we consider the running index convention $m=1,...,M$ from (\autoref{eq:runningindex}) and we omit denoting parameter-dependencies on the control points. Let us highlight that also the basis functions in this ansatz depend on the floating regulation point coordinates which are unknown variables.

In the PDE-based reparametrization concept the differential-type Laplace operator is applied to this ansatz in physical space
\begin{equation}
\dfrac{\partial}{\partial \bm{x}} \cdot \dfrac{\partial}{\partial \bm{x}} \xi^h\left(\bm{x};\mathcal{H}\right) = 0\qquad \textrm{in}\ \Omega,
\label{eq:laplacestrong}
\end{equation}
complemented by the Dirichlet boundary conditions
\begin{align}
\begin{split}
\xi^h &= 0\qquad \textrm{on}\ \tilde{\xi}=0,\\
\xi^h &= 1\qquad \textrm{on}\ \tilde{\xi}=1,
\end{split}
\label{eq:laplacedirichlet}
\end{align}
and by the Neumann boundary conditions
\begin{align}
\begin{split}
\dfrac{\partial}{\partial \bm{x}}\xi^h\left(\bm{x};\mathcal{H}\right)\cdot \bm{n} &= 0\qquad \textrm{on}\ \eta=0,\\
\dfrac{\partial}{\partial \bm{x}}\xi^h\left(\bm{x};\mathcal{H}\right)\cdot \bm{n} &= 0\qquad \textrm{on}\ \eta=1,
\end{split}
\label{eq:laplaceneumann}
\end{align}
where $\bm{n}$ is the outward normal unit vector at the boundary. Let us now express (\autoref{eq:laplacestrong}) in weak form as
\begin{equation}
R_i = \sum^{2J-2}_{l=1}\sum^{n^{QP,l}}_{g=1}\sum^M_{m=1}\dfrac{\partial}{\partial \bm{x}} \hat{B}_i\left(\tilde{\bm{\xi}}^{gl}_{sn}\left(\argumentbullet;\mathcal{H}\right);\mathcal{H}\right) \cdot \dfrac{\partial}{\partial \bm{x}} \hat{B}_{m}\left(\tilde{\bm{\xi}}^{gl}_{sn}\left(\argumentbullet;\mathcal{H}\right);\mathcal{H}\right)\ h_m\ W^{gl}_s\left(\argumentbullet;\mathcal{H}\right) \overset{!}{=} 0\qquad \forall\ i,
\label{eq:laplace}
\end{equation}
where (\autoref{eq:laplacedirichlet}) is incorporated in the solution space (by setting $h_{1j}=0,\ h_{(I_j)(j)}=1\ \forall j=1,...,J$) and (\autoref{eq:laplaceneumann}) is satisfied naturally. Here we make use of numerical integration by the concept introduced in \autoref{ssec:quadrature}. Obviously, the obtained problem is nonlinear in $h_m$ (recall that $\mathcal{H}$ also contains the floating regulation points) and requires a solution e.g. by the iterative Newton-Raphson procedure (with tangent stiffness provided in \ref{laplaceNewtonRaphson}). We directly propose the solution $\mathcal{H}$ as suitable floating regulation point coordinates.

Note that (\autoref{eq:laplacestrong}) acts as a smoothing operator on the parametrization, while (\autoref{eq:laplacedirichlet}) ensures that the new parametrization fits the parametric domain. (\autoref{eq:laplaceneumann}) forces the levels of $\xi^h$ perpendicular to the physical domain boundaries along $\xi$, cf. \autoref{fig:levelfunctions} (b). Hence we expect the procedure to determine floating regulation point positions such as to restore a regular ascent of $\xi$ along the characteristic elements in physical space.\footnote{Note that we here assume that the characteristic element interfaces inside the domain always reasonably align to the outer physical domain boundaries, and have spared the imposition of additional constraints on $\xi^h$ at such interfaces.} Importantly, the physical positions of the novel quadrature points along the characteristic element interfaces (\autoref{eq:quadLag}) are independent of the floating regulation point positions, such that no history data projection is required upon reparametrization.

\begin{figure}[t!]
\centering
\includegraphics[trim={12cm 4.75cm 5.5cm 4cm},clip,width=0.4775\textwidth]{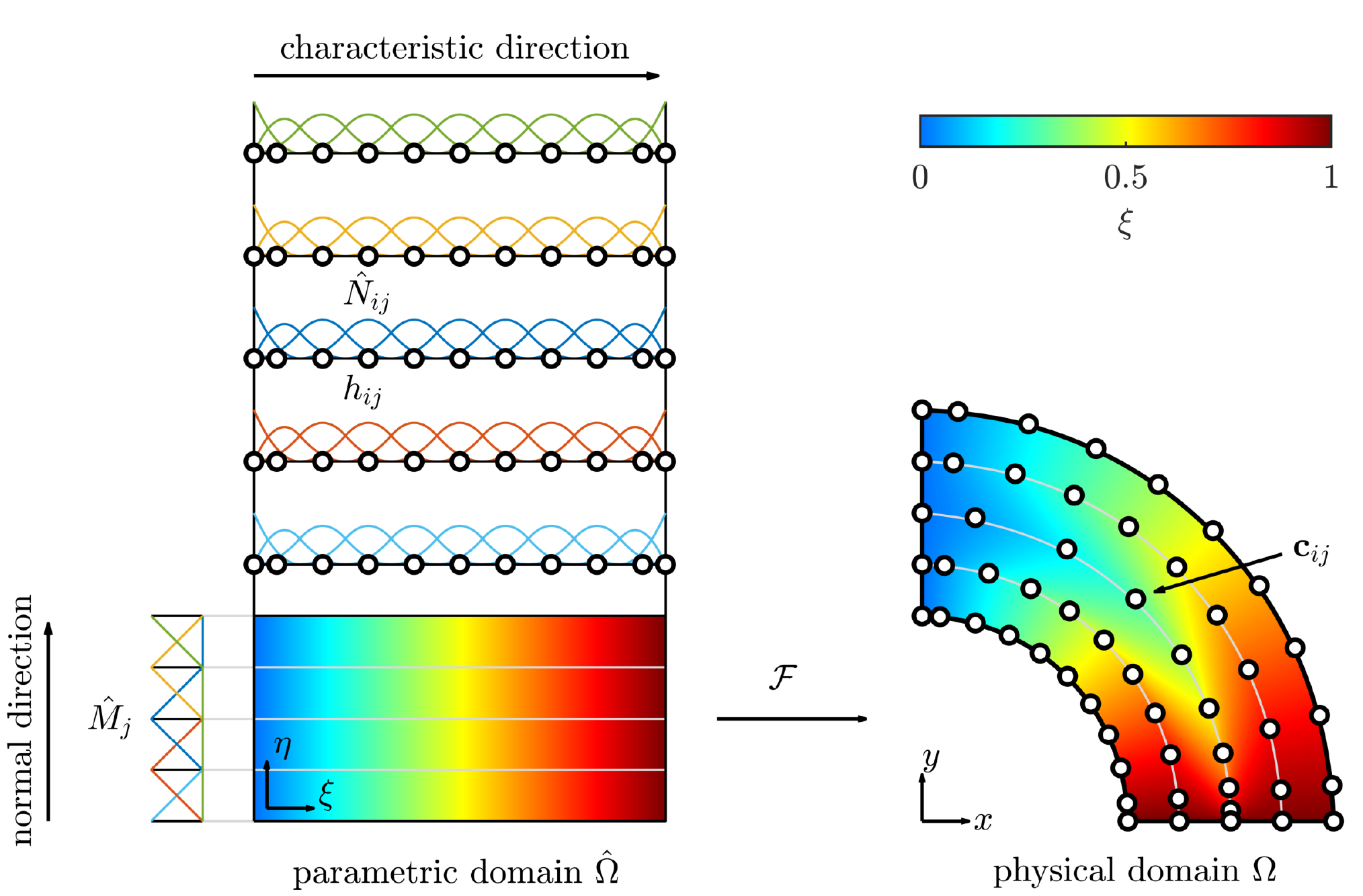}\qquad
\includegraphics[trim={12cm 4.75cm 5.5cm 4cm},clip,width=0.4775\textwidth]{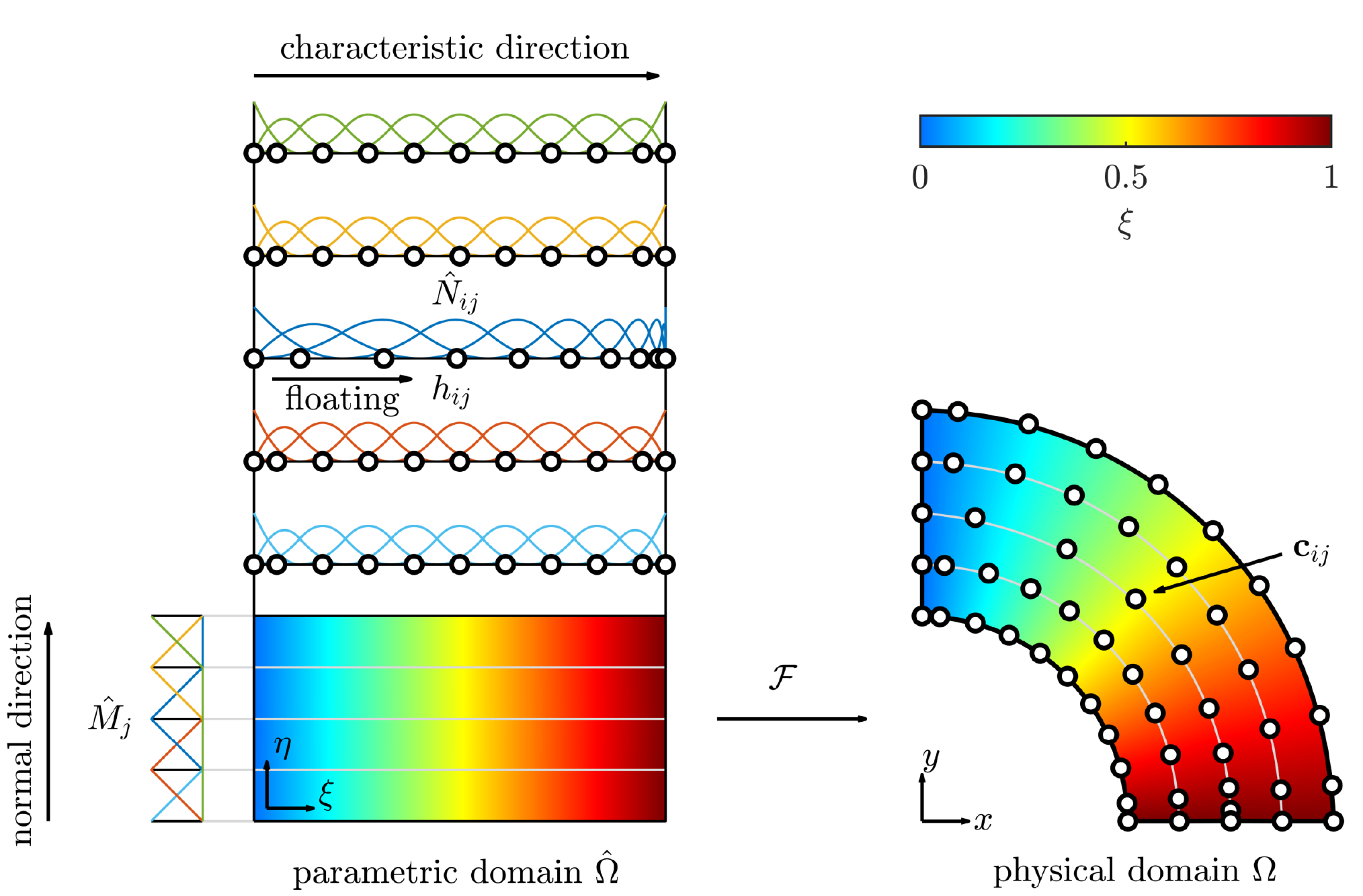}
\caption{Restoring an analysis-suitable parametrization by automated adjustment of floating regulation points.}
\label{fig:floatingrefinement}
\end{figure}

In \autoref{fig:floatingrefinement}, we illustrate the restoration of an undistorted parameterization by the above procedure. The distorted (left) and reparametrized (right) B-Spline discretizations only differ in terms of the floating regulation points, whose appropriate choice resolves the problem of shear distortion of the basis functions along the characteristic direction. However, a distortion of the basis functions in terms of elongation or compression (seen e.g. from the control point distances) remains. This is addressed in the next section.

\subsubsection{Local $h$-refinement}
\label{ssec:refinement}
In the following, we describe an adaptive procedure for knot insertion and/or removal, which solves the  problem of distortion of the basis functions in terms of elongation or compression. We finally outline the effect of this procedure on the quadrature scheme described in \autoref{ssec:quadrature}.
\paragraph{Adaptive knot insertion}
We propose the standard \textit{knot insertion} strategy \citep{PiegTill96} for $h$-refinement along the characteristic direction of a floating B-Spline geometry:
\begin{itemize}
\item Choose a $j\in\left\lbrace 1,...,J\right\rbrace$ specifying the characteristic basis to be refined;
\item choose a knot $\tilde{\xi}^+$ to insert which lies within knot span $\left(\tilde{\xi}_{k}, \tilde{\xi}_{k+1}\right)$ of parent knot vector $\tilde{\Xi}_{j}$;
\item for each $k-p+1\leq i \leq k$: Compute the ratio $a_i = \dfrac{\tilde{\xi}^+-\tilde{\xi}_i}{\tilde{\xi}_{i+p}-\tilde{\xi}_i}$ with $\tilde{\xi}_i, \tilde{\xi}_{i+p}\in\tilde{\Xi}_j$ as well as the
\subitem new floating regulation point $h^+_{ij}=(1-a_i)h_{(i-1)j}+a_i h_{ij}$
\subitem new control point $\bm{c}^+_{ij}=(1-a_i)\bm{c}_{(i-1)j}+a_i\bm{c}_{ij}$
\subitem new control variable for each field approximation, e.g. $\bm{d}^+_{ij}=(1-a_i)\bm{d}_{(i-1)j}+a_i \bm{d}_{ij}$;
\item for each $k-p+1\leq i \leq k-1$: remove the
\subitem old floating regulation point $h_{ij}$
\subitem old control point $\bm{c}_{ij}$
\subitem old control variable for each field approximation, e.g. $\bm{d}_{ij}$;
\item[] and insert in the gap the respective $p$ new quantities;
\item insert $\tilde{\xi}^+$ in $\tilde{\Xi}_j$ obtaining the refined parent knot vector $\tilde{\Xi}^+_j=\left\lbrace...,\ \tilde{\xi}_k,\ \tilde{\xi}^+,\ \tilde{\xi}_{k+1},\ ...\right\rbrace$.
\end{itemize}
Note that the number of characteristic basis functions associated to the specific $j$ as well as the global number of basis functions hereby increase by one. This refinement has \textit{local} nature since we construct the other characteristic bases from different parent knot vectors (recall \autoref{ssec:construction}). For spline order $q=1$ in $\eta$ direction all parametrizations $\mathcal{G}_{j}$ and $\mathcal{F}$ are preserved by the proposed refinement concept, hence also the material point character of the quadrature points.

We determine the knots to be inserted adaptively during the analysis by evaluating for each (non--zero) parent knot span $\tilde{\Omega}_{kj}=\left[\tilde{\xi}_{k},\right.\left.\tilde{\xi}_{k+1}\right]$ with $\tilde{\xi}_{k},\tilde{\xi}_{k+1}\in\tilde{\Xi}_j$ the length of its physical representation as curve segment
\begin{equation}
L_{kj}=\int_{\Omega_{kj}} 1\cdot\mathrm{d}s \approx \sum_{g} L^{gl} =\sum_g\tilde{w}^{gl}\sum^{I_j}_{i=1}\bm{c}_{ij}\dfrac{\partial}{\partial\tilde{\xi}}\tilde{N}_{ij}\left(\tilde{\xi}^{gl}\right),
\label{eq:segmentlength}
\end{equation} 
where $\Omega_{kj}=\mathcal{F}\left(\mathcal{G}_j\left(\tilde{\Omega}_{kj}\right)\right)$; we choose $l: \mathcal{S}(l)=j$ (inner interfaces are associated to two quadrature knot vectors, we can choose either one); and summation is over all $g: \tilde{\xi}^{gl}\in\mathcal{Q}^{Leg,G,l}\cap\tilde{\Omega}_{kj}$. We set a length threshold $T_{ins}$, and if $L_{kj}>T_{ins}$ we insert a new knot
\begin{equation}
\tilde{\xi}^+=0.5\cdot\left(\tilde{\xi}_k+\tilde{\xi}_{k+1}\right),
\label{eq:centralknotinsertion}
\end{equation}
at the center of $\tilde{\Omega}_{kj}$.

\paragraph{Adaptive knot removal}
The inverse process of knot insertion is called \textit{knot removal}. Note that unlike knot insertion, removal of knots conserves the exact parametrization only under specific circumstances \citep{PiegTill96}. For this reason, we follow one of the strategies proposed in \citep{ECK1995} for standard B-Spline curves, minimizing the change of geometry and parametrization in the sense of a certain $L^{\infty}$ norm. Its transfer to the floating tensor product structure yields an algorithm as follows:
\begin{itemize}
\item Choose a $j\in\left\lbrace 1,...,J\right\rbrace$ specifying the characteristic basis to be coarsened;
\item choose a knot $\tilde{\xi}^-=\tilde{\xi}_{k}$ to remove which bounds the knot span $\left[\tilde{\xi}_{k}, \tilde{\xi}_{k+1}\right)$ of parent knot vector $\tilde{\Xi}_{j}$;
\item if $p>1:$
\subitem compute or set:
\subsubitem $l_r = \dfrac{\tilde{\xi}^- - \tilde{\xi}_{k-p+r-1}}{\tilde{\xi}_{k+r} - \tilde{\xi}_{k-p+r-1}} \qquad \mathrm{for}\ r=1,...,p+1$,
\subsubitem $\gamma_{\infty}=\sum^p_{t=1}\begin{bmatrix}
p\\
t+1\\
\end{bmatrix},$
\subsubitem $\mu_r = \dfrac{1}{\gamma_{\infty}} \sum^{r-1}_{t=1}\begin{bmatrix}
p\\
t+1\\
\end{bmatrix} \qquad \mathrm{for}\ r=1,...,p-1$,
\subsubitem with bracket operator
\subsubitem \qquad $\begin{bmatrix}
b\\
a\\
\end{bmatrix} := \begin{cases}
\dfrac{1}{l_{a}}\prod^b_{r=a}\dfrac{1-l_{r+1}}{l_{r+1}}, & \mathrm{for}\ a\leq b\\
\dfrac{1}{l_{a}}, & \mathrm{for}\ a = b+1\\
\dfrac{1}{1-l_{a}}, &\mathrm{for}\ a=b+2,\\
\end{cases}$
\subsubitem $\bm{c}^{\mathrm{I}}_{1} = \bm{c}_{(k-p-1)(j)};\qquad\bm{c}^{\mathrm{I}}_{r}  = \dfrac{1}{l_r} \bm{c}_{(k+r-2)(j)} + \left(1-\dfrac{1}{l_r}\right)\bm{c}^{\mathrm{I}}_{r-1}\quad\mathrm{for}\ r=2,...,p,$
\subsubitem $\bm{c}^{\mathrm{II}}_{p+1} = \bm{c}_{(k)(j)};\qquad\bm{c}^{\mathrm{II}}_{r}  = \dfrac{1}{1-l_{r+1}} \bm{c}_{(k+p-r-1)(j)} + \left(1-\dfrac{1}{l_{r+1}}\right)\bm{c}^{\mathrm{II}}_{r+1}\quad\mathrm{for}\ r=p,p-1,...,2$;
\subitem for $r=1,...,p-1$: replace the old control point $\bm{c}_{(k-p+r-1)(j)}=\left(1-\mu_r\right)\bm{c}^{\mathrm{I}}_{r+1} + \mu_r \bm{c}^{\mathrm{II}}_{r+1}$;
\item remove the old control point $\bm{c}_{(k-1)(j)}$;
\item remove $\tilde{\xi}^-$ from $\tilde{\Xi}_j$ obtaining the coarsened parent knot vector $\tilde{\Xi}^-_j=\left\lbrace...,\ \tilde{\xi}_{k-1},\ \tilde{\xi}_{k+1},\ ...\right\rbrace$.
\end{itemize}
\begin{figure}[t!]
\centering
\includegraphics[trim={5.5cm 4.75cm 4.6cm 4cm},clip,width=1.0\textwidth]{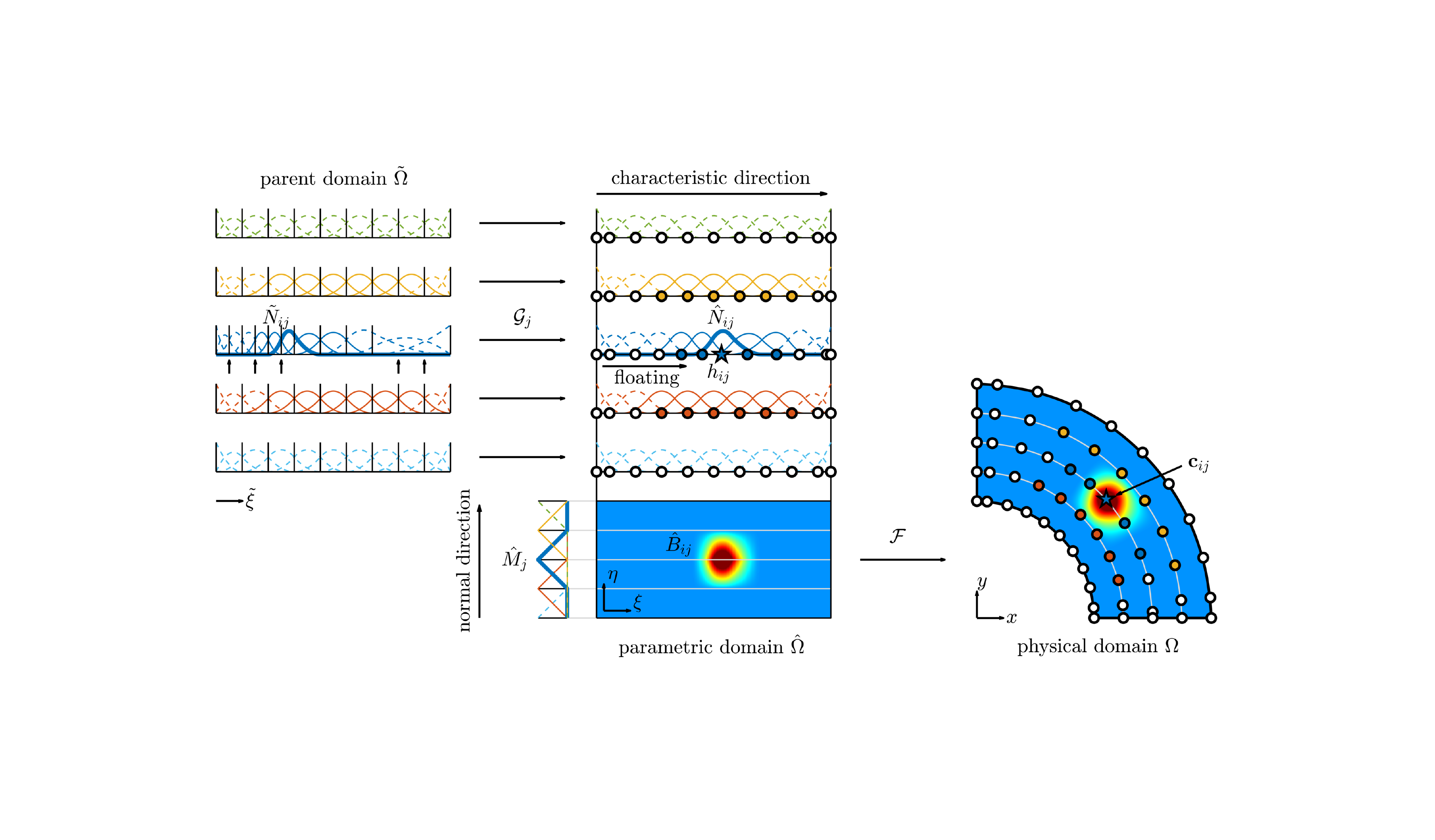}
\caption{Removed shear \textit{and} dilatational mesh distortion in FLIGA.}
\label{fig:refinement}
\end{figure}
The procedure for deriving the new floating regulation points and control variables of each field approximation is fully equivalent. Like knot insertion, knot removal along the characteristic direction is a local process for floating B-Splines, initiated for any parent knot span with physical span length $L_{kj}<T_{rem}$ where $T_{rem}$ is the removal threshold. \autoref{fig:refinement} demonstrates how the elongational/compressive mesh distortion that remained in \autoref{fig:meshdistortion2} is cured by adaptive knot insertion and removal in $\tilde{\Xi}_j$ (with $j=3$) at the locations indicated by arrows.

\paragraph{Effect on quadrature}
The quadrature knot vectors $\tilde{\Xi}^l$ introduced in \autoref{ssec:quadrature} are unaffected by knot insertion or removal in the parent knot vectors along the course of analysis. While we cannot precisely predict in advance which parent knot spans will be split how often, we simply initialize the quadrature point set by sufficiently dense quadrature knot vectors $\tilde{\Xi}^l$. The insertion of new quadrature points along with the mapping of history data is thereby avoided even upon adaptive refinement. To this end, before the analysis, we estimate (in terms of knot span density) the maximum refinement of the parent knot vectors occurring during the analysis, and accordingly construct all $\tilde{\Xi}^l$. With the convention to insert new parent knots only at \textit{centers} of parent knot spans (\autoref{eq:centralknotinsertion}) the one-time construction of $\tilde{\Xi}^l$ is straightforward and preserves all beneficial properties of the novel quadrature concept. To build the quadrature vectors, we propose a simple split of each knot span of the (assumed uniform\footnote{The idea is readily generalized to the non-uniform case as well, but $\tilde{\Xi}^l$ must be a superset of $\tilde{\Xi}_j$.}) initial parent knot vectors into a certain number of subspans, according to
\begin{equation}
\tilde{\Xi}^l=\mathcal{K}\left(\mathrm{max}\left(\rho_{\mathcal{S}(l)}n^S_{\mathcal{S}(l)},\ 2\rho_{\mathcal{N}(l)}n^S_{\mathcal{N}(l)}\right),p\right),
\label{eq:quadraturesetrefinement}
\end{equation}
where $\mathcal{K}(n^S,p)$ denotes an open uniform knot vector with $n^S$ (non-zero) knot spans for polynomial order $p$. Further, $n^S_j$ denotes the number of knot spans in the initial (unrefined) $\tilde{\Xi}_j$, and $\rho_j\geq1$ (chosen as a power of $2$) is the assumed maximum parent knot span relative shrinkage due to (possibly nested) refinement in $\tilde{\Xi}_j$. We recommend to multiply a factor of two to $\rho_{\mathcal{N}(l)}$ in (\autoref{eq:quadraturesetrefinement}), in order to improve quadrature accuracy for integration of the \textit{neighbor} characteristic basis (as its quadrature is not of Gauss character, recall \autoref{ssec:quadrature}). Note that an overestimation of the quadrature knot span density does not lead to any difficulties (except for an increase in computational cost), whereas an underestimation may cause problems due to under-integration after repeated knot insertion in the parent knot vectors.

\begin{figure}[t!]
\centering
\includegraphics[trim={5.75cm 5.1cm 4.25cm 4cm},clip,width=1.0\textwidth]{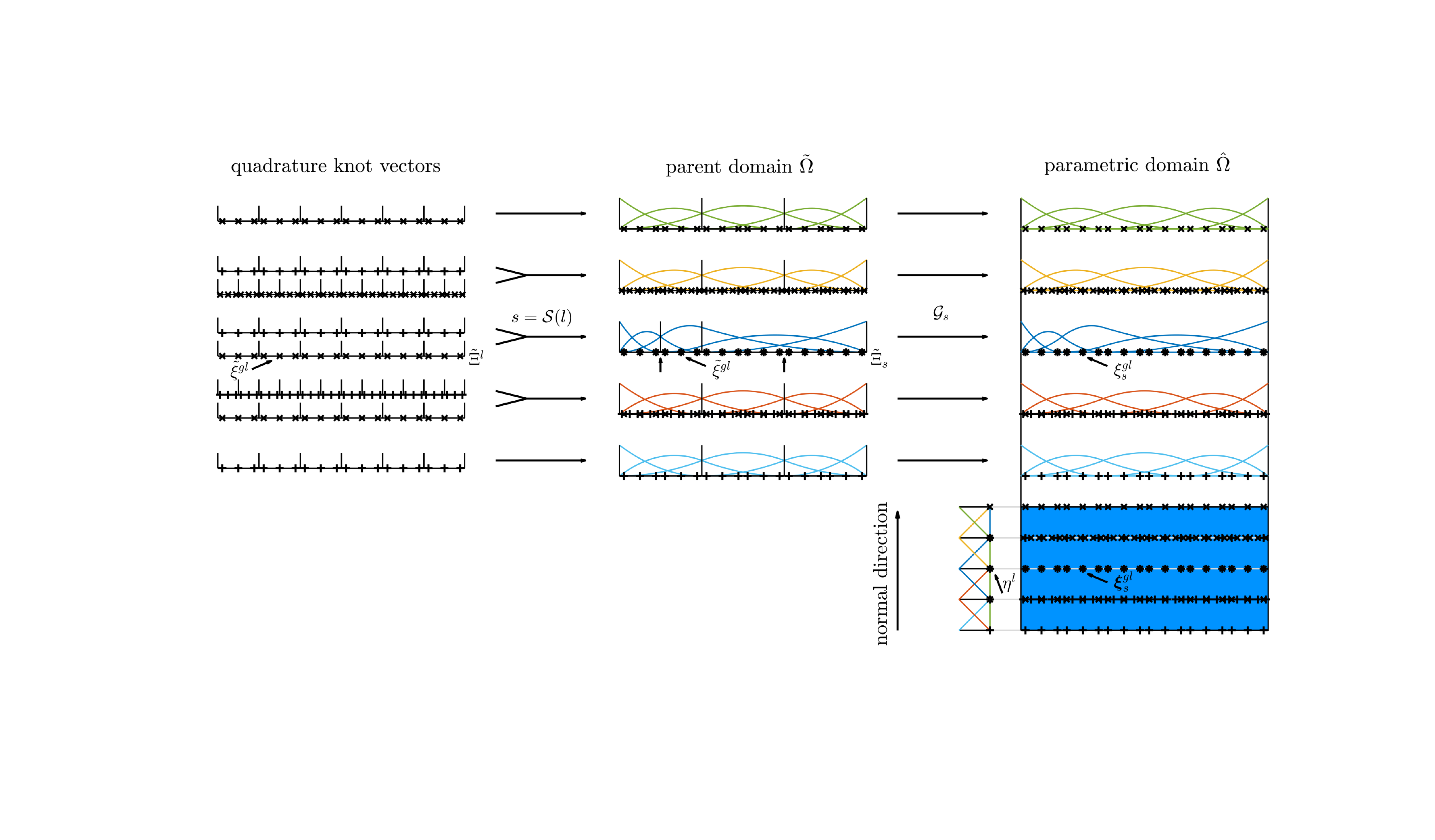}
\caption{Quadrature and refinement.}
\label{fig:quadratureconcept2}
\end{figure}

To make this clearer, let us give an example of how to construct the quadrature knot vectors for adaptive refinement, assuming that the course of refinement is known in advance. To this end, we consider once more the example of \autoref{fig:quadratureconcept1}, for simplicity in an ``unfloated'' configuration without loss of generality. In the figure, as refinement was not yet introduced in \autoref{ssec:quadrature} ($\rho_j=1$), we had equal and uniform parent knot vectors $\tilde{\Xi}_j = \tilde{\Xi}=\mathcal{K}(n^S,p)$ and equal and uniform quadrature knot vectors $\tilde{\Xi}^l=\mathcal{K}\left(2n^S,p\right)$, with $n^S=3$. Now we assume that the knot spans of parent knot vector $\tilde{\Xi}_3$ are expected to undergo a split to half the size, i.e. $\rho_3=2$, while for $j=1,2,4,5$ no refinement is expected, $\rho_{j\neq 3}=1$, see \autoref{fig:quadratureconcept2}. In this setting, according to (\autoref{eq:quadraturesetrefinement}), we construct the quadrature knot vectors:

\begin{tabular}{lll}
$\bullet\ l=1,2,7,8$: & $\tilde{\Xi}^l=\mathcal{K}\left(2\rho_{j\neq 3}n^S,p\right)$&$=\mathcal{K}\left(6,p\right)$\\
$\bullet\ l=3,6$: & $\tilde{\Xi}^l=\mathcal{K}\left(2\rho_{j=3}
n^S,p\right)$&$=\mathcal{K}\left(12,p\right)$\\
$\bullet\ l=4,5$: & $\tilde{\Xi}^l=\mathcal{K}\left(2\rho_{j\neq 3}n^S,p\right)=\mathcal{K}\left(\rho_{j=3}n^S,p\right)$&$=\mathcal{K}\left(6,p\right)$.
\end{tabular}

We now see the motivation for the introduction of quadrature knot vectors related to $l$ and not to $j$ in \autoref{ssec:quadrature}, in that superfluous points are avoided in $\tilde{\Xi}^2$ and $\tilde{\Xi}^7$. These are associated to either one of the two outer characteristic elements in which the refined basis functions of $\left\lbrace\hat{B}_{i3}\right\rbrace_{i=1,...,I_3}$ are not supported.

In many cases, the exact refinement is \textit{a priori} unknown, hence $\rho_j$ is just estimated. We highlight once more that the quadrature stencil is designed for the most refined state and employed at all load steps. Mapping of history data is not required.\footnote{For history-independent behavior the quadrature knot vectors can be adapted to the current parent knot vectors for cost reduction.}

\subsubsection{Function properties}

Among others, enhanced floating B-Splines possess the following analysis-suitable properties inherited from conventional B-Splines:
\begin{itemize}
\item partition of unity
\item weak Kronecker-delta property at the boundaries
\item first-order consistency
\item boundary preservation
\item interface preservation
\end{itemize}
In \cite{HKDL} we provided proofs for the first four of these properties whose extensions to the enhanced splines are trivial and therefore omitted here. The interface preservation property follows immediately from the Lagrangian character of the interface points which was shown in \autoref{ssec:quadrature}.

\subsubsection{Enhancements summary}

Let us conclude the overview of the enhancements with a brief summary. Firstly, the novel quadrature concept resolves all the drawbacks of the previous advection procedure in \cite{HKDL}, however, it solves the problem of loss of Gauss point character only to a large extent. Secondly, we have proposed an automated procedure for floating regulation to comfortably overcome shear distortion of the basis functions along the characteristic direction. Thirdly, adaptive knot insertion and removal prevent excessive basis function elongation/compression in the same direction. The combination of distortion-free approximation spaces with fully Lagrangian, yet highly systematic quadrature provides an ideal basis for numerical analysis of extreme, history-dependent deformations. As a limitation, the effectiveness of the proposed concepts relies on the orientation of the extreme deformations along one characteristic direction. Hence, the application to viscoelastic extrusion-based processes (\autoref{sec:Introduction}) is a natural choice. The related analysis concepts are introduced next.

\section{Continuum model, IGA and enhanced FLIGA}\label{sec:analysis}

In this section we first present the continuum mechanical model for viscoelastic finite deformation problems following the Lagrangian viewpoint of motion, and then illustrate the conventional IGA and enhanced FLIGA discretizations.

\subsection{Continuum problem}
Consider the quasi-static balance of linear momentum
\begin{equation}
\dfrac{\partial}{\partial \bm{x}} \cdot \bm{\sigma} + \bm{f} = \bm{0} \qquad\mathrm{in} \ \Omega,
\label{eq:PVPcontinuous}
\end{equation}
where $\bm{\sigma}$ is the Cauchy stress  and $\bm{f}$ the external body force, with Dirichlet boundary conditions on the velocity $\bm{v}$
\begin{equation}
\bm{v} = \bm{v}_D \qquad\mathrm{on} \ \partial\Omega_D,
\end{equation}
as well as Neumann boundary conditions
\begin{equation}
\bm{\sigma} \bm{n} = \bm{h}_N \qquad \mathrm{on} \ \partial\Omega_N.
\end{equation}
Here, $\bm{h}_N$ is the surface traction vector. Further, the velocity is derived as $\bm{v}=\dot{\bm{u}}$ from the displacement $\bm{u}$. We also enforce the incompressibility condition
\begin{equation}
\dfrac{\partial}{\partial \bm{x}} \cdot \bm{v} = 0 \qquad \mathrm{in} \ \Omega,
\label{eq:DIVcontinuous}
\end{equation}
and in preparation for viscoelastic material models, we split the Cauchy stress in a hydrostatic, a solvent, and a polymeric extra stress
\begin{equation}
\bm{\sigma} = -p \bm{\mathrm{I}} + 2\eta_s\nabla^s \bm{v} + \bm{\tau}_p.
\end{equation}
Here, $p$ is the pressure, $\nabla^s\left(\bullet\right)=\dfrac{1}{2}\left(\dfrac{\partial\left(\bullet\right)}{\partial \bm{x}}\right)+\dfrac{1}{2}\left(\dfrac{\partial\left(\bullet\right)}{\partial \bm{x}}\right)^T$ the symmetric gradient of vector $\left(\bullet\right)$, and $\eta_s$ the solvent viscosity. The polymeric extra stress $\bm{\tau}_p$ evolves according to an ordinary differential equation in time
\begin{equation}
\dot{\bm{\tau}}_p = \dot{\bm{\tau}}_p \left(\dfrac{\partial\bm{v}}{\partial \bm{x}},\bm{\tau}_p;\left\lbrace C_k\right\rbrace\right),
\label{eq:POLcontinuous}
\end{equation}
starting from the initial value $\bm{\tau}^0_p$. The set of material parameters $\left\lbrace C_k\right\rbrace$ may include parameters such as relaxation time $\lambda$ and polymeric viscosity $\eta_p$. A detailed introduction to viscoelastic modelling is given in \citep{BIRD1987}. In our work, we focus only on the classical Oldroyd-B model where
\begin{equation}
\dot{\bm{\tau}}_p = \dfrac{\partial\bm{v}}{\partial \bm{x}}\ \bm{\tau}_p + \bm{\tau}_p \ \left(\dfrac{\partial\bm{v}}{\partial \bm{x}}\right)^T-\dfrac{1}{\lambda}\left(\bm{\tau}_p - 2\eta_p\nabla^s\bm{v}\right).
\label{eq:OLDBcontinuous}
\end{equation}
Upon loading the body deforms over time, so that 
\begin{equation}
\bm{x}\left(\bm{X},t\right) = \bm{X} + \bm{u}\left(\bm{X},t\right) = \bm{X}+\int^t_0 \bm{v}\left(\bm{X},\tau \right) d\tau,
\label{eq:LAGcontinuous}
\end{equation}
where $\bm{X}$ is a generic coordinate in the initial configuration. By referring the description of motion to the initial configuration, we apply the Lagrangian viewpoint of motion. Note that the (material) time derivative $\dot{(\ )}$ in (\autoref{eq:POLcontinuous}) and (\autoref{eq:OLDBcontinuous}) describes the change of stress over time focusing on a material point $\bm{X}$.

For problems involving contact, we devise a penalty formulation with respect to wall penetration and penetration rate, augmented by another penalty formulation to regularize the no-slip constraint, $\bm{v}=0$. We arrive at the contact boundary condition at the wall interface $\partial\Omega_C$
\begin{equation}
\bm{\sigma} \bm{n} = -\kappa^*_P P \bm{n}-\kappa^*_R \dot{P} \bm{n} - \kappa^*_S\bm{v} \qquad \mathrm{on} \ \partial\Omega_C,
\label{eq:CONTACTcontinuous}
\end{equation}
where $P$ is the wall penetration, $\dot{P}$ the wall penetration rate and $\bm{n}$ the outward normal unit vector. Finally, the active penalty parameters, $\kappa^*_P,\ \kappa^*_R,\ \kappa^*_S$ are,
\begin{align}
\kappa^*_P = \begin{cases}\kappa_P, & \mathrm{for\ } P>0,\\ 0, & \mathrm{otherwise}, \end{cases}\qquad \kappa^*_R = \begin{cases}\kappa_R, & \mathrm{for\ }\dot{P}>0,\\ 0, & \mathrm{otherwise}, \end{cases}\qquad \kappa^*_S = \begin{cases}\kappa_S, & \mathrm{for\ }P>0,\\ 0, & \mathrm{otherwise,} \end{cases}
\label{eq:penaltyparameters}
\end{align}
based on the constant penalty coefficients, $\kappa_P,\ \kappa_R,\ \kappa_S$, respectively.

\subsection{Isogeometric discretization}
\label{ssec:IGA}

For classical IGA discretization of the problem presented above, we describe the geometry of the undeformed domain based on isogeometric basis functions, e.g. standard B-Splines (see \autoref{sec:bsplines} and references therein). With the geometry representation available, discretization in space (indicated by index $h$) and in time (indicated by time step index $n$ for time step $t^n$) of the unknown velocity and pressure fields follows in case of a mixed method, using (\autoref{eq:physicalpushforwardiga}), as
\begin{equation}
\bm{v}^{h,n}\left(\bm{x}\right) = \sum^M_{m=1} B^{TP}_m\left(\bm{x}\right)\bm{d}^n_m = \sum^M_{m=1} \hat{B}^{TP}_m\left(\bm{\xi}\right)\bm{d}^n_m = \bm{v}^{h,n}\left(\bm{\xi}\right),
\label{eq:VELapproximation}
\end{equation}
\begin{equation}
p^{h,n}\left(\bm{x}\right) = \sum^Z_{z=1} A^{TP}_z\left(\bm{x}\right)q^n_z = \sum^Z_{z=1} \hat{A}^{TP}_z\left(\bm{\xi}\right)q^n_z = p^{h,n}\left(\bm{\xi}\right),
\label{eq:PRESSapproximation}
\end{equation}
where $\left\lbrace \hat{B}^{TP}_m \right\rbrace_{m=1,...,M}$ and $\left\lbrace \hat{A}^{TP}_z \right\rbrace_{z=1,...,Z}$ are classical B-Spline bases and $\bm{d}^n_m$ and $q^n_z$ are the current velocity and pressure control variables, respectively. Further, the isoparametric concept is applied, meaning that $\left\lbrace \hat{B}^{TP}_m \right\rbrace_{m=1,...,M}$ here is the same basis that describes the mapping between parametric and physical space, (\autoref{eq:physicalmapiga}). 
Inserting the velocity ansatz in the weak form of (\autoref{eq:PVPcontinuous}) leads to the equilibrium of internal and external force vectors at each control point $\bm{c}^n_m$ and each time step $n$
\begin{equation}
\bm{F}^n_{int,m} = \sum_{g=1}^{n^{QP}} \left[\left. \left(\bm{\sigma}^{h,n}\ \dfrac{\partial}{\partial \bm{x}} \hat{B}^{TP}_m\right)\right|_{\bm{\xi}^g} W^{g,n}\right] \overset{!}{=} \sum_{g=1}^{n^{QP}} \left[\left. \left(\bm{f}^n \hat{B}^{TP}_m \right)\right|_{\bm{\xi}^g} W^{g,n}\right] = \bm{F}^n_{ext,m} .
\label{eq:PVPdiscrete}
\end{equation}
Here we assume $\bm{h}^n_N=\bm{0}$ on the Neumann boundary, which is satisfied naturally in the weak formulation. Dirichlet boundary conditions can be incorporated \textit{a priori} in the solution space. Numerical integration is applied over Gauss-Legendre quadrature points with coordinates $\bm{\xi}^g$ and physical weights $W^{g,n}$, where $g=1,...,n^{QP}$. Note that each $\bm{\xi}^g$ is constant over time in standard IGA.

Likewise, we insert the velocity ansatz in the weighted residual form of (\autoref{eq:DIVcontinuous}), obtaining one equation for each pressure control variable at each time step
\begin{equation}
Q^n_z = -\sum_{g=1}^{n^{QP}} \left[\left. \left(\dfrac{\partial}{\partial \bm{x}} \cdot \bm{v}^{h,n} \ \hat{A}^{TP}_z\right)\right|_{\bm{\xi}^g} W^{g,n}\right] \overset{!}{=} 0.
\label{eq:DIVdiscrete}
\end{equation}
We multiply here by $-1$ to obtain a symmetric stiffness matrix later. The discretized Cauchy stress evaluated at $\bm{\xi}^g$ is given by
\begin{equation}
\bm{\sigma}^{h,g,n} = - p^{h,g,n} \bm{\mathrm{I}} + 2\eta_s\nabla^s \bm{v}^{h,g,n} + \bm{\tau}^{h,n,g}_{p}\left(\dfrac{\partial\bm{v}^{h,g,n-1}}{\partial \bm{x}},\bm{\tau}^{h,g,n-1}_{p}; \left\lbrace C_k\right\rbrace; \Delta t\right),
\label{eq:CAUCHYdiscrete}
\end{equation}
where $\Delta t$ is the time step size. This expression makes use of the time discretization of (\autoref{eq:POLcontinuous}) by the forward Euler method, so that the polymeric extra stress is independent of the current unknown solution. In the Lagrangian framework, history data at each Gauss-Legendre point are available from the previous time step $n-1$. For the Oldroyd-B model in (\autoref{eq:OLDBcontinuous}), we obtain the current polymeric extra stress by
\begin{equation}
\bm{\tau}^{h,g,n}_{p} = \bm{\tau}^{h,g,n-1}_{p} + \Delta t\left[\dfrac{\partial\bm{v}^{h,g,n-1}}{\partial \bm{x}}\ \bm{\tau}^{h,g,n-1}_{p} + \bm{\tau}^{h,g,n-1}_{p} \ \left(\dfrac{\partial\bm{v}^{h,g,n-1}}{\partial \bm{x}}\right)^T-\dfrac{1}{\lambda}\left(\bm{\tau}^{h,g,n-1}_{p} - 2\eta_p\nabla^s\bm{v}^{h,g,n-1}\right)\right].
\label{eq:OLDBdiscrete}
\end{equation}
Let us also discretize the rate form of (\autoref{eq:LAGcontinuous}) in space  and time
\begin{equation}
\dot{\bm{x}}^{h,n}\left(\mathcal{F}^0\left(\bm{\xi}\right)\right)=\bm{v}^{h,n}\left(\mathcal{F}^0\left(\bm{\xi}\right)\right),
\label{eq:Lagrangediscrete}
\end{equation}
where $\mathcal{F}^0$ is the physical mapping at the initial time step $n=0$, or equivalently
\begin{equation}
\dot{\bm{x}}^{h,n}\left(\bm{\xi}\right)=\bm{v}^{h,n}\left(\bm{\xi}\right).
\label{eq:Lagrangediscretesimple}
\end{equation}
Inserting the basis functions (\autoref{eq:physicalmapiga}) and (\autoref{eq:VELapproximation}) yields
\begin{equation}
\sum^M_{m=1} \dot{\bm{c}}^n_m \hat{B}^{TP}_m\left(\bm{\xi}\right)=\sum^M_{m=1} \bm{d}^n_m \hat{B}^{TP}_m\left(\bm{\xi}\right),
\label{eq:Lagrangewithbasisfunctions}
\end{equation}
which is then discretized again by the forward Euler method, resulting in
\begin{equation}
\bm{c}^{n}_m = \bm{c}^{n-1}_m+\Delta t\ \bm{d}^{n-1}_m.
\label{eq:igaeuler}
\end{equation}
Hence the construction of basis functions at the current configuration is independent of the current velocity solution. Collecting the individual equations from (\autoref{eq:PVPdiscrete}) and (\autoref{eq:DIVdiscrete}) for each time step, we obtain the global residual vectors
\begin{equation}
\bm{R}^n(\bm{U}^n)=\bm{F}^n_{int}(\bm{U}^n)-\bm{F}^n_{ext}\overset{!}{=}\bm{0},
\label{eq:PVWglobalsystem}
\end{equation}
and
\begin{equation}
\bm{Q}^n(\bm{U}^n)\overset{!}{=}\bm{0},
\label{eq:DIVglobalsystem}
\end{equation}
respectively. $\bm{U}^n$ contains the control variables $\bm{d}^n_m$ and $q^n_z$. Here, the linearization of $\left[{\bm{R}^n}^T, {\bm{Q}^n}^T\right]^T$ w.r.t. $\bm{U}^n$ gives a symmetric global stiffness matrix $\bm{K}^n$ (\ref{tangentstiffness}) which is independent of $\bm{U}^n$ and involves only current physical basis function values $\dfrac{\partial}{\partial \bm{x}}\hat{B}^{TP}_m\left(\bm{\xi}^g\right)$ and $\hat{A}^{TP}_z\left(\bm{\xi}^g\right)$ and physical quadrature weights $W^{g,n}$. Note that linearity of the discrete system is due to the explicit Lagrangian treatment of (\autoref{eq:OLDBdiscrete}) and (\autoref{eq:igaeuler}). 

For contact problems the translation of (\autoref{eq:CONTACTcontinuous}) into weighted residual form and discretization gives the nodal contact force at a boundary $b=1,2,3,4$ of the patch
\begin{align}
\bm{F}^n_{C,m} = -\sum_{g=1}^{n^{QP}_b} \left[\left(\kappa^{h,*}_P P^{h,n}\bm{n}^{h,n} + \kappa^{h,*}_R \dot{P}^{h,n}\bm{n}^{h,n} + \kappa^{h,*}_S\bm{v}^{h,n}\right) \hat{B}^{TP}_m\right]_{\bm{\xi}=\bm{\xi}^{g}} L^{g,n},
\label{eq:CONTACTdiscrete}
\end{align}
where $P^{h,n}$ and $\dot{P}^{h,n}=\bm{v}^{h,n}\cdot\bm{n}^{h,n}$ are the discrete penetration and penetration rate; $\bm{n}^{h,n}$ is the discrete current outward normal unit vector; and numerical quadrature is applied over $g=1,...,n^{QP}_b$ boundary Gauss-Legendre points with weight $L^{g,n}$ along the respective physical boundary. We define the active penalty parameters, $\kappa^{h,*}_P,\ \kappa^{h,*}_R,\ \kappa^{h,*}_S$ in the discrete setting as
\begin{align}
\kappa^{h,*}_P = \begin{cases}\kappa_P, & \mathrm{for\ } P^h>0,\\ 0, & \mathrm{otherwise}, \end{cases}\qquad \kappa^{h,*}_R = \begin{cases}\kappa_R, & \mathrm{for\ }\dot{P}^h>0,\\ 0, & \mathrm{otherwise,} \end{cases}\qquad \kappa^{h,*}_S = \begin{cases}\kappa_S, & \mathrm{for\ }P^h>0,\\ 0, & \mathrm{otherwise}. \end{cases}
\label{eq:penaltyparametersdiscrete}
\end{align}
The nodal contact forces are added to the external force vector
\begin{equation}
\bm{F}^n_{ext,C,m}=\bm{F}^n_{ext,m}+\bm{F}^n_{C,m},
\end{equation}
leading to the nodal balance
\begin{equation}
\bm{F}^n_{int,m}=\bm{F}^n_{ext,C,m}.
\end{equation}
The final global system becomes
\begin{equation}
\bm{R}^n_C(\bm{U}^n)=\bm{F}^n_{int}(\bm{U}^n)-\bm{F}^n_{ext,C}(\bm{U}^n)\overset{!}{=}\bm{0},
\label{eq:PVWglobalsystemcontact}
\end{equation}
supplemented with (\autoref{eq:DIVglobalsystem}). It is nonlinear in $\bm{U}^n$ due to contact and we solve it with the Newton-Raphson method. The again symmetric global tangent matrix $\bm{K}^n_C(\bm{U}^n)$ is assembled from the nodal tangent matrix contributions as formulated in \ref{contacttangentstiffness}.

Let us add the remark that, since we assume incompressible material behavior as per (\autoref{eq:DIVdiscrete}), it is important to properly select the discretization spaces for velocity and pressure to preclude effects of volumetric locking. Two important candidates are \textit{(i)} the so-called \textit{Taylor-Hood} pair where the pressure basis $\left\lbrace\hat{A}\right\rbrace_{z=1,...,Z}$ is of one polynomial degree less than the velocity basis (see \cite{BUFFA11}), and \textit{(ii)} the construction of the velocity basis $\left\lbrace\hat{B}\right\rbrace_{m=1,...,M}$ as a subdivision of the pressure basis (see \cite{RUBERG12}). Due to the tensor product structure of the B-Spline bases, these concepts are applicable at the level of the constituent univariate bases.

The reviewed IGA procedure is conceptually straightforward. Unfortunately, due to the movement of the control points, the mapping of the parametrically static basis functions undergoes a distortion, recall \autoref{fig:meshdistortion1}. When extreme deformations take place, this may lead to complete failure of the analysis.

\subsection{Enhanced FLIGA}
\label{ssec:FLIGA}

\begin{figure}
\centering
\includegraphics[trim={0cm 0cm 0cm 0cm},clip,width=10.5cm]{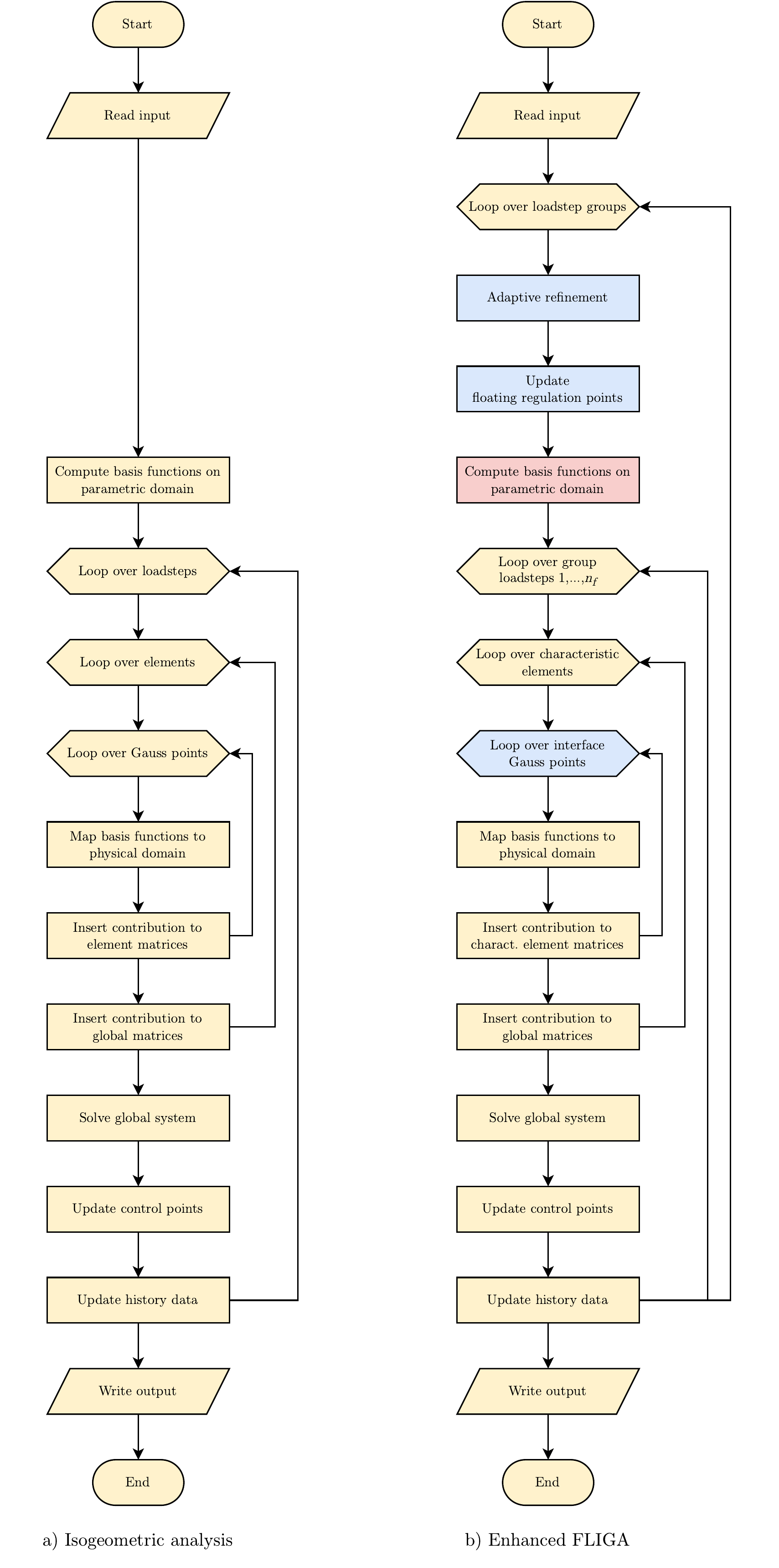}
\caption{Algorithm flow chart comparison.}
\label{fig:flowchart}
\end{figure}

We now outline a discretization method following IGA (\autoref{ssec:IGA}) but with the floating tensor product B-Splines and the enhancements (improved quadrature, automated floating regulation, adaptive refinement) from \autoref{sec:bsplines}. The most important modifications to the IGA discretization described in \autoref{ssec:IGA} are the following: 
\begin{itemize}
\item Enrich the approximation bases by floating regulation points (\autoref{ssec:construction}-\autoref{ssec:floatingeffect}).
\item Use linear polynomial basis functions in $\eta$ (\autoref{ssec:normalbasis}).
\item Place all quadrature points at the characteristic element boundaries along $\xi$ (\autoref{ssec:quadrature}).
\item Update floating regulation points after regular time step increments (\autoref{ssec:updates}).
\item Introduce an adaptive local refinement procedure along $\xi$ (\autoref{ssec:refinement}).
\end{itemize}
The governing equations are easily transferred to FLIGA by replacing
\begin{align}
\begin{split}
\hat{B}^{TP}_m&\leftarrow\hat{B}^n_m\\
\bm{\xi}^g&\leftarrow\bm{\xi}^{gl,n}_s\\
W^{g,n}&\leftarrow W^{gl,n}_s,
\end{split}
\end{align}
in (\autoref{eq:PVWglobalsystem}) or (\autoref{eq:PVWglobalsystemcontact}), respectively, and in (\autoref{eq:DIVglobalsystem}). Here we need the current floating B-Splines $\hat{B}^n_m$, novel quadrature points $\bm{\xi}^{gl,n}_s$ and weights $W^{gl,n}_s$.

Further, we propose the construction of stable mixed approximation spaces in enhanced FLIGA following the subdivision strategy in the normal direction, as well as the Taylor-Hood or subdivision approach along the characteristic direction. Subdivision in normal direction is preferred as we have $q=1$ and, by nesting two velocity spans within one pressure span, we can conveniently reuse the floating maps $\mathcal{G}_{j=1,3,5,...,J}$ applied to the velocity basis for floating regulation of the pressure basis. This facilitates the overall procedure as we avoid the introduction of additional floating regulation points for pressure, and we can even use the same quadrature coordinates to compute all velocity and pressure contributions. The concepts for combining the sets of univariate bases to a bivariate floating basis (\autoref{ssec:construction}) are then correspondingly employed. (As in \autoref{ssec:IGA}, we apply the physical map $\mathcal{F}$ for the velocity basis also to the pressure basis.)

In \autoref{fig:flowchart} we schematically overview the procedures followed by IGA (a) and enhanced FLIGA (b). The introduction of floating regulation points in FLIGA mainly leads to modification of the basis computation in parametric coordinates (red block). Moreover, the operations for the three blocks highlighted in blue correspond to the three enhancements of this work. Let us add a remark on optimizing the computational cost. In the flow chart, we group a certain number of time steps and perform the new ``floating'' operations on the parametric bases only after each group, i.e., say, every $n_f=5$ time steps. During the $5$ time steps of each group, the mesh distorts (slightly), which is then ``relaxed'' after the subsequent floating operations. Groups of $5+$ time steps are often unproblematic (this obviously also depends on the choice of the time step size) and quadrature points remain fully Lagrangian. Following this strategy the increase of average cost per load step introduced by the floating procedures is quite limited, for more details see \autoref{ssec:taylorcouette}.

\section{Numerical examples}
\label{sec:results}

In this section, we conduct a numerical study of enhanced FLIGA. The patch test, Taylor-Couette flow, planar extrusion and a small set of extrusion-based AM processes serve as examples. Where possible, we directly compare results to those of the first version of FLIGA proposed in \citep{HKDL}, denoted as FLIGA 1.0 in the following. Accordingly, we denote enhanced FLIGA also as FLIGA 2.0.

Numerical approximations are assessed in terms of the logarithmic $L^2$ error norm
\begin{equation}
L^2\left(v^h\right)=\log_{10} \left(\dfrac{\sqrt{\int_{\Omega^h} \left(v-v^h\right)^2\ \mathrm{d}\Omega}}{\sqrt{\int_{\Omega^h} v^2\ \mathrm{d}\Omega}}\right),
\label{eq:L2}
\end{equation}
where $v$ and $v^h$ are components of the analytical and numerical solutions, respectively.

\subsection{Patch test}
\label{ssec:patchtest}

\begin{figure}[t]
\centering
\includegraphics[trim={5.25cm 2.9cm 2.5cm 2.25cm},clip,width=0.575\textwidth]{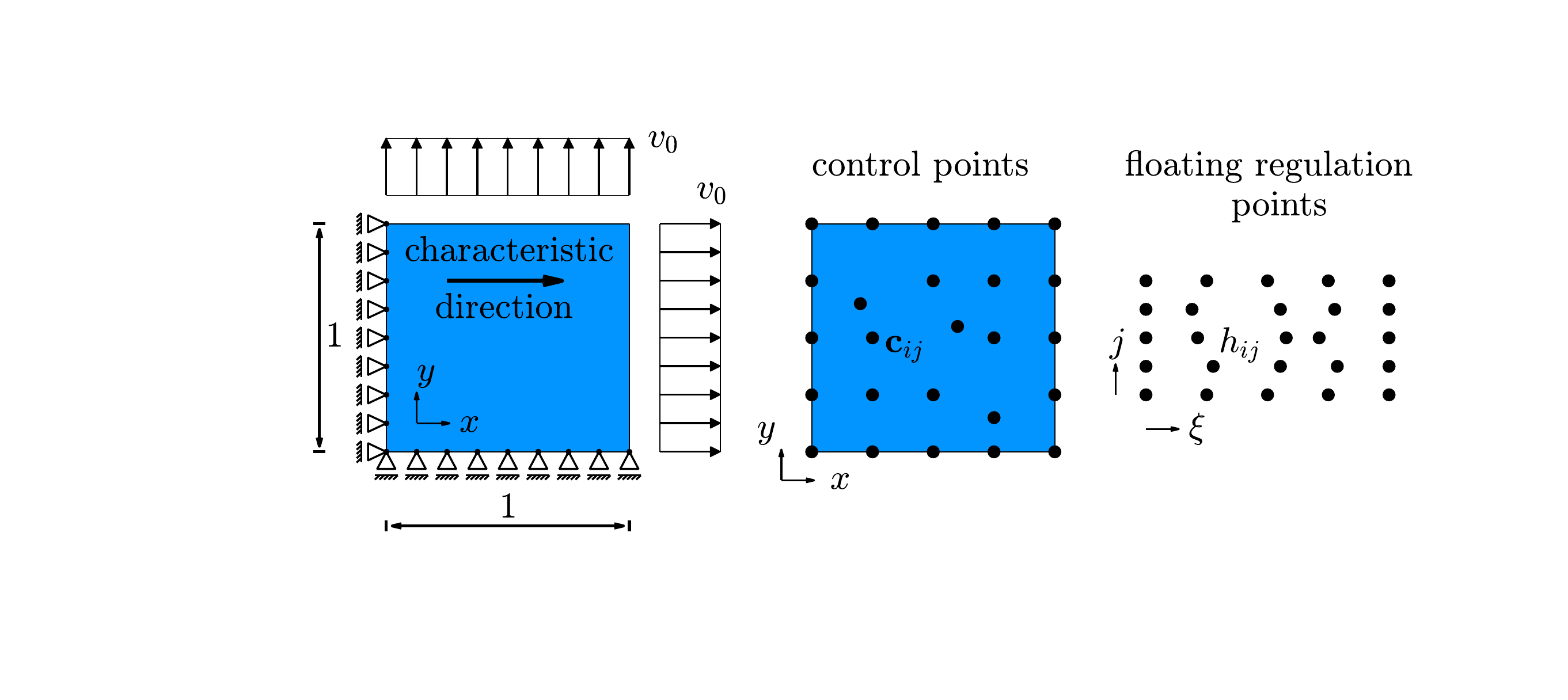}
\caption{Patch test problem.}
\label{fig:patchtest}
\end{figure}
We start with a two-dimensional patch test, similar to the one in \citep{HKDL}. Here we consider a square domain occupied by Newtonian fluid ($\eta_s=50$, $\eta_p=0$) subjected to Dirichlet boundary conditions in the normal directions as shown in \autoref{fig:patchtest}. The imposed velocity is chosen as $v_0=1$. The pressure is set to $p=0$ and the incompressibility constraint is removed, so as to enable a simple volumetric expansion of the domain with analytical solution
\begin{equation}
\bm{v}=
\begin{pmatrix}
v_x\\
v_y
\end{pmatrix}
=
\begin{pmatrix}
x\\
y\\
\end{pmatrix},
\label{eq:patchtestanalytical}
\end{equation}
with a gradient of
\begin{equation}
\dfrac{\partial \bm{v}}{\partial \bm{x}} =
\begin{pmatrix}
1 & 0\\
0 & 1
\end{pmatrix}.
\label{eq:patchtestanalyticalgrad}
\end{equation}
Hence, the velocity component in $x$ (in $y$) has a constant gradient pointing in the $x$ (in the $y$) direction.

In order to enable an easier comparison with \citep{HKDL}, we adopt the control and floating regulation points employed there and the automated placement procedure on the floating regulation points introduced in \autoref{ssec:updates} is not applied. We consider three test cases $p=1,2,3$ (while, as always, $q=1$). Note that due to the fixed number of control points, the open uniform parent knot vectors $\tilde{\Xi}_j=\mathcal{K}\left(n^S_p,p\right)$ are different for each $p$ (as we have the knot span counts $n^S_1=4,n^S_2=3,n^S_3=2$), and we construct the quadrature knot vectors $\tilde{\Xi}^l=\mathcal{K}\left(2n^S_p,p\right)$ to be twice as dense. Each knot span of any $\tilde{\Xi}^l$ collects $G=p+1$ quadrature points and a total of roughly $n^{QP}=100$ quadrature points is obtained for each $p$, allowing for a fair comparison to the FLIGA 1.0 patch test, where we employed a similar number.

An overview of the patch test results for the $L^2$ errors (\autoref{eq:L2}) w.r.t. the analytical velocity components (\autoref{eq:patchtestanalytical}) is provided in \autoref{tab:patchtest}, where we also reproduce the errors obtained previously for IGA and FLIGA 1.0, however with $q=p$ \citep{HKDL}.
\begin{table}[tb]
\centering
\begin{tabular}{l l c l l l l} 
 \toprule
 Polynomial order & Velocity component & $p=1$ & $p=2$ & $p=3$ \\
                                    & & $q=1$ & $q=2$ & $q=3$ \\ [0.5ex] 
 \hline
 \multirow{2}{10em}{IGA} & $x$ & $-16.28$ & $-15.67$ & $-15.73$ \\
  & $y$ & $-16.02$ & $-15.85$ & $-15.63$ \\
  \hline
  \multirow{2}{10em}{FLIGA 1.0} & $x$ (characteristic) & $-1.93$ & $-2.46$ & $-2.88$ \\
 & $y$ (normal) & $-2.86$ & $-3.78$ & $-3.94$ \\
  \hline
  \hline\\ [-1.5ex] 
 \multicolumn{2}{l}{Polynomial order} & $p=1$ & $p=2$ & $p=3$ \\
                                    & & $q=1$ & $q=1$ & $q=1$ \\ [0.5ex] 
 \hline
  \multirow{2}{10em}{FLIGA 2.0} & $x$ (characteristic) & $-3.79$ & $-4.80$ & $-7.13$ \\
  & & $(-4.73)$ & $(-6.67)$ & $(-8.50)$ & \\
 & $y$ (normal) & $-3.04$ & $-4.45$ & $-6.18$ \\
  & & $(-4.32)$ & $(-5.95)$ & $(-8.11)$ & \\
 \bottomrule
\end{tabular}
\caption{$L^2$ errors for the patch test. The bracketed values are obtained for quadrature knot spans \textit{eight} times smaller than the parent knot spans.}
\label{tab:patchtest}
\end{table}
As expected, IGA passes the patch test up to machine precision. With FLIGA 1.0 we had observed a limited accuracy for the small number of integration points. Conversely, for FLIGA 2.0 the patch test reveals distinctive reductions of the error by employment of the novel quadrature stencil. The improvements are most pronounced for higher polynomial orders and solution field gradients along the characteristic direction (corresponding to the solution component $v^h_x$, see (\autoref{eq:patchtestanalyticalgrad})). The numerical error further decreases as the knot span density in $\tilde{\Xi}^l$ is increased. E.g., the values in brackets in \autoref{tab:patchtest} are obtained for $\tilde{\Xi}^l=\mathcal{K}\left(8n^S_p,p\right)$. This error reduction is possible up to machine precision due to the first-order consistency property of floating B-Splines.

\subsection{Taylor-Couette flow}
\label{ssec:taylorcouette}

\begin{table}[t!]
\centering
\begin{tabular}{l l l l l l l l l}
\hline
\textit{Problem parameter} & \textit{Variable} & \textit{Value} & \textit{Unit} & & \textit{Simulation parameter} & \textit{Variable} & \textit{Value} & \textit{Unit}\\
\hline
radius & & & & & polynomial order & & &  \\
\quad inner cylinder & $R_I$ & $0.1$ & m & & \quad characteristic axis& $p$ & $\left\lbrace 2,3\right\rbrace$ & \\
\quad outer cylinder & $R_O$ & $0.2$ & m & & \quad normal axis& $q$ & $1$ & \\
angular frequency & & & & & element count & & \\
\quad inner cylinder & $\Omega_I$ & $0$ & rad/s & & \quad velocity & & $36 \times \left\lbrace 24,72\right\rbrace$ & \\
\quad outer cylinder & $\Omega_O$ & $7.5$ & rad/s & & \quad pressure & & $18 \times \left\lbrace 12,36\right\rbrace$ & \\
relaxation time & $\lambda$ & $0.1$ & s & & time step Newtonian & $\Delta t_{Newt}$ & $5$e$-5$ & s\\
polymeric viscosity & $\eta_p$ & $1.5$ & Pa$\cdot$s & & time step Oldroyd-B & $\Delta t_{Old}$ & $\left\lbrace5\mathrm{e}-5,1\mathrm{e}-5\right\rbrace$ & s\\
solvent viscosity & $\eta_s$ & $0.5$ & Pa$\cdot$s & & floating update interval & $n_f$ & $20$ & \\
polymer stress at $t=0$& $\bm{\tau}^0_{p}$ & $\bm{0}$ & Pa & & & & & \\
\hline
\end{tabular}
\caption{Parameters for the Taylor-Couette problem. Value pairs in braces refer to $p=2$ and $p=3$, respectively.}
\label{tab:taylorcouetteparameters}
\end{table}

\begin{figure}[t!]
    \centering
    \includegraphics[trim={0cm 0cm 0cm 0cm},clip,width=0.8\textwidth]{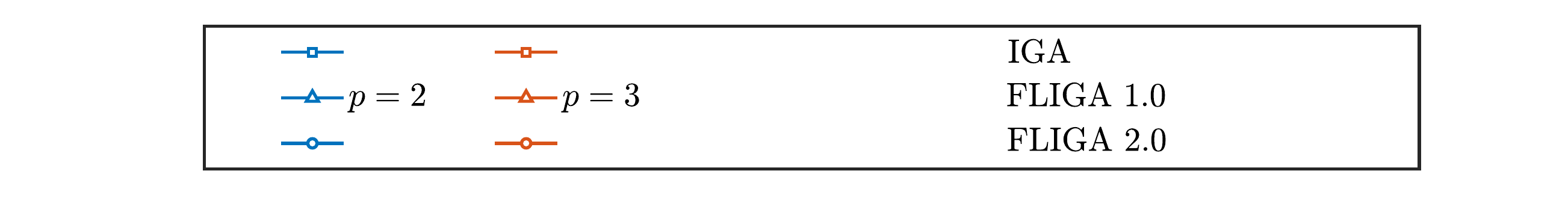}
    \qquad
    \subfigure[\centering Newtonian, horizontal velocity.]{\includegraphics[trim={0cm 0cm 0cm 0cm},clip,width=0.8\textwidth]{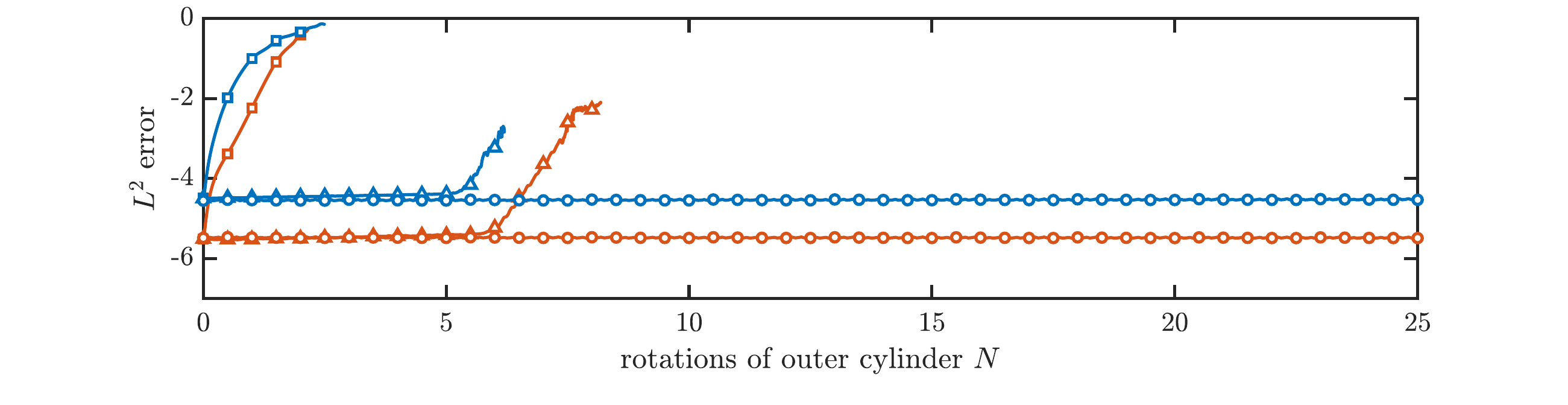}  }
    \qquad
    \subfigure[\centering Oldroyd-B, horizontal velocity.]{\includegraphics[trim={0cm 0cm 0cm 0cm},clip,width=0.8\textwidth]{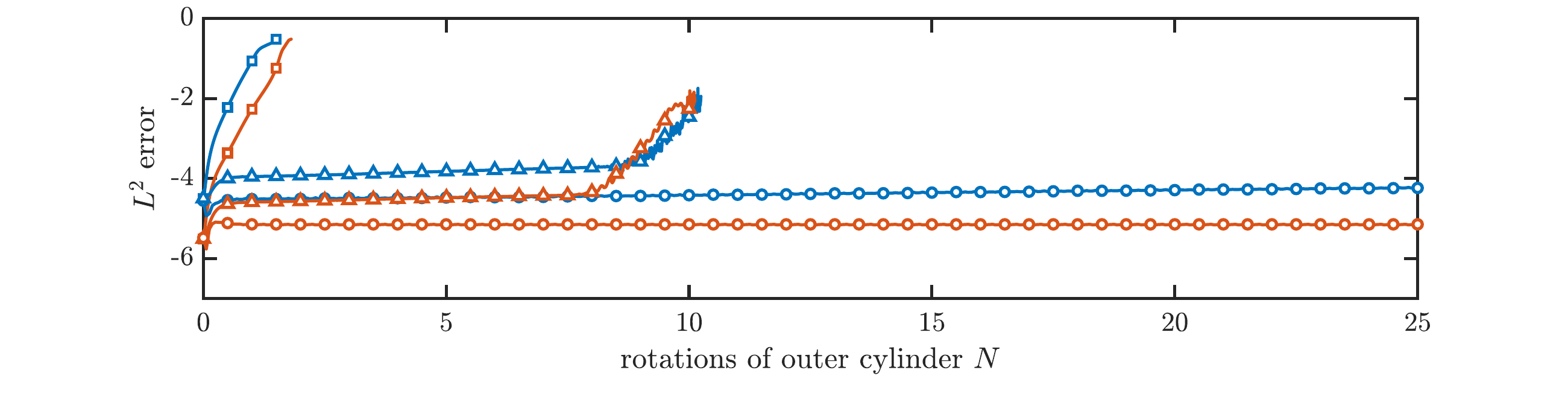} }
    \qquad
    \subfigure[\centering Oldroyd-B, pressure.]{\includegraphics[trim={0cm 0cm 0cm 0cm},clip,width=0.8\textwidth]{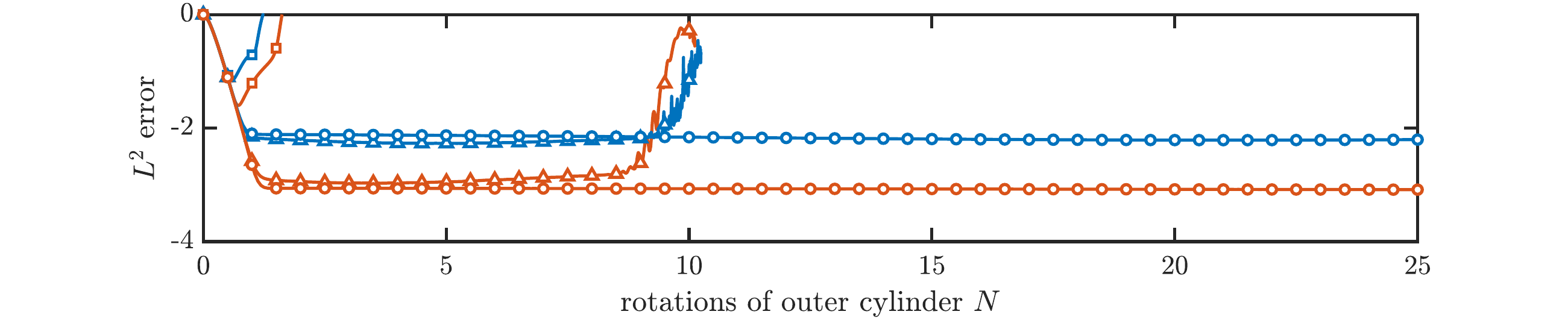} }
    \caption{Transient $L^2$ errors w.r.t. the analytical steady-state solutions for the Taylor-Couette problem.}
    \label{fig:tcL2}
\end{figure}

Next, we revisit the two-dimensional problem of Taylor-Couette flow for Newtonian as well as viscoelastic Oldroyd-B fluid behavior, which we investigated with FLIGA 1.0 in \citep{HKDL}. We represent the annulus of fluid by a single patch, treating the circumferential direction as the characteristic one and employing a periodic boundary condition to close the ring structure\footnote{In \cite{HKDL} we provide the details about how to construct periodic floating B-Spline bases.}. While with FLIGA 1.0 we had $q\geq1$, here $q=1$, thus we increase the number of elements in normal direction in order to retain a comparable accuracy. Quadrature is performed by the novel procedure constructing $\tilde{\Xi}^l=\mathcal{K}\left(2n^S,p\right)$ twice as dense as $\tilde{\Xi}_j = \mathcal{K}\left(n^S,p\right)$ with $n^S=36$. For the study of the Oldroyd-B model, we double the time step size w.r.t. \cite{HKDL}. In addition, we reduce the computational cost by applying the floating operations only every $n_f=20$ time steps. All other parameters are given in \autoref{tab:taylorcouetteparameters}.

We compute the $L^2$ errors (\autoref{eq:L2}) of horizontal velocity and pressure for different rotation states of the outer cylinder (with the inner one fixed). The analytical (steady-state) solutions are
\begin{equation}
v_{x,Newt}=v_{x,Old}= \sin (\varphi)\left( \alpha r + \dfrac{\beta}{r} \right),\qquad p_{Newt}=0,\qquad p_{Old}= 2\pi\beta^2\lambda\left(\dfrac{1}{r^4}-\dfrac{1}{R^4_O}\right),
\end{equation}
with
\begin{equation}
\alpha = \Omega_O\cdot \dfrac{R^2_O}{R^2_O-R^2_I},\qquad \beta = -\alpha R^2_I,
\end{equation}
and the polar coordinates $r=\sqrt{x^2+y^2}$, $\varphi=\mathrm{atan}\left(\frac{y}{x}\right)$. Results are provided in \autoref{fig:tcL2} and demonstrate the significantly improved stability of enhanced FLIGA over FLIGA 1.0. We emphasize that adaptive refinement is not applied here and that very similar results are also obtained when using the old manual procedure for updating floating regulation points (instead of the new automated one). Consequently, we believe that the improvements are due to the novel quadrature concept.

We study the computation cost for enhanced FLIGA. To this end, let us consider the case $p=2,\ q=1$ with spatial discretization according to \autoref{tab:taylorcouetteparameters} and viscoelastic behavior. Different floating update intervals are tested,  $n_f\in\left\lbrace1,2,4,8,16,32,64,128,256,512,1024\right\rbrace$. For each $n_f$ we measure the computation time needed to process the first $1024$ analysis load steps for the time discretization $\Delta t=1\mathrm{e}-5\mathrm{s}$. Additionally, we test the two cases of, one, quadrature knot vectors identical to the parent knot vectors, $\tilde{\Xi}^l=\mathcal{K}\left(n^S_p,p\right)$, and two, twice as dense, $\tilde{\Xi}^l=\mathcal{K}\left(2n^S_p,p\right)$; and it holds $G=3$. All tests are performed in fivefold replication. We compare the computation times relative to the one of IGA with identical discretization, except that classical Gauss-Legendre quadrature ($3\times 2$ points per velocity element) is employed (and no floating operations are performed).

\begin{figure}[t!]
\centering
\includegraphics[trim={0cm 0cm 0cm 0cm}, clip,width=0.6\textwidth]{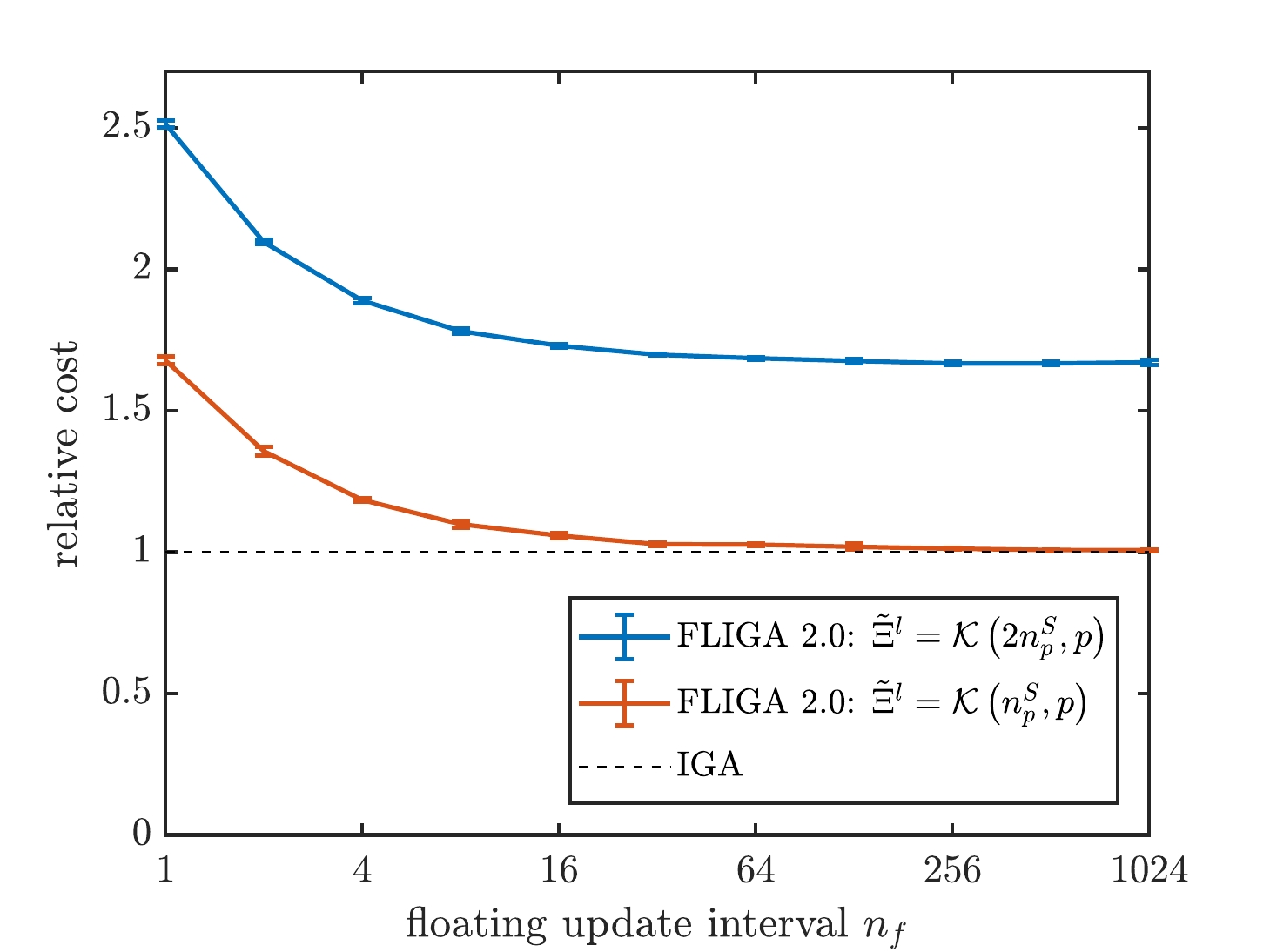}
\caption{Computation cost for $p=2,\ q=1$ with standard deviations.}
\label{fig:costs}
\end{figure}
Results for the relative computation cost are provided in \autoref{fig:costs}. We observe a decreasing trend of the cost for increasing floating update intervals $n_f$. For the case $\tilde{\Xi}^l=\mathcal{K}\left(n^S_p,p\right)$ the total quadrature point count in enhanced FLIGA and IGA is identical, and the cost converges towards $1$ in the limit of no floating operations being applied. Conversely, for the case $\tilde{\Xi}^l=\mathcal{K}\left(2n^S_p,p\right)$ some extra cost remains in that limit due to a larger quadrature point set. We confirm that a suitable adjustment of $n_f$ notably reduces the cost overhead due to the floating operations (e.g., in the previous computations we had $n_f=20$). Based on the current study we conclude that the majority of such overhead is avoided, even with smaller floating update intervals; as a rule of thumb, $n_f\approx5$.

\subsection{Planar extrusion}

\begin{figure}[t!]
\centering
\includegraphics[trim={5.125cm 4.75cm 7.5cm 4cm},clip,width=0.9\textwidth]{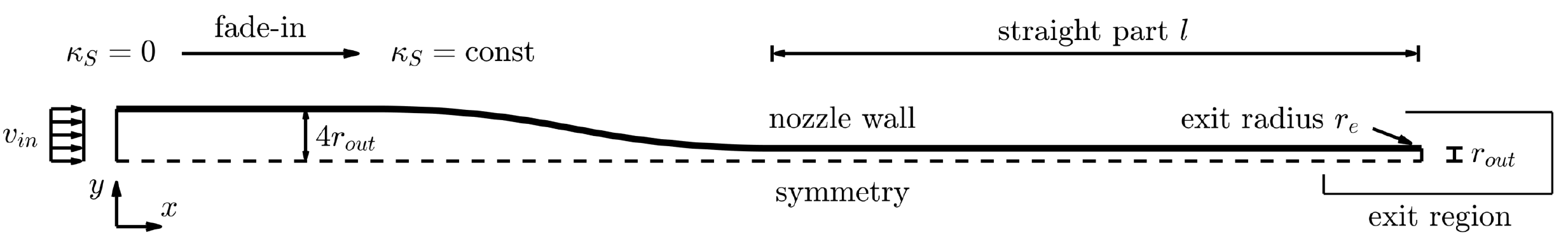}
\caption{Planar extrusion problem.}
\label{fig:planarextrusionproblem}
\end{figure}

The problem of planar extrusion from a straight nozzle is investigated. We adopt as reference the computational fluid dynamics (CFD) steady-state study in \cite{COMMINAL2018b}. Due to the Lagrangian treatment in our work, we modify the problem setup as follows (\autoref{fig:planarextrusionproblem}).
To circumvent the inflow condition, a convergent segment ($4:1$) is attached to the nozzle and fluid motion is driven by a Dirichlet boundary condition ($v_{in}=0.5\mathrm{mm/s}$); the length of the straight part is much longer ($l=10\mathrm{mm}$) compared to the radius ($r_{out}=0.2\mathrm{mm}$); the strict contact and no-slip conditions are relaxed by penalty enforcement; the penalty coefficients are set as $\kappa_P=7.5\mathrm{e}5$, $\kappa_R=1.5\mathrm{e}3$, $\kappa_S=1\mathrm{e}3$, respectively, with a ramp $(0\rightarrow 1\mathrm{e}3)$ on $\kappa_S$ in the region indicated in the figure; a small radius $r_e=\dfrac{r_o}{8}$ is added at the nozzle exit; all discretization parameters are listed in \autoref{tab:planarextrusionparameters}. Note that, in our setting, we are investigating a transient problem.

\begin{table}[t!]
\centering
\begin{tabular}{l l l l l l l l l}
\hline
\textit{Simulation parameter} & \textit{Variable} & \textit{Value} & & & \textit{Simulation parameter} & \textit{Variable} & \textit{Value} \\
\hline
polynomial order & & & & &  element count & & \\
\quad characteristic axis& & & & & \quad velocity & & (adaptive) $\times\ 16$ \\
\quad \quad velocity& $p$ & $2$ & & & \quad pressure & & \quad (adaptive) $\times\ 8$ \\
\quad \quad pressure& $p$ & $1$ & & &  quadrature point count & & (adapt. initialized) \\
\quad normal axis & & &  && \ \quad per initial element & & $\quad\times\ (q+1)$ \\
\quad \quad velocity& $q$ & $1$ & & & time step & $\Delta t$ & $5$e$-4$s \\
\quad \quad pressure& $q$ & $1$ & & & floating update interval & $n_f$ & $10$ \\
\hline
\end{tabular}
\caption{Simulation parameters for the planar extrusion problem.}
\label{tab:planarextrusionparameters}
\end{table}

We focus on the Oldroyd-B model with solvent viscosity ratio
\begin{equation}
\beta = \dfrac{\eta_s}{\eta_s+\eta_p}=\dfrac{1}{9},
\end{equation}
obtained by the choice $\eta_s = 1000\mathrm{Pa}\cdot\mathrm{s}$, $\eta_p = 8000\mathrm{Pa}\cdot\mathrm{s}$. The relaxation time is adjusted so as to produce a given Weissenberg number $Wi$ according to
\begin{equation}
\lambda=\mathrm{Wi}\cdot\dfrac{1}{\dot{\gamma}}=\mathrm{Wi}\cdot\dfrac{r_{out}}{12v_{in}},
\end{equation}
where $\dot{\gamma}$ is a maximum shear rate estimate in the straight part of the nozzle. The initial condition for the polymer stress is $\bm{\tau}^0_{p}=\bm{0}$.

\begin{figure}[t!]
    \centering
    \subfigure[\centering $t=0\mathrm{s}$]{\includegraphics[trim={0cm 2.25cm 0cm 0.5cm},clip,width=0.7\textwidth]{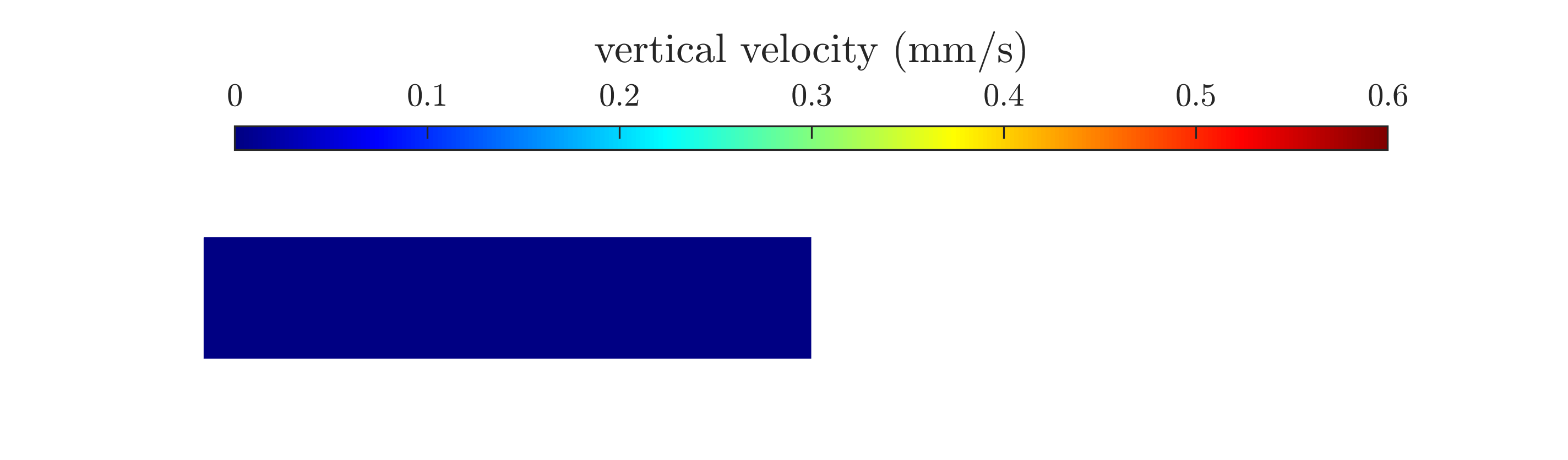} }
    \qquad
    \subfigure[\centering $t=0.30\mathrm{s}$]{\includegraphics[trim={0cm 2.25cm 0cm 3.5cm},clip,width=0.7\textwidth]{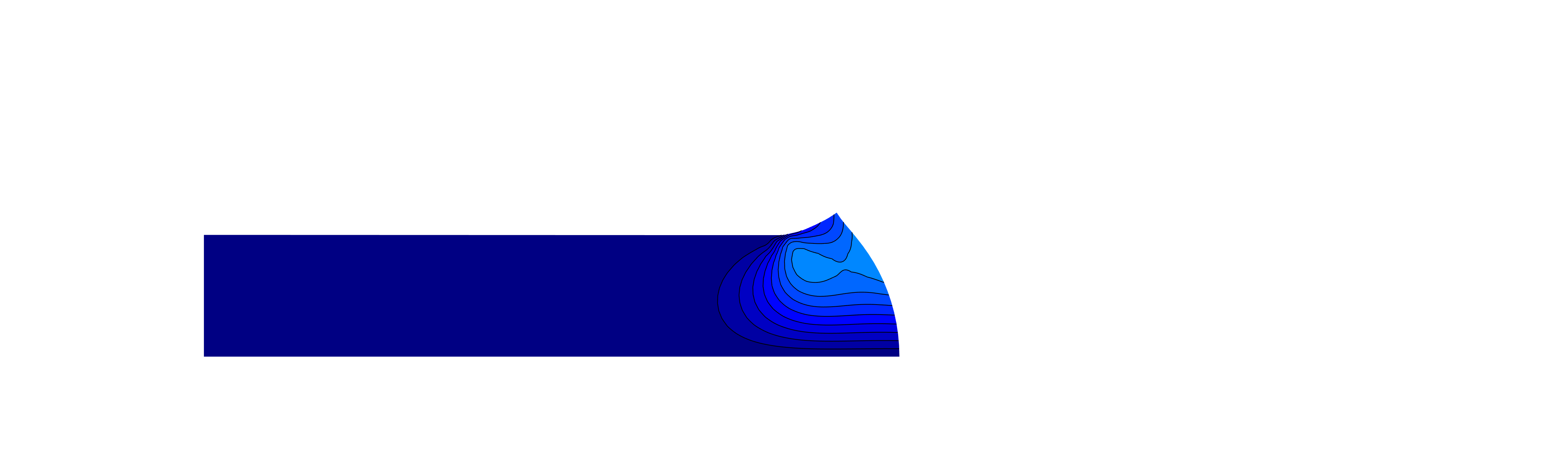}  }
    \qquad
    \subfigure[\centering $t=0.75\mathrm{s}$]{\includegraphics[trim={0cm 2.25cm 0cm 3.5cm},clip,width=0.7\textwidth]{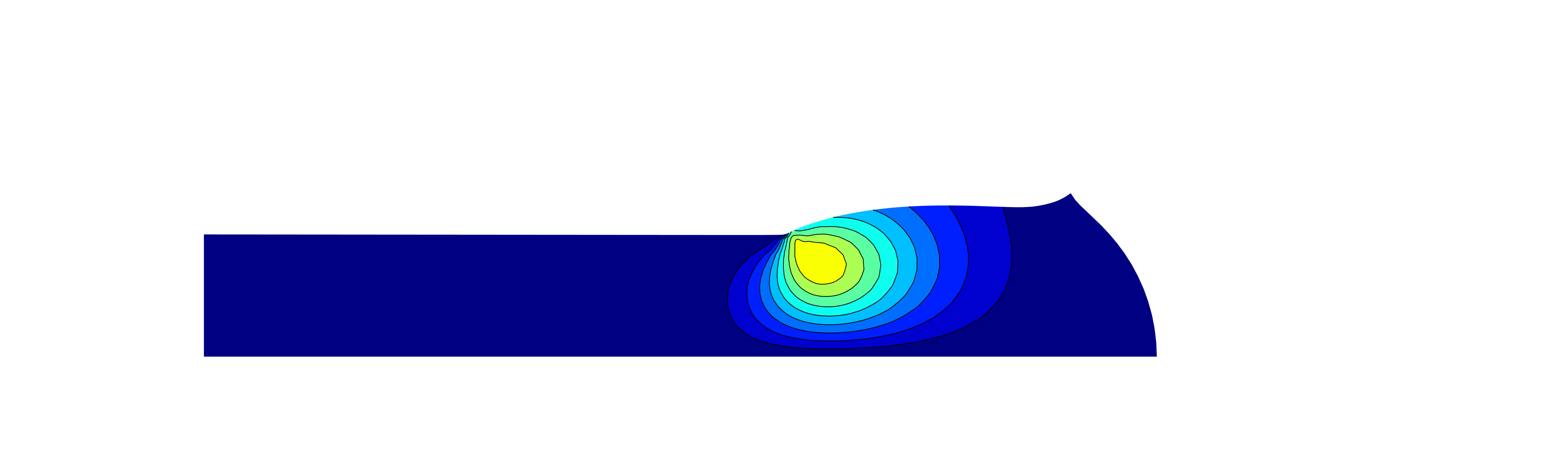} }
    \qquad
    \subfigure[\centering $t=1.15\mathrm{s}$]{\includegraphics[trim={0cm 2.25cm 0cm 3.5cm},clip,width=0.7\textwidth]{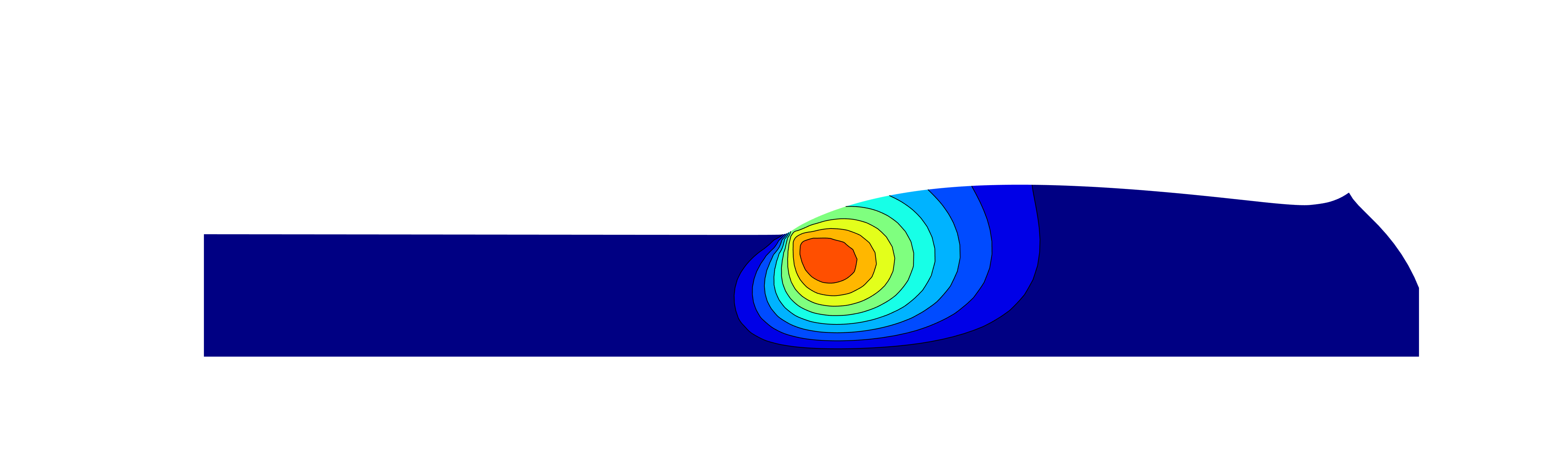} }
    \qquad
    \subfigure[\centering $t=7.50\mathrm{s}$]{\includegraphics[trim={0cm 2.25cm 0cm 3.5cm},clip,width=0.7\textwidth]{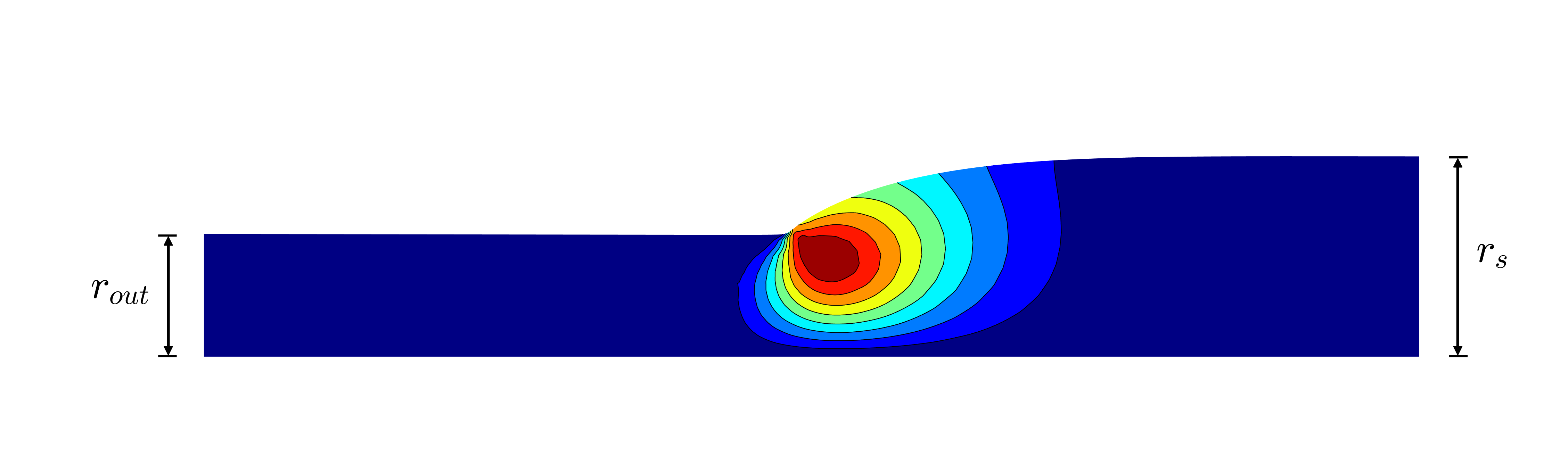} }
    \caption{Vertical velocity contours during transient strand formation (a-d) and steady-state (e) in the exit region .}
    \label{fig:planarextrusiontransient}
\end{figure}
\begin{figure}[t!]
    \centering
    \subfigure[Field contours]{\includegraphics[trim={0cm 0.5cm 0cm 0cm},clip,width=0.7\textwidth]{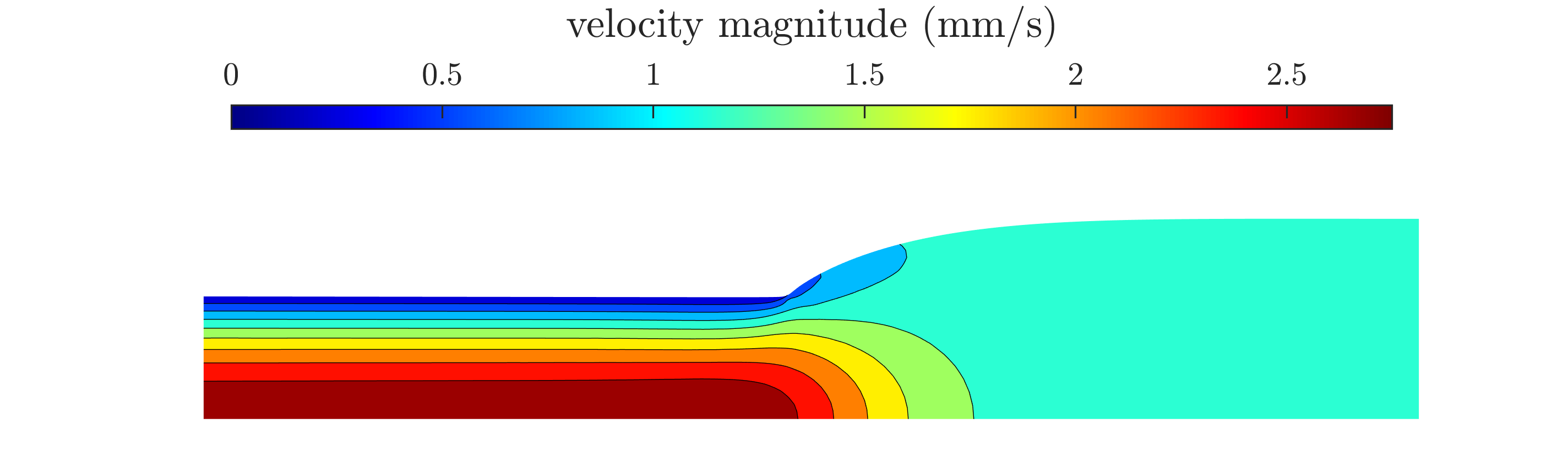} }
    \qquad
    \subfigure[Control points]{\includegraphics[trim={0cm 1.5cm 0cm 3cm},clip,width=0.7\textwidth]{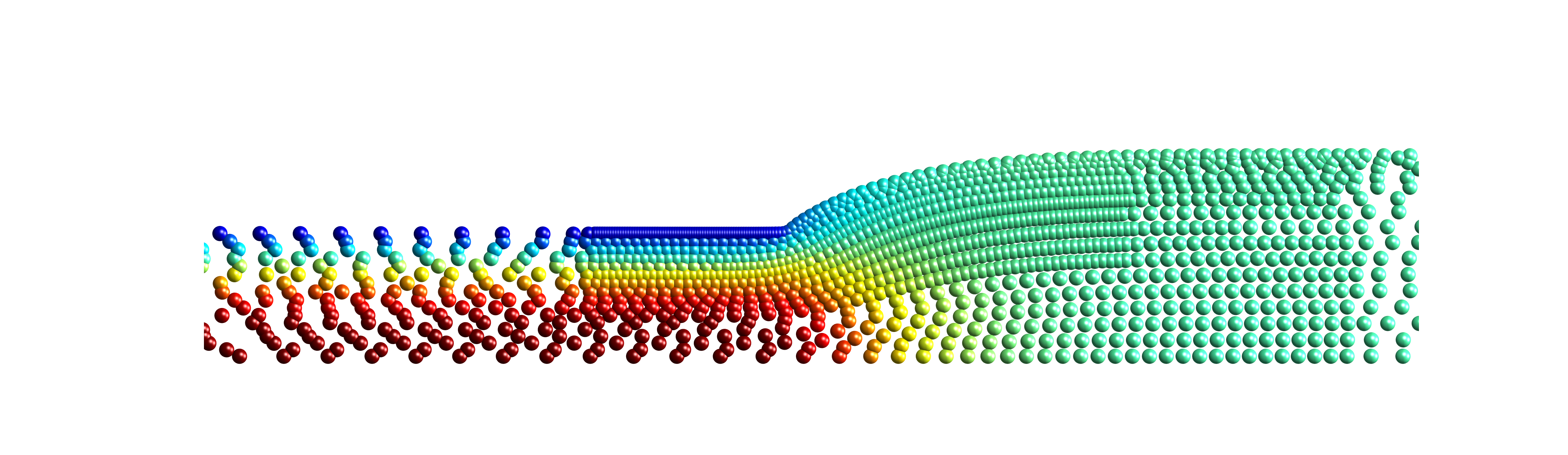} }
    \caption{Velocity magnitude during steady state in the exit region ($t=7.50\mathrm{s}$).}
    \label{fig:planarextrusionsteady}
\end{figure}

In \autoref{fig:planarextrusiontransient} (a-d) we visualize for $Wi=2.5$ the build-up of the free surface behind the nozzle exit at different time steps, which is controlled by viscoelastic swelling. Contours of vertical velocities in the nozzle exit area concentrate close to the wall tip. After $t=7.5\mathrm{s}$, steady-state behavior is reached in the region, see \autoref{fig:planarextrusiontransient} (e). Moreover, \autoref{fig:planarextrusionsteady} (a) shows the velocity magnitude contours during steady state at $t=7.5\mathrm{s}$, indicating a parabolic-type velocity profile inside the nozzle as well as a transition towards plug profile behind the exit. Our qualitative observations agree well with the description in \cite{COMMINAL2018b}. The corresponding control point set is illustrated in \autoref{fig:planarextrusionsteady} (b) and adaptive refinement is recognized from the nodal distances.

\begin{figure}[t!]
\centering
\includegraphics[trim={0cm 0cm 0cm 0.375cm},clip,width=0.6\textwidth]{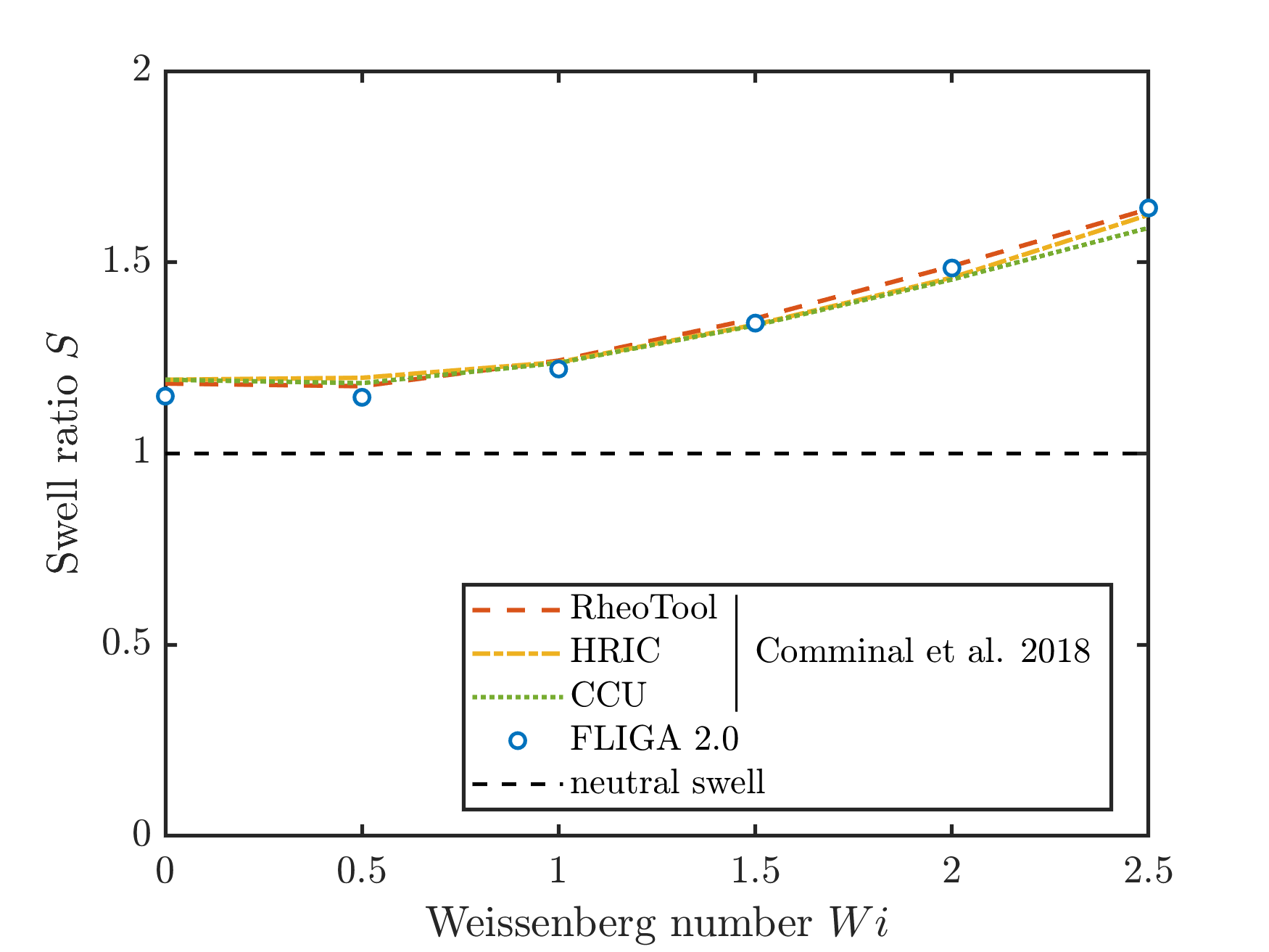}
\caption{Die swell ratios.}
\label{fig:planarextrusiondieswell}
\end{figure}

Finally, we provide a quantitative comparison to \cite{COMMINAL2018b} for the swell ratios
\begin{equation}
S=\dfrac{r_s}{r_{out}},
\end{equation}
at different $Wi\in\left\lbrace0,0.5,1,1.5,2,2.5\right\rbrace$.\footnote{For the Newtonian fluid, $Wi=0$, we adjust $\kappa_S=250$.} Here $r_s$ is the steady-state extrudate radius, see \autoref{fig:planarextrusiontransient} (e). Results are depicted in \autoref{fig:planarextrusiondieswell} and compared to the results obtained in \cite{COMMINAL2018b} using a selection of CFD techniques. In view of the deviations between results of the different CFD techniques, our results show very satisfactory agreement for high $Wi$. The small underestimation of swell ratios for (fluid-like) low $Wi$ may due to our approximate enforcement of the no-slip condition at the wall.

\subsection{Extrusion-based AM}

Lastly, we apply enhanced FLIGA to the simulation of extrusion-based AM. The problem setup is adopted from the previous benchmark of planar extrusion, however with the following modifications: the complete setup is rotated such that the nozzle opens in downwards vertical direction; the symmetry assumption is removed and the entire plane represented; the lengths of the contraction segment and of the straight part ($l=0.2\mathrm{mm}$) are decreased; the nozzle exit radius is increased ($r_e=r_o/4$); and the penalty coefficients are adjusted ($\kappa_P=1\mathrm{e}4$, $\kappa_R=1\mathrm{e}3$, $\kappa_S=100$). In addition, we model a substrate below the nozzle exit by imposing homogeneous Dirichlet boundary conditions for the vertical velocity on all control points penetrating a user-defined substrate shape. We simulate the horizontal movement of the nozzle by imposing a constant Dirichlet boundary condition for the horizontal velocity ($v_n=2.4\mathrm{mm/s}$) to the control points in contact with the substrate. We proceed similarly to simulate a vertical movement of the nozzle. For simplicity, a renewed material detachment from the substrate is not considered here. The discretization parameters are summarized in \autoref{tab:additivemanufacturingparameters}.

\begin{table}[t!]
\centering
\begin{tabular}{l l l l l l l l l}
\hline
\textit{Simulation parameter} & \textit{Variable} & \textit{Value} & & & \textit{Simulation parameter} & \textit{Variable} & \textit{Value} \\
\hline
polynomial order & & & & &  element count & & \\
\quad characteristic axis& & & & & \quad velocity & & (adaptive) $\times\ 16$ \\
\quad \quad velocity& $p$ & $2$ & & & \quad pressure & & \quad (adaptive) $\times\ 8$ \\
\quad \quad pressure& $p$ & $1$ & & &  quadrature point count & & (adapt. initialized) \\
\quad normal axis & & &  && \ \quad per initial element & & $\quad\times\ (q+1)$ \\
\quad \quad velocity& $q$ & $1$ & & & time step & $\Delta t$ & $1.\overline{6}$e$-3$s \\
\quad \quad pressure& $q$ & $1$ & & & floating update interval & $n_f$ & $5$ \\
\hline
\end{tabular}
\caption{Simulation parameters for the extrusion-based AM problem.}
\label{tab:additivemanufacturingparameters}
\end{table}

\begin{figure}[h!]
    \centering
    \subfigure[\centering Straight deposition]{\includegraphics[trim={0cm 0cm 0cm 0cm},clip,width=0.9\textwidth]{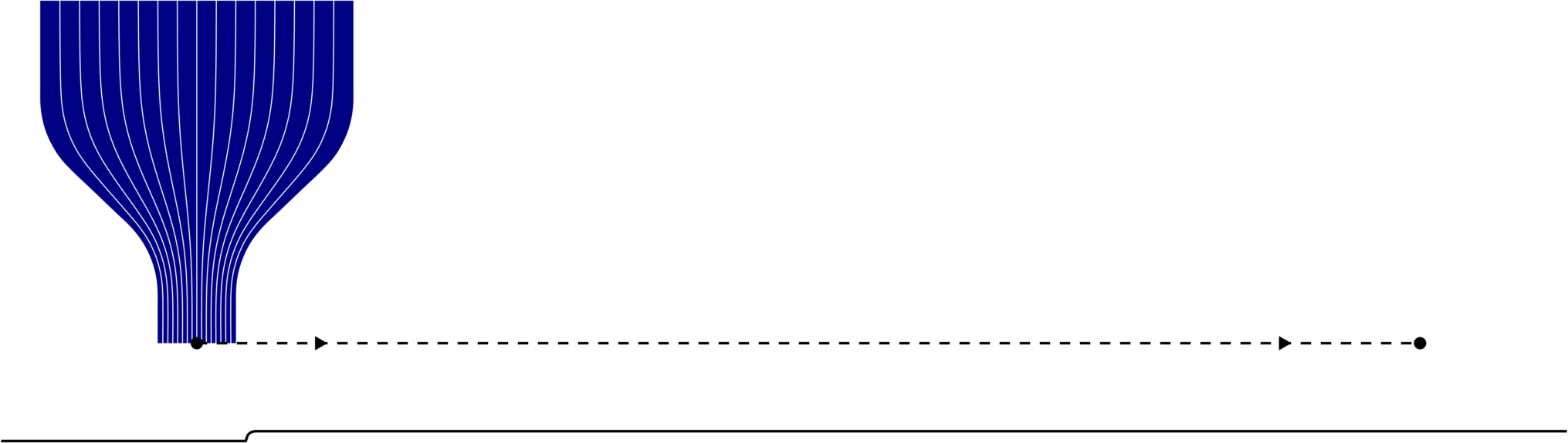} }
    \qquad
    \subfigure[\centering Nozzle vibration]{\includegraphics[trim={0cm 0cm 0cm 0cm},clip,width=0.9\textwidth]{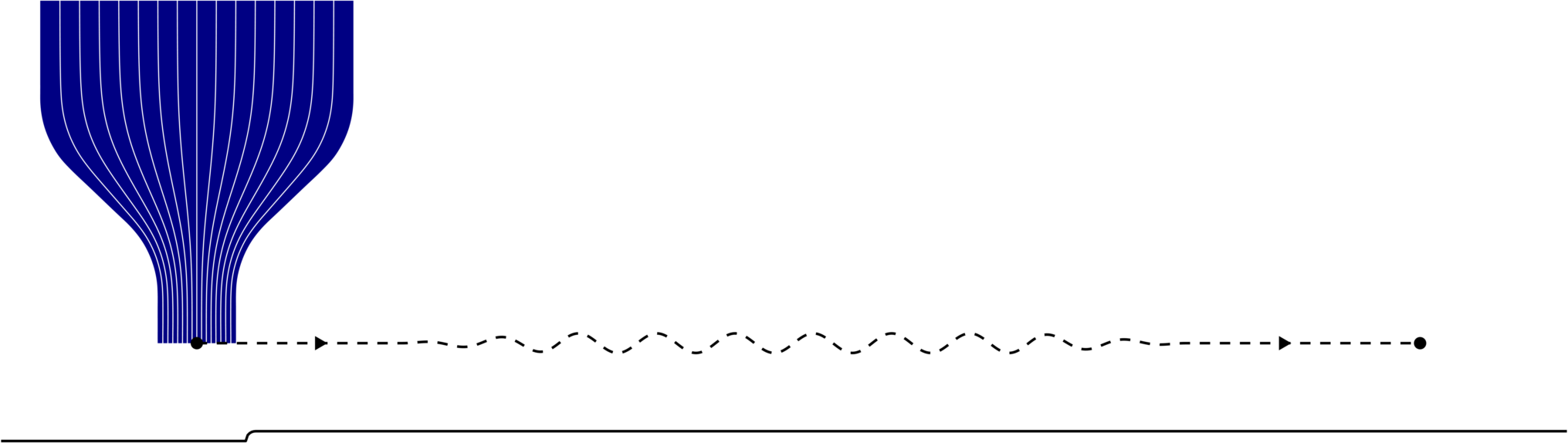}  }
    \qquad
    \subfigure[\centering Uneven substrate]{\includegraphics[trim={0cm 0cm 0cm 0cm},clip,width=0.9\textwidth]{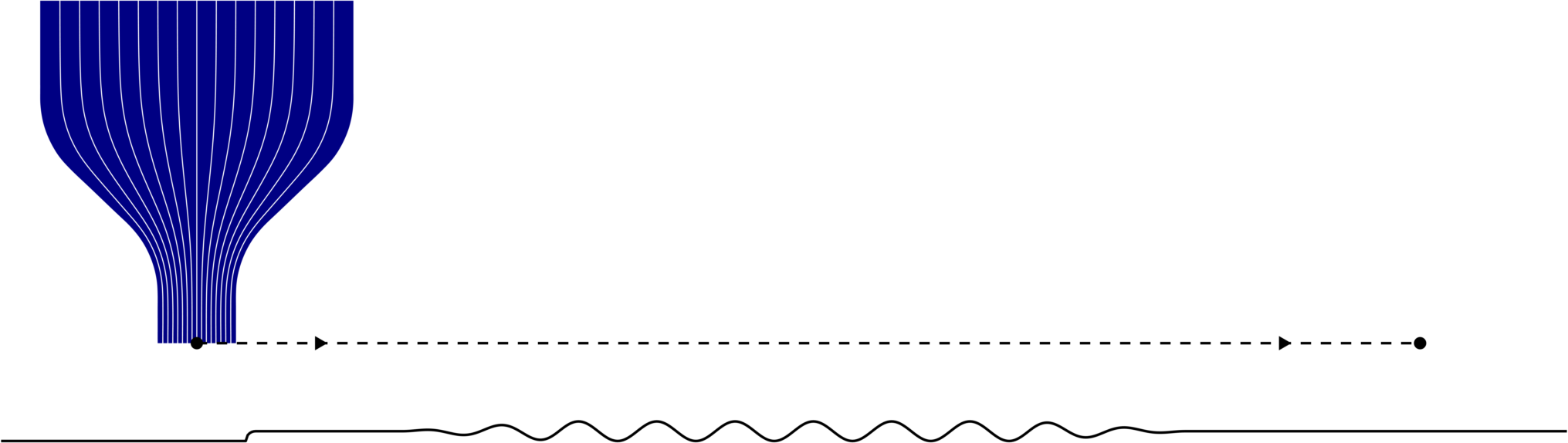}  }
    \qquad
    \subfigure[\centering Obstacle]{\includegraphics[trim={0cm 0cm 0cm 0cm},clip,width=0.9\textwidth]{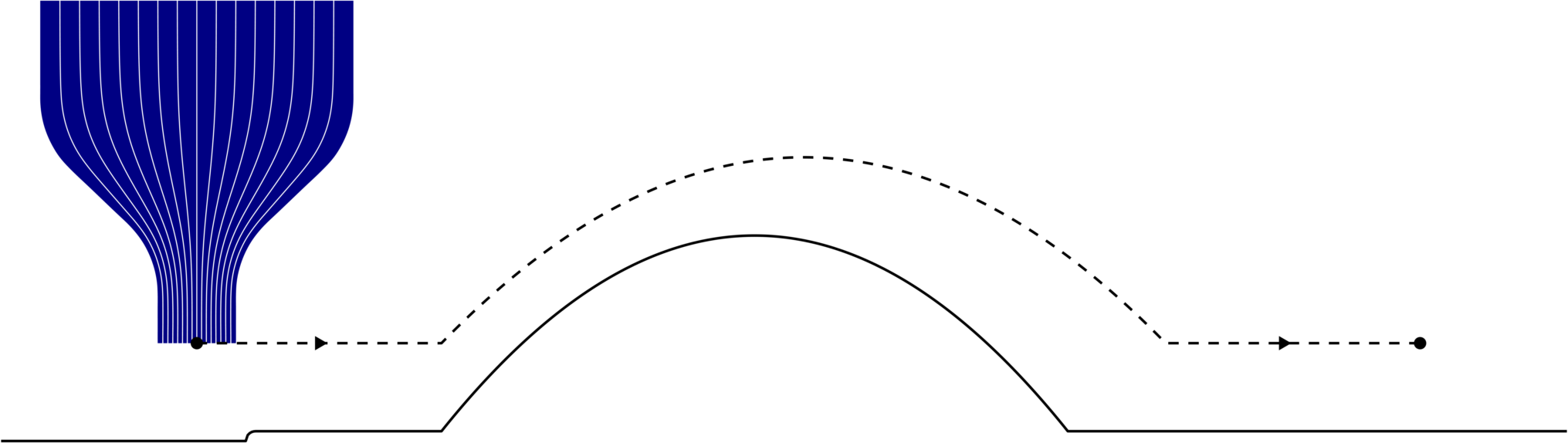}  }
    \caption{Extrusion-based AM problems.}
    \label{fig:additivemanufacturing1}
\end{figure}

\begin{figure}[h!]
    \centering
    \subfigure[\centering Straight deposition]{\includegraphics[trim={0cm 0cm 0cm 0cm},clip,width=0.9\textwidth]{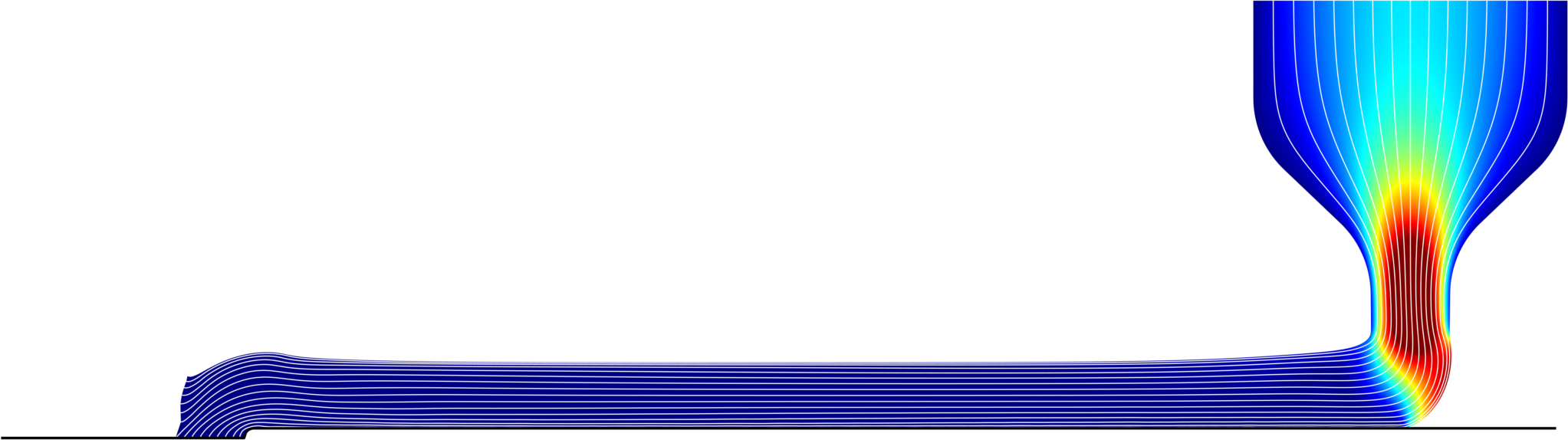} }
    \qquad
    \subfigure[\centering Nozzle vibration]{\includegraphics[trim={0cm 0cm 0cm 0cm},clip,width=0.9\textwidth]{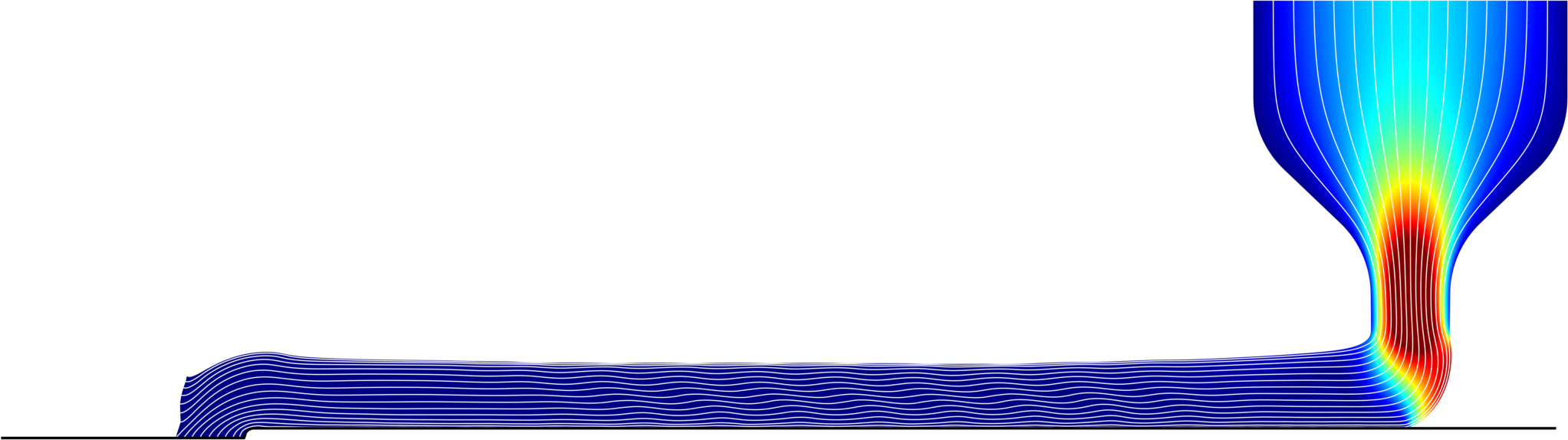}  }
    \qquad
    \subfigure[\centering Uneven substrate]{\includegraphics[trim={0cm 0cm 0cm 0cm},clip,width=0.9\textwidth]{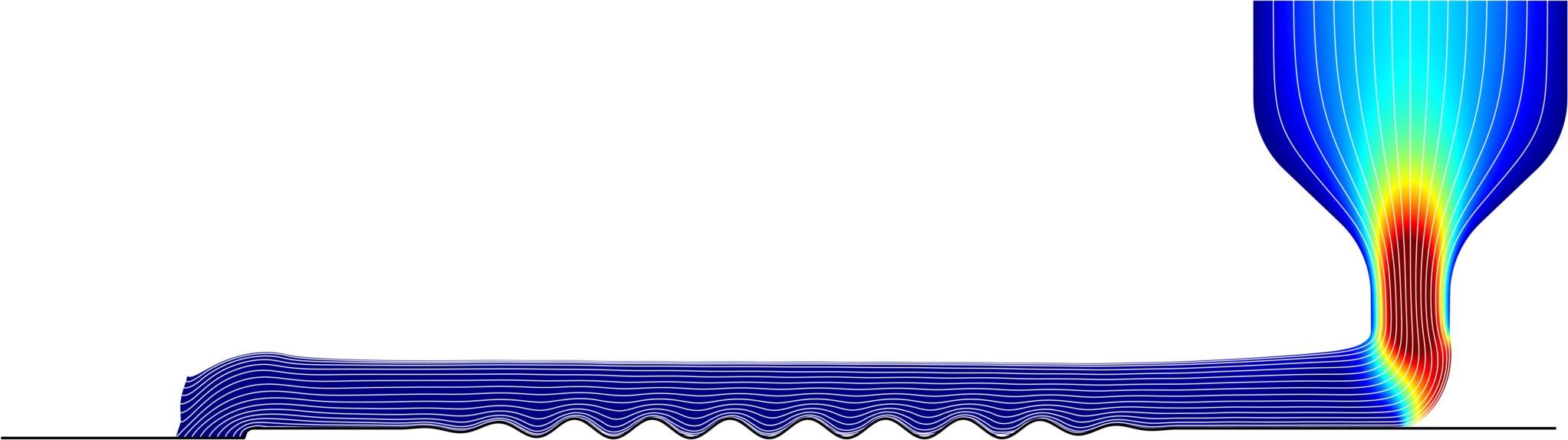}  }
    \qquad
    \subfigure[\centering Obstacle]{\includegraphics[trim={0cm 0cm 0cm 0cm},clip,width=0.9\textwidth]{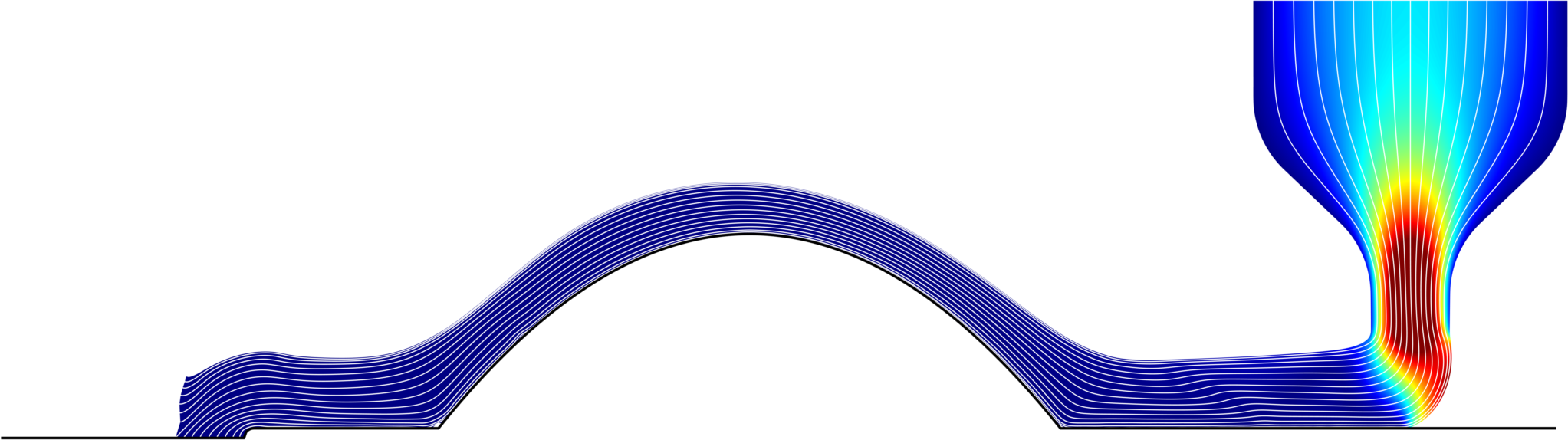}  }
    \caption{Final strand geometries.}
    \label{fig:additivemanufacturing2}
\end{figure}

We investigate the viscoelastic Oldroyd-B material with the same parameters as in the planar extrusion example for the case $Wi=1.5$. Four different combinations of substrate shape and nozzle movement are tested, see \autoref{fig:additivemanufacturing1}. The first case (a) corresponds to an idealized setup with mere horizontal nozzle movement along a linear path and deposition on a planar substrate. Secondly (b), we investigate an imperfection of the nozzle movement in terms of a superposed harmonic vibration. The third case (c) explores substrate imperfections in terms of surface variations. And finally (d), we put an obstacle of parabolic shape and adjust the nozzle path. In all examples, the center of the nozzle exit starts at the left black circles in the figure, following the dashed path until the final state, represented by the black circles on the right. Also, we add a small step to the front segment of the substrates so as to avoid a singularity in the physical map of the rectangular parametric domain at the respective corner. Finally, the characteristic element division in the undeformed state is illustrated by the white curves in \autoref{fig:additivemanufacturing1}.

Simulation results are illustrated in \autoref{fig:additivemanufacturing2}. For the first printing scenario (a), the characteristic element interfaces (white curves) align evenly along the extrusion strand. A vibration of the nozzle (b) causes deviations in the alignment leading to a periodic, wavy pattern of material arrangement. For the uneven substrate (c), the extrudate adapts to the substrate shape. Interestingly, the effect is local and almost completely fades out up to the top edge of the strand, resulting in an even surface. The nozzle path we have designed for case (d) allows to overcome the obstacle with the material being pressed also in the two corners. For all examples (a-d), we show the vertical velocity magnitudes at the final configuration by contour plots. The approximately enforced no-slip condition is recognized from the blue colors at the nozzle walls, indicating low material speed. This is an important enhancement with respect to to the extrusion-based AM problem investigated for FLIGA 1.0 in \cite{HKDL}, where the resolution of the resulting extreme elongation rates behind the nozzle exit would not have been possible due to the lack of adaptive refinement.

We confirm a good overall stability and robustness of enhanced FLIGA limited to the validity of the assumption from \autoref{ssec:updates}, i.e., that inner characteristic element interfaces reasonably align to the outer patch boundaries. From the element topology in the above AM test cases, we see the assumption fulfilled even when modeling moderate process imperfections.

\section{Conclusions}
\label{sec:conclusions}

In this paper, we have proposed enhancements for Floating Isogeometric Analysis with respect to the initial formulation in \cite{HKDL}. A new quadrature concept conserves the material point character of quadrature points, while reducing the relative errors in the patch test up to four orders of magnitude and improving the range of numerical stability in the Taylor-Couette benchmark from five to more than $25$ turns. The novel procedure for floating regulation removes shear mesh distortion along the characteristic direction in a fully automatic fashion and allows for the prediction of complicated strand geometries in extrusion-based additive manufacturing simulation, where the manually designed floating regulation of \cite{HKDL} would have been prohibitively complicated. Moreover, dilatational mesh distortion is overcome by adaptive refinement, enabling the solution of extrusion problems featuring extremely extensional flow.

While offering an effective solution to extreme mesh distortion along the characteristic direction, enhanced FLIGA shares many discretization concepts with Lagrangian IGA, as well as many of its benefits. These include excellent stability, direct imposition of boundary conditions, easy construction of stable mixed ansatz spaces, adjustable smoothness of the basis functions (in the characteristic direction), absence of advection terms, and natural treatment of free surface evolution. On the other hand, there are still some limitations. It is not possible to apply the strategy to problems with critical mesh distortion along the normal direction. The conceptual generalization to multi-patch geometries appears limited to the separate floating within each patch coming along with a distortion of patch interfaces. 
Finally, the polynomial order is fixed to $q=1$ in the normal direction. These limitations may well be addressed in future investigations. In addition, it would be useful to thoroughly investigate stability and accuracy from a numerical analysis perspective. Also, while we focused on spatial discretization, further investigations may consider the aspect of time discretization in more detail. Finally, further significant applications of FLIGA beyond viscoelastic extrusion are worth exploring.

\FloatBarrier

\appendix

\section{Parametrization ansatz}
\label{xiansatz}
The exact representation of $\xi$ by the ansatz in (\autoref{eq:xiansatz}) is shown as
\begin{align*}
\xi^h\left(\bm{x};\mathcal{H}\right) &= \sum^M_{m=1} h_m B_m\left(\bm{x};\mathcal{H}\right)= \sum^M_{m=1} h_m\hat{B}_m\left(\mathcal{F}^{-1}\left(\bm{x};\mathcal{H}\right);\mathcal{H}\right)= \sum^M_{m=1} h_m\hat{B}_m\left(\bm{\xi}\left(\argumentbullet;\mathcal{H}\right);\mathcal{H}\right)\\
&= \sum^J_{j=1}\left(\sum^{I_j}_{i=1} h_{ij} \hat{N}_{ij}\left(\xi\left(\argumentbullet;\mathcal{H}\right);\mathcal{H}\right)\right)\hat{M}_j(\eta\left(\argumentbullet;\mathcal{H}\right))= \sum^J_{j=1}\left(\sum^{I_j}_{i=1} h_{ij} \tilde{N}_{ij}\left(\mathcal{G}^{-1}_j\left(\xi\left(\argumentbullet;\mathcal{H}\right);\mathcal{H}\right)\right)\right)\hat{M}_j(\eta\left(\argumentbullet;\mathcal{H}\right))\\
&=\sum^J_{j=1}\mathcal{G}_j\left(\mathcal{G}^{-1}_j\left(\xi\left(\argumentbullet;\mathcal{H}\right);\mathcal{H}\right);\mathcal{H}\right)\hat{M}_j(\eta\left(\argumentbullet;\mathcal{H}\right)) = \sum^J_{j=1}\xi\left(\argumentbullet;\mathcal{H}\right)\hat{M}_j(\eta\left(\argumentbullet;\mathcal{H}\right))=\xi\left(\argumentbullet;\mathcal{H}\right)\sum^J_{j=1}\hat{M}_j(\eta\left(\argumentbullet;\mathcal{H}\right))\\
&=\xi\left(\argumentbullet;\mathcal{H}\right).\\
\end{align*}

\section{Tangent matrix for solving the parametrization equation}
\label{laplaceNewtonRaphson}

Let us denote
\begin{equation*}
\bm{J}^{gl}_{sn}\left(\argumentbullet;\mathcal{H}\right):=\bm{J}\left(\tilde{\bm{\xi}}^{gl}_{sn}\left(\argumentbullet;\mathcal{H}\right);\mathcal{H}\right).
\end{equation*}
Then the global tangent to \autoref{eq:laplace} is given as
\begin{align*}
K^R_{ik} =\dfrac{\partial}{\partial h_k}R_i=&\sum^{2J-2}_{l=1}\sum^{n^{QP,l}}_{g=1}\sum^M_{m=1}\sum^2_{a=1}\Bigg[\dfrac{\partial^2}{\partial h_k \partial x_a}\hat{B}_{i}\left(\tilde{\bm{\xi}}^{gl}_{sn}\left(\argumentbullet;\mathcal{H}\right);\mathcal{H}\right)\ \dfrac{\partial}{\partial x_a}\hat{B}_{m}\left(\tilde{\bm{\xi}}^{gl}_{sn}\left(\argumentbullet;\mathcal{H}\right);\mathcal{H}\right)\ h_m\ W^{gl}\left(\argumentbullet;\mathcal{H}\right)\\
&\qquad +\dfrac{\partial}{\partial x_a}\hat{B}_{i}\ \left(\tilde{\bm{\xi}}^{gl}_{sn}\left(\argumentbullet;\mathcal{H}\right);\mathcal{H}\right)\dfrac{\partial^2}{\partial h_k \partial x_a} \hat{B}_{m}\left(\tilde{\bm{\xi}}^{gl}_{sn}\left(\argumentbullet;\mathcal{H}\right);\mathcal{H}\right)\ h_m\ W^{gl}\left(\argumentbullet;\mathcal{H}\right)\\
&\qquad + \dfrac{\partial}{\partial x_a}\hat{B}_{i}\left(\tilde{\bm{\xi}}^{gl}_{sn}\left(\argumentbullet;\mathcal{H}\right);\mathcal{H}\right)\ \dfrac{\partial}{\partial x_a}\hat{B}_{m}\left(\tilde{\bm{\xi}}^{gl}_{sn}\left(\argumentbullet;\mathcal{H}\right);\mathcal{H}\right)\ \delta_{mk}\ W^{gl}\left(\argumentbullet;\mathcal{H}\right)\\
&\qquad +\dfrac{\partial}{\partial x_a}\hat{B}_{i}\left(\tilde{\bm{\xi}}^{gl}_{sn}\left(\argumentbullet;\mathcal{H}\right);\mathcal{H}\right)\ \dfrac{\partial}{\partial x_a}\hat{B}_{m}\left(\tilde{\bm{\xi}}^{gl}_{sn}\left(\argumentbullet;\mathcal{H}\right);\mathcal{H}\right)\ h_m\ \dfrac{\partial}{\partial h_k}W^{gl}\left(\argumentbullet;\mathcal{H}\right)\Bigg],\\
\dfrac{\partial^2}{\partial h_k \partial x_a}\hat{B}_{i}\left(\tilde{\bm{\xi}}^{gl}_{sn}\left(\argumentbullet;\mathcal{H}\right);\mathcal{H}\right)=&\Bigg[-\sum^M_{o=1}\sum^2_{b=1}\sum^2_{p=1}\sum^2_{q=1}\left(J^{gl}_{sn}\right)^{-1}_{bp}\left(\argumentbullet;\mathcal{H}\right)\ \left(J^{gl}_{sn}\right)^{-1}_{qa}\left(\argumentbullet;\mathcal{H}\right)\ c_{op}\ \dfrac{\partial}{\partial \xi_{b}}\hat{B}_{i}\left(\tilde{\bm{\xi}}^{gl}_{sn}\left(\argumentbullet;\mathcal{H}\right);\mathcal{H}\right) \\
&\qquad \dfrac{\partial^2}{\partial h_k \partial \xi_q}\hat{B}_{o}\left(\tilde{\bm{\xi}}^{gl}_{sn}\left(\argumentbullet;\mathcal{H}\right);\mathcal{H}\right)\Bigg] +\sum^2_{b=1}\left(J^{gl}_{sn}\right)^{-1}_{ba}\left(\argumentbullet;\mathcal{H}\right)\ \dfrac{\partial^2}{\partial h_k \partial \xi_{b}}\hat{B}_{i}\left(\tilde{\bm{\xi}}^{gl}_{sn}\left(\argumentbullet;\mathcal{H}\right);\mathcal{H}\right),\\
\dfrac{\partial^2}{\partial h_k \partial \bm{\xi}}\hat{B}_i\left(\tilde{\bm{\xi}}^{gl}_{sn}\left(\argumentbullet;\mathcal{H}\right);\mathcal{H}\right)=&
\begin{cases}
\begin{pmatrix}
-\dfrac{\partial}{\partial\tilde{\xi}}\tilde{N}_{ts}\left(\tilde{\xi}^{gl}\right)\ \hat{M}_t\left(\eta^l\right)\ J_s\left(\tilde{\xi}^{gl};\mathcal{H}\right)^{-2}\ \tilde{N}_{us}\left(\tilde{\xi}^{gl}\right)\\
0\\
\end{pmatrix}, &\mathrm{for}\ j=s \wedge v=s,
\\
\begin{pmatrix}
0\\
0\\
\end{pmatrix}, &\mathrm{for}\ j=s \wedge v=n,
\\
\begin{pmatrix}
0\\
\tilde{N}_{us}\left(\tilde{\xi}^{gl}\right)\ \dfrac{\partial}{\partial\eta}\hat{M}_v\left(\eta^l\right)\ J_n\left(\tilde{\xi}^{gl}_{sn}\left(\argumentbullet;\mathcal{H}\right);\mathcal{H}\right)^{-1}\ \dfrac{\partial}{\partial\tilde{\xi}}\tilde{N}_{tn}\left(\tilde{\xi}^{gl}_{sn}\left(\argumentbullet;\mathcal{H}\right)\right)\\
\end{pmatrix}, &\mathrm{for}\ j=n \wedge v=s,
\\
\begin{pmatrix}
0\\
-\tilde{N}_{uv}\left(\tilde{\xi}^{gl}_{sn}\left(\argumentbullet;\mathcal{H}\right)\right)\ \dfrac{\partial}{\partial\eta}\hat{M}_n\left(\eta^l\right)\ J_n\left(\tilde{\xi}^{gl}_{sn}\left(\argumentbullet;\mathcal{H}\right);\mathcal{H}\right)^{-1}\ \dfrac{\partial}{\partial\tilde{\xi}}\tilde{N}_{tn}\left(\tilde{\xi}^{gl}_{sn}\left(\argumentbullet;\mathcal{H}\right)\right)\\
\end{pmatrix}, &\mathrm{for}\ j=n \wedge v=n,\\
\begin{pmatrix}
0\\
0
\end{pmatrix}, &\mathrm{otherwise},
\end{cases}\\
\dfrac{\partial}{\partial h_k}W^{gl}\left(\argumentbullet;\mathcal{H}\right)=&\begin{cases}\Bigg[\dfrac{\partial}{\partial \tilde{\xi}}\tilde{N}_{us}\left(\tilde{\xi}^{gl}\right)\ \mathrm{det}\left(\bm{J}^{gl}\left(\argumentbullet;\mathcal{H}\right)\right)+& \\ \qquad \sum^M_{o=1}\sum^2_{a=1}\sum^2_{b=1}J_s\left(\tilde{\xi}^{gl};\mathcal{H}\right)\ \mathrm{det}\left(\bm{J}^{gl}\left(\argumentbullet;\mathcal{H}\right)\right)\ \left(\bm{J}^{gl}\left(\argumentbullet;\mathcal{H}\right)^{-1}\right)_{ba}\ c_{oa}& \\ \qquad\dfrac{\partial^2}{\partial h_k \partial \xi_{b}}\hat{B}_o\left(\tilde{\bm{\xi}}^{gl}_{sn}\left(\argumentbullet;\mathcal{H}\right);\mathcal{H}\right)\Bigg]\tilde{w}^{gl}w^l, & \mathrm{for}\ v=s,\\
& \\
\sum^M_{o=1}\sum^2_{a=1}\sum^2_{b=1}J_s\left(\tilde{\xi}^{gl};\mathcal{H}\right)\ \mathrm{det}\left(\bm{J}^{gl}\left(\argumentbullet;\mathcal{H}\right)\right)\ \left(\bm{J}^{gl}\left(\argumentbullet;\mathcal{H}\right)^{-1}\right)_{ba}\ c_{oa}& \\\qquad \dfrac{\partial^2}{\partial h_k \partial \xi_{b}}\hat{B}_o\left(\tilde{\bm{\xi}}^{gl}_{sn}\left(\argumentbullet;\mathcal{H}\right);\mathcal{H}\right)\ \tilde{w}^{gl}w^l, & \mathrm{for}\ v=n,\\
& \\
0, &\mathrm{otherwise},
\end{cases}
\end{align*}
for $(t,j)\leftarrow i$ and $(u,v)\leftarrow k$ according to (\autoref{eq:runningindex}) as well as $s=\mathcal{S}(l)$ and $n=\mathcal{N}(l)$ according to (\autoref{eq:lsmap}) and (\autoref{eq:lnmap}), respectively.

\section{Tangent matrix for solving the governing equations}
\label{tangentstiffness}
The node-wise linearizations of the discrete governing equations (\autoref{eq:PVPdiscrete}) and (\autoref{eq:DIVdiscrete}) are:
\begin{align*}
\bm{K}^n_{(m)(j)} = \dfrac{\partial \bm{F}^n_{int,m}}{\partial \bm{d}^n_j} &= \sum^{n^{QP}}_{g=1} \eta_s\left[\left(\dfrac{\partial}{\partial \bm{x}}\hat{B}^{TP}_m\cdot \dfrac{\partial}{\partial \bm{x}}\hat{B}^{TP}_j\right) \bm{I} +  \dfrac{\partial}{\partial \bm{x}}\hat{B}^{TP}_m\left(\dfrac{\partial}{\partial \bm{x}}\hat{B}^{TP}_j\right)^T \right]_{\bm{\xi}=\bm{\xi}^g}W^{g,n} && (2 \times 2)\\
\bm{K}^n_{(m)(M+\zeta)} = \dfrac{\partial \bm{F}^n_{int,m}}{\partial q^n_{\zeta}} &= -\sum^{n^{QP}}_{g=1} \left[\dfrac{\partial}{\partial \bm{x}}\hat{B}^{TP}_m\ \hat{A}^{TP}_{\zeta}\right]_{\bm{\xi}=\bm{\xi}^g} W^{g,n} && (2 \times 1)\\
\bm{K}^n_{(M+z)(j)} = \dfrac{\partial Q^n_z}{\partial \bm{d}^n_j} &= -\sum^{n^{QP}}_{g=1} \left[\hat{A}^{TP}_z\ \left(\dfrac{\partial}{\partial \bm{x}}\hat{B}^{TP}_j\right)^T\right]_{\bm{\xi}=\bm{\xi}^g} W^{g,n} && (1 \times 2)\\
\bm{K}^n_{(M+z)(M+\zeta)} &= 0. && (1 \times 1)
\end{align*}

\section{Tangent matrix for solving contact problems}
\label{contacttangentstiffness}
The linearization of the discrete nodal contact force (\autoref{eq:CONTACTdiscrete}) in addition to \ref{tangentstiffness} leads to the node-wise stiffness tangent for contact problems
\begin{align*}
\bm{K}^{n}_{C,(m)(j)} = \dfrac{\partial \bm{R}^n_{C,m}}{\partial \bm{d}^n_j}  = \bm{K}^{n}_{(m)(j)} - \dfrac{\partial \bm{F}^n_{C,m}}{\partial \bm{d}^n_j} 
 &= \bm{K}^{n}_{(m)(j)}+\sum_{g=1}^{n^{QP}_b} \left[\left(\kappa^{h,*}_R \bm{n}^{n}_W{\bm{n}^{n}_W}^T + \kappa_S^*\bm{I}\right) \hat{B}^{TP}_m \hat{B}^{TP}_j\right]_{\bm{\xi}=\bm{\xi}^g} L^{g,n} && (2 \times 2)\\
\bm{K}^n_{C,(m)(M+\zeta)} &= \bm{K}^n_{(m)(M+\zeta)} && (2 \times 1)\\
\bm{K}^n_{C,(M+z)(j)} &= \bm{K}^n_{(M+z)(j)} && (1 \times 2)\\
\bm{K}^n_{C,(M+z)(M+\zeta)} &= 0. && (1 \times 1)
\label{eq:contactforce}
\end{align*}

\FloatBarrier
%\nocite{*}
\bibliographystyle{elsarticle-harv}
\bibliography{Bib}
%%%%%%%%%%%%%%%%%%%%%%%%%%%%%%%%%%%%%%%%%%%%%%%%%%%%%%%%%%%%%%%%%%%%%%%%%%%%%%%%%%%%%%%%%%%%%%%%%%%%%%%%%%%%%%%%%%%%%%%

\end{document}